\documentclass[11pt]{article}

\usepackage[margin=1in]{geometry}
\usepackage[T1]{fontenc}
\usepackage{lmodern}
\usepackage{amsmath,amssymb,amsthm,mathtools}
\usepackage{microtype}
\usepackage{enumitem}
\usepackage{xcolor}
\usepackage{booktabs}
\usepackage{tikz}
\usetikzlibrary{shapes.geometric}
\usetikzlibrary{calc,arrows.meta}
\usepackage{aliascnt}
\usepackage{hyperref}
\usepackage[nameinlink,capitalize,noabbrev]{cleveref}

\definecolor{linkblue}{RGB}{0,75,180}
\definecolor{accent}{RGB}{0,120,105}
\definecolor{warm}{RGB}{200,90,30}
\hypersetup{colorlinks=true,linkcolor=linkblue,citecolor=linkblue,urlcolor=linkblue}

\newtheorem{theorem}{Theorem}[section]
\newaliascnt{lemma}{theorem}
\newtheorem{lemma}[lemma]{Lemma}
\aliascntresetthe{lemma}
\newaliascnt{proposition}{theorem}
\newtheorem{proposition}[proposition]{Proposition}
\aliascntresetthe{proposition}
\newaliascnt{corollary}{theorem}
\newtheorem{corollary}[corollary]{Corollary}
\aliascntresetthe{corollary}
\theoremstyle{definition}
\newaliascnt{definition}{theorem}
\newtheorem{definition}[definition]{Definition}
\aliascntresetthe{definition}
\newaliascnt{remark}{theorem}

\aliascntresetthe{remark}
\newtheorem{claim}{Claim}
\newenvironment{claimproof}[1]%
  {\begin{proof}[Proof of #1]}{\end{proof}}

\crefname{theorem}{Theorem}{Theorems}\Crefname{theorem}{Theorem}{Theorems}
\crefname{lemma}{Lemma}{Lemmas}\Crefname{lemma}{Lemma}{Lemmas}
\crefname{proposition}{Proposition}{Propositions}\Crefname{proposition}{Proposition}{Propositions}
\crefname{corollary}{Corollary}{Corollaries}\Crefname{corollary}{Corollary}{Corollaries}
\crefname{definition}{Definition}{Definitions}\Crefname{definition}{Definition}{Definitions}
\crefname{remark}{Remark}{Remarks}\Crefname{remark}{Remark}{Remarks}
\crefname{figure}{Figure}{Figures}\Crefname{figure}{Figure}{Figures}
\crefname{table}{Table}{Tables}\Crefname{table}{Table}{Tables}

\numberwithin{equation}{section}
\allowdisplaybreaks
\setlist{itemsep=2pt,topsep=4pt}

\newcommand{\eps}{\varepsilon}
\newcommand{\yes}{\textup{\textsc{yes}}}
\newcommand{\no}{\textup{\textsc{no}}}
\newcommand{\Dyes}{\mathcal D_{\mathsf y}}
\newcommand{\Dno}{\mathcal D_{\mathsf n}}
\newcommand{\nb}{\mathrm{nb}}
\newcommand{\dtv}{d_{\mathrm{TV}}}
\newcommand{\KL}{\mathrm{KL}}
\newcommand{\trans}{\mathsf{trans}}
\newcommand{\OPT}{\mathrm{OPT}}

\newcommand{\comp}{\operatorname{comp}}

\newcommand{\softO}{\widetilde O}
\newcommand{\ind}{\mathbf 1}
\newcommand{\Ber}{\operatorname{Ber}}
\newcommand{\VO}{\mathsf{VO}}
\newcommand{\EO}{\mathsf{EO}}
\DeclareMathOperator*{\Ex}{\mathbb E}
\DeclareMathOperator{\Prob}{\mathbb P}
\DeclareMathOperator{\poly}{poly}

\newcommand{\SC}{\mathrm{SC}}

\title{Tight Bounds for Sublinear Metric Steiner Tree}
\author{%
  Sepideh Mahabadi\thanks{Microsoft Research. Email: \texttt{smahabadi@microsoft.com}.}
  \and
  Mohammad Roghani\thanks{Google Research. Email: \texttt{roghani@google.com}.}
  \and
  Jakub Tarnawski\thanks{Microsoft Research. Email: \texttt{jatarnaw@microsoft.com}.}
  \and
  Ali Vakilian\thanks{Rutgers University. Email: \texttt{a.vakilian@rutgers.edu}. Supported in part by NSF CAREER Award CCF-2542925.}
}
\date{}

\begin{document}
\pagenumbering{roman}
\maketitle
\thispagestyle{empty}

\begin{abstract}
In the metric Steiner tree problem we are given an $n$-point metric space $(V,w)$
and a set $T\subseteq V$ of terminals, and we seek the smallest weight $\mathrm{OPT}$ of
a tree that spans $T$ and may use points of $V\setminus T$.
We study sublinear algorithms
in the distance-oracle model,
where the algorithm
knows $V$ and $T$ upfront
and can query the distance function $w$.
The objective is to estimate $\mathrm{OPT}$
while querying significantly fewer than $n^2$ distances.

The minimum spanning tree on $T$ estimates $\mathrm{OPT}$ within a factor $2$, and
its weight can be approximated with $\widetilde{O}(n)$ queries. The natural question that has been studied is
therefore the query complexity of estimating $\mathrm{OPT}$ within a factor strictly
below $2$.
Chen, Khanna, and Tan [SODA 2023] gave such an algorithm using $\widetilde{O}(n^{13/7})$ queries,
and a lower bound of $\widetilde\Omega(n^{6/5})$.
Later  Mahabadi, Roghani, Tarnawski, and Vakilian [ITCS 2025] improved the algorithm to $\widetilde{O}(n^{5/3})$,
and Mahabadi, Tarnawski, and Vakilian [ITCS 2027]
improved the lower bound to $\widetilde{\Omega}(n^{7/5})$.

% The best known algorithm uses
% %$\widetilde{O}(n^{13/7})$ and then
% $\widetilde{O}(n^{5/3})$ queries, and the best known lower bound is
% $\widetilde\Omega(n^{6/5})$.

In this work we settle the above question.
We give an algorithm
that
estimates $\OPT$ within a factor $2-\eta_0$, for an absolute constant
$\eta_0>0$,
with query (and time) complexity $\widetilde{O}(n^{3/2})$.
We also show that the exponent $3/2$ is tight.

%\jakub{TODO: decide whether to leave it at that (probably a good option), or add a little bit about the algorithm too}
%\mohammad{let's not add anything about the techniques here}

%{\color{red} option for lower bound}
% Option 1 (one sentence)

%Our lower bound replaces the structured gadgets of earlier constructions by
%randomness: every Steiner point is close to a random set of terminals drawn
%using hidden random terminal weights, and the two hard distributions differ
%only in the distribution of these weights, whose low-order moments agree.

%\ali{Link to Lean's github repo: https://github.com/vakiliana/sublinear-steiner-tree}

\end{abstract}

\clearpage
% \setcounter{tocdepth}{2}
% \tableofcontents
% \clearpage
\pagenumbering{arabic}

\section{Introduction}\label{sec:intro}

The Steiner tree problem asks for a minimum-weight network connecting a
prescribed set of terminals, possibly through additional points. It is one of
Karp's original NP-complete problems~\cite{Karp72}; the best known
polynomial-time approximation ratio is $\ln4+\eps\approx1.39$~\cite{BGRS13},
and no ratio better than $96/95$ is achievable unless
$\mathrm P=\mathrm{NP}$~\cite{CC08}. Steiner problems have also been studied in
streaming~\cite{CJKV22}, dynamic~\cite{IW91,GK14,LOPSZ15},
online~\cite{IW91,GuGK13,AAB96,BC97,BDM22}, and sublinear-query
models~\cite{CKT23,MRTV25}.

An instance of the \emph{metric Steiner tree} problem consists of a finite set
$V$ of $n$ points, a metric $w$ on $V$, and a set $T\subseteq V$ of $k$
\emph{terminals}. A feasible solution is a tree in the complete graph on $V$
whose vertex set contains $T$; we write $\OPT$ for the least weight of a
feasible solution. In the \emph{distance-oracle model} an algorithm knows $V$
and $T$ and may query any pair $\{u,v\}$ to learn $w(u,v)$; its cost is the
number of queries. Let $L$ be the weight of a minimum spanning tree (MST) of
the metric restricted to $T$. Doubling a Steiner tree and shortcutting gives
$\OPT\le L\le2\OPT$, and Czumaj and Sohler~\cite{CS09} showed that $L$ can be estimated
within a factor $1+\eps$ using $\softO(k)$ queries, so a $(2+\eps)$-estimate
needs only a nearly linear number of queries. The natural question that prior works have focused on is the cost
of estimating $\OPT$ within a factor strictly below~$2$.

Chen, Khanna, and Tan~\cite{CKT23} initiated this study and settled much of the
landscape: estimating $\OPT$ within a factor $5/3-\eps$ requires $\Omega_\eps(n^2)$ queries, even when $k=O(1)$,
and explicitly \emph{returning} a tree of weight below $2 \cdot \OPT$ requires $\Omega(nk)$
queries.
This explains why we can only aim for estimation algorithms (that do not return a tree).
They also gave an estimator with a factor $2-\eps_0$ using
$\softO(n^{12/7}+kn^{6/7})=\softO(n^{13/7})$ queries.
Mahabadi, Roghani,
Tarnawski, and Vakilian~\cite{MRTV25} improved this to $\softO(n^{5/3})$ using
improved bounds for sublinear set cover. On the lower-bound side, \cite{CKT23}
showed that $\widetilde\Omega(n+k^{6/5})$ queries are needed, which is
$\widetilde\Omega(n^{6/5})$ when $k=\Theta(n)$.
Mahabadi, Tarnawski, and Vakilian~\cite{MTV26} then gave an improved lower bound of $\widetilde{\Omega}(n^{7/5})$.
See also \cref{tab:prior}.

This problem belongs to a
broader line of work on estimating optimization problems with distance
or adjacency queries, including minimum spanning
tree~\cite{chazelle2005approximating,CS09}, maximal independent
set~\cite{nguyen2008constant,yoshida2009improved,mahabadi2026sublinear},
matching~\cite{yoshida2009improved,kapralov2020space,Behnezhad21,BehnezhadRR23a,BehnezhadRRS-SODA23,behnezhad2024approximating,bhattacharya2023sublinear,MahabadiRT25,azarmehr2025tight},
Steiner forest~\cite{mahabadi2026sublinear}, and metric
TSP~\cite{CKK20,chen2023sublinear}.

\subsection{Our results}

We determine the exponent of the query complexity of estimation below
factor~$2$. Following~\cite{CKT23,MRTV25}, we define an estimate within factor $2-\eta$
to be a number $\widehat C$ with
\begin{equation}\label{eq:window}
  \OPT\ \le\ \widehat C\ \le\ (2-\eta)\,\OPT .
\end{equation}

\begin{theorem}[Upper bound]\label{thm:upper}
There is an absolute constant $\eta_0>0$ and a randomized algorithm that, given
an $n$-point instance, outputs with probability at least $2/3$ a number
$\widehat C$ satisfying \eqref{eq:window} with $\eta=\eta_0$. In the worst
case it makes $\softO(n^{3/2})$ distance queries and uses $\softO(n^{3/2})$ time.
\end{theorem}

% \lean{
% \paragraph{Lean variant (upper bound): scope of the verification.}
% The formalization proves this existential theorem by giving a particular
% causal distance-query program, not by proving that the implementation described
% in black below and that program are identical. The program retains the same
% cover--saving--band--discounted-MST architecture, but replaces some subroutines
% and uses more conservative parameters. The blue notes below describe those
% changes and their proofs; parameters from the two implementations should not
% be mixed. In particular, the formalized program uses
% $\eta_0=1/(2\cdot10^{12})$, succeeds with probability at least $9/10$, and, on
% every execution, has queries and charged work at most
% \[
%  C(n+1)\sqrt{n+1}\bigl(1+\lfloor\log_2(n+2)\rfloor\bigr)^{206}
% \]
% for an absolute constant $C$. Its input is a public enumeration of the
% terminals; all metric-dependent information used at run time comes from
% distance queries. No theorem about an earlier MST estimator or local matching
% algorithm is assumed: the required replacements are proved internally. The
% model is the unit-cost arithmetic RAM specified below, including indexed
% memory and random sampling, not bit complexity or a running-time claim about
% Lean's evaluator. Ordinary Lean/mathlib foundations and library mathematics
% are used. Thus this verifies an explicit algorithm achieving the stated
% result, while the sharper auxiliary claims replaced below are not thereby
% verified for the original implementation.
% }

We also show that the exponent $3/2$ is tight by giving an improved hardness for sublinear algorithms. See \cref{cor:exponent}.

\begin{theorem}[Lower bound]\label{thm:lower}
For every $\eta\in(0,1)$ and every integer $s\ge1$ there are constants
$c_{\eta,s}>0$ and $n_{\eta,s}$ such that the following holds for every
$n\ge n_{\eta,s}$. Every randomized algorithm that, on every $n$-point
instance, outputs with probability at least $2/3$ a number $\widehat C$
satisfying \eqref{eq:window} makes at least $c_{\eta,s}\,n^{b_s}$ queries in
expectation on some instance, where
\[
  b_s=\frac{3s+7}{2s+5}=\frac32-\frac1{4s+10}.
\]
The hard instances have $\Theta_{\eta,s}(n)$ terminals, and every distance
between two distinct points lies in $\{\Theta_{\eta,s}(n^{-3}),1,2\}$.
\end{theorem}

\begin{corollary}\label{cor:exponent}
For every fixed $\eta\in(0,1)$ and $\eps>0$, estimating the metric Steiner
tree cost within a factor $2-\eta$ requires $\Omega_{\eta,\eps}(n^{3/2-\eps})$
queries. In particular, for the factor $2-\eta_0$ of \cref{thm:upper} the
query complexity is $n^{3/2\pm o(1)}$.
\end{corollary}

\begin{table}[t]
\centering\small
\begin{tabular}{@{}llll@{}}
\toprule
Task & Lower bound & Upper bound & Reference\\
\midrule
Estimation within $2+\eps$ & $\Omega(k)$ & $\softO(k)$ & \cite{CS09}\\
Estimation within $2-\eps$ & $\widetilde\Omega(n+k^{6/5})$
  & $\softO(n^{13/7})$ & \cite{CKT23}\\
\quad ~
  & ~
  & $\softO(n^{5/3})$ &  \cite{MRTV25} \\[2pt]
\quad ~
  & $\widetilde\Omega(n^{7/5})$
  & ~ &  \cite{MTV26} \\[2pt]
\quad\emph{\textbf{this paper}}
  & $\boldsymbol{\Omega_{\eps,\eps'}(n^{3/2-\eps'})}$
  & $\boldsymbol{\softO(n^{3/2})}$ & \cref{thm:upper,thm:lower}\\[2pt]
Estimation within $\tfrac53-\eps$ & $\Omega_\eps(n^2)$ & $O(n^2)$ & \cite{CKT23}\\
\midrule
Return a tree of weight $\le(2-\eps)\OPT$ & $\Omega(nk)$
  & $O(nk)$ & \cite{CKT23}\\
\bottomrule
\end{tabular}
\caption{Bounds for metric Steiner tree in the distance-oracle model. The upper
bounds below factor $2$ hold for one particular (very small) constant $\eps_0>0$; our lower
bound holds for every fixed $\eps\in(0,1)$ and every fixed $\eps'>0$, and the
instances have $k=\Theta(n)$.}\label{tab:prior}
\end{table}

The $n^{3/2}$ query complexity is thus tight
up to subpolynomial factors.
The lower
bound is stated for expected query complexity; in particular it also holds for
worst-case complexity.

\paragraph{Lean formalization}

Both of our results are formalized in Lean. The formalization is available under
\url{https://github.com/vakiliana/sublinear-steiner-tree}.

\medskip \medskip
In the following \cref{sec:tech_overview_algorithm,sec:overview-lb} we give an intuitive overview of the techniques underlying our results.

\subsection{Technical Overview of the Algorithm}
\label{sec:tech_overview_algorithm}

In this section, we provide an overview of the techniques used to design our algorithm of \Cref{thm:upper}. We begin with the framework of Chen, Khanna, and Tan~\cite{CKT23} and its implementation by Mahabadi, Roghani, Tarnawski, and Vakilian~\cite{MRTV25}, which reduces estimation below factor 2 to estimating a set cover quantity and estimates that quantity by estimating the size of a maximal matching in an auxiliary graph. We then describe the new idea that brings the cost of this set cover step from $\widetilde{O}(n^{5/3})$ down to $\widetilde{O}(n^{3/2})$. As we will see, this idea alone does not yield an $\widetilde{O}(n^{3/2})$ algorithm for Steiner tree, and we describe the remaining challenges and how we overcome them step by step. Throughout, $n$ is the number of points, $k\le n$ the number of terminals, and for simplicity we assume $k=\Theta(n)$, which is the hardest case. A nonterminal point is called a \emph{center}.

\subsubsection{Overview of the Previous Algorithms}

\paragraph{Starting point: the terminal MST and the set cover reduction.} Let $L$ be the weight of a minimum spanning tree on the terminals. It is well known that $\OPT\le L\le 2\,\OPT$, and $L$ can be estimated within a factor $1+\varepsilon$ using $\widetilde{O}(k)$ queries~\cite{CS09}. Therefore the whole difficulty of beating factor~$2$ is to certify, with few queries, a \emph{discount} $D\le L-\OPT$ that is a constant fraction of $L$ whenever such a discount exists. The algorithm then outputs $L-D$. The framework of~\cite{CKT23} looks for two types of improvements over the MST. For the first type, fix a scale $a$ and consider the \emph{clusters} of terminals at this scale, i.e., the connected components of the graph on terminals in which two terminals are adjacent if their distance is less than $a$. The MST spends roughly $a$ to connect each of these clusters to the rest. Now, if a center $v$ is close to three or more clusters, we can instead connect all of them to $v$ using short edges, which is cheaper. For the second type, we look for a small group of terminals (at most four) inside one cluster that can be connected through a single center more cheaply than the MST connects them. The main structural result of~\cite{CKT23} is that these two types of improvements are the only ones that matter: if neither type yields a significant improvement at any scale, then $L\le(2-\eta)\,\OPT$, and it suffices to output an estimate of $L$.

The first type of improvement is naturally a set cover problem. Fix a scale $a$ and let the universe $U$ be the set of clusters at scale $a$. For each center $v$, let $S_v\subseteq U$ be the clusters that have a terminal within distance $5a/8$ of $v$. A set $S_v$ is useful only if $|S_v|\ge 3$, since connecting two clusters through a center is never cheaper than connecting them directly. We call such sets \emph{eligible} (see \Cref{fig:capping}(a)). Let $\mathcal F$ be the collection of all sets $S_v$, and write $\SC(U,\mathcal F)$ for the least number of eligible sets and singletons needed to cover $U$. The total improvement of the first type at scale $a$ is captured by the \emph{saving}
\[
  \Phi \;=\; |U| - \SC(U,\mathcal F).
\]
Intuitively, the MST has $|U|-1$ edges of length at least $a$. Once centers are allowed, only $\SC(U,\mathcal F)-1$ of them are still needed, as the others can be replaced by short edges to centers. Thus $\Phi$ counts the long MST edges at this scale that we save. \cite{CKT23} show that it suffices to estimate $\Phi$ at every scale within a constant factor plus an additive $\varepsilon|U|$, using only \emph{membership queries} ``is cluster $C$ in $S_v$?'', each of which is implemented by a few distance queries. They give such an estimator with $\widetilde{O}(|\mathcal F|^{7/4})$ membership queries, which leads to their $\widetilde{O}(n^{13/7})$ bound.

\paragraph{Estimating the saving via maximal matching.} Both~\cite{CKT23} and~\cite{MRTV25} estimate the saving through a maximal matching in an auxiliary graph, and we follow the more efficient implementation of~\cite{MRTV25}. Consider the multigraph $H$ whose vertices are the elements of $U$, with one edge between $e$ and $e'$ for every eligible set containing both (\Cref{fig:capping}(b)). Let $\mathcal M$ be any maximal matching of $H$. Then
\[
  |\mathcal M| \;\le\; \Phi \;\le\; 2|\mathcal M| .
\]
The left inequality holds because the $|\mathcal M|$ sets labelling the matched edges cover $2|\mathcal M|$ elements. The right one holds because the unmatched vertices form an independent set, so no eligible set contains two of them, and each of them needs its own set in any cover. The size of a \emph{random greedy} maximal matching can be estimated by the local algorithm of~\cite{nguyen2008constant, yoshida2009improved, Behnezhad21}: sample a few uniformly random vertices and, for each, decide whether it is matched by exploring its incident edges in random rank order and recursing on the other endpoints. For a uniformly random starting vertex the exploration visits about $\bar d$ vertices in expectation, where $\bar d$ is the average degree of $H$. Each visited vertex costs about $n$ membership queries, since one has to query all $n$ sets to learn which of them contain the element.\footnote{This is not the only cost. The local algorithm also needs the incident edges of a visited vertex $v$ in random order, and in the membership model a random incident edge is found by querying random pairs of a set containing $v$ and an element, which costs about $n\cdot(\text{number of sets containing } v)/\deg(v)$ queries per edge. A substantial part of~\cite{MRTV25} is devoted to making this work. They extend the analysis of~\cite{Behnezhad21} to multigraphs, so that such pairs yield uniformly random incident edges, and they show that for a uniformly random starting vertex the exploration asks for only about $\deg(v)/n$ incident edges of $v$ in expectation, so that the degree cancels. We reuse this machinery, and since the resulting cost turns out to be dominated by the cost of learning the sets of the visited vertices, we focus on the latter.} Thus, one local query costs roughly $\bar d\cdot n$, and
\[
  \bar d \;\le\; (\text{number of sets containing an element})\times(\text{number of elements in a set}).
\]
Everything is therefore about making both factors small, cheaply. \cite{MRTV25} sparsify both. Sets with more than $n^{x}$ elements are detected by testing each set against $\widetilde{O}(n^{1-x})$ random elements, taken into the cover, and their elements deleted. Symmetrically, elements lying in more than $n^{x}$ sets are detected by testing $\widetilde{O}(n^{1-x})$ random sets against all elements, covered by those sets, and deleted. Each step costs $\widetilde{O}(n^{2-x})$ queries and adds $\widetilde{O}(n^{1-x})$ sets to the additive error. Then, we have $\bar d\le n^{2x}$, one local query costs $\widetilde{O}(n^{1+2x})$, and balancing against the sparsification cost $n^{2-x}$ gives $x=1/3$, i.e.\ $\widetilde{O}(n^{5/3})$.

\subsubsection{Improving the Set Cover Estimator: Capping Instead of Removing}

\paragraph{Challenge (1): the exploration cost is a product of two thresholds.} \cite{MRTV25} bring both factors of $\bar d$ down to $n^{x}$, and each of the two sparsification steps costs $\widetilde{O}(n^{2-x})$ queries. The exploration cost $n^{1+2x}$ grows with the product of the two thresholds, so at the balance point $x=1/3$, lowering either threshold further costs more queries than it saves. To go below $n^{5/3}$ we need to make one of the two factors small without paying for it. We show that this is possible for the frequency of elements: it can be replaced by a constant at no cost in queries and with a negligible loss in the estimate.

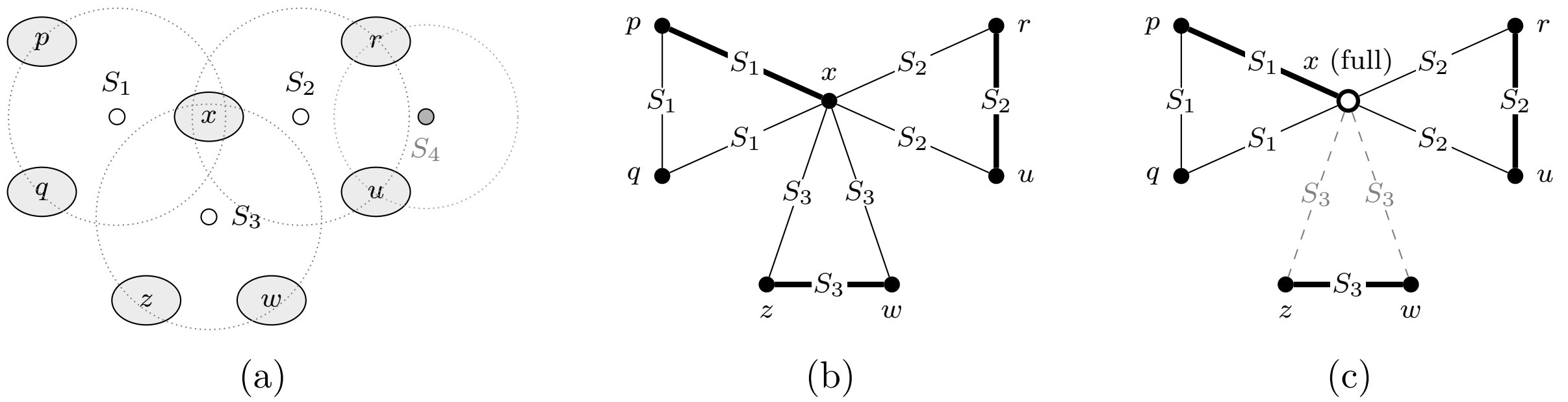
\begin{figure}[t]
  \centering
  % ---------- (a): one scale, clusters and centers ----------
  \begin{minipage}[b]{0.36\textwidth}\centering
  \begin{tikzpicture}[scale=0.85,
      cl/.style={ellipse,draw,fill=gray!15,minimum width=7mm,minimum height=5mm,inner sep=1pt,font=\scriptsize},
      ce/.style={circle,draw,fill=white,inner sep=1.6pt},
      ball/.style={draw,densely dotted,gray}]
    \node[cl] (x) at (0,0) {$x$};
    \node[cl] (p) at (-2.0,0.9) {$p$};   \node[cl] (q) at (-2.0,-0.9) {$q$};
    \node[cl] (r) at (2.0,0.9) {$r$};    \node[cl] (u) at (2.0,-0.9) {$u$};
    \node[cl] (z) at (-0.75,-2.2) {$z$}; \node[cl] (w) at (0.75,-2.2) {$w$};
    \node[ce,label={[font=\scriptsize]above:$S_1$}] (c1) at (-1.1,0) {};
    \node[ce,label={[font=\scriptsize]above:$S_2$}] (c2) at (1.1,0) {};
    \node[ce,label={[font=\scriptsize]right:$S_3$}] (c3) at (0,-1.2) {};
    \node[ce,fill=gray!60,label={[font=\scriptsize,gray]below:$S_4$}] (c4) at (2.6,0) {};
    \draw[ball] (c1) circle (1.3); \draw[ball] (c2) circle (1.3);
    \draw[ball] (c3) circle (1.35); \draw[ball,gray!70] (c4) circle (1.1);
  \end{tikzpicture}\par\smallskip (a)
  \end{minipage}\hfill
  % ---------- (b): the multigraph H ----------
  \begin{minipage}[b]{0.30\textwidth}\centering
  \begin{tikzpicture}[scale=0.85,
      v/.style={circle,draw,fill=black,inner sep=1.5pt,label distance=1pt},
      lab/.style={font=\scriptsize,fill=white,inner sep=0.5pt},
      mt/.style={line width=1.6pt}]
    \node[v,label=above:{\scriptsize $x$}] (x) at (0,0) {};
    \node[v,label=left:{\scriptsize $p$}] (p) at (-2.0,0.9) {};
    \node[v,label=left:{\scriptsize $q$}] (q) at (-2.0,-0.9) {};
    \node[v,label=right:{\scriptsize $r$}] (r) at (2.0,0.9) {};
    \node[v,label=right:{\scriptsize $u$}] (u) at (2.0,-0.9) {};
    \node[v,label=below:{\scriptsize $z$}] (z) at (-0.75,-2.2) {};
    \node[v,label=below:{\scriptsize $w$}] (w) at (0.75,-2.2) {};
\draw[mt] (x)--(p) node[lab,pos=.5]{$S_1$}; \draw (x)--(q) node[lab,pos=.5]{$S_1$}; \draw (p)--(q) node[lab,pos=.5]{$S_1$};
    \draw (x)--(r) node[lab,pos=.5]{$S_2$}; \draw (x)--(u) node[lab,pos=.5]{$S_2$}; \draw[mt] (r)--(u) node[lab,pos=.5]{$S_2$};
    \draw (x)--(z) node[lab,pos=.5]{$S_3$}; \draw (x)--(w) node[lab,pos=.5]{$S_3$}; \draw[mt] (z)--(w) node[lab,pos=.5]{$S_3$};
  \end{tikzpicture}\par\smallskip (b)
  \end{minipage}\hfill
  % ---------- (c): the truncated multigraph H_b with b=2 ----------
  \begin{minipage}[b]{0.30\textwidth}\centering
  \begin{tikzpicture}[scale=0.85,
      v/.style={circle,draw,fill=black,inner sep=1.5pt,label distance=1pt},
      full/.style={circle,draw,very thick,fill=white,inner sep=2pt},
      lab/.style={font=\scriptsize,fill=white,inner sep=0.5pt},
      mt/.style={line width=1.6pt},
      gone/.style={dashed,gray}]
    \node[full,label=above:{\scriptsize $x$ (full)}] (x) at (0,0) {};
    \node[v,label=left:{\scriptsize $p$}] (p) at (-2.0,0.9) {};
    \node[v,label=left:{\scriptsize $q$}] (q) at (-2.0,-0.9) {};
    \node[v,label=right:{\scriptsize $r$}] (r) at (2.0,0.9) {};
    \node[v,label=right:{\scriptsize $u$}] (u) at (2.0,-0.9) {};
    \node[v,label=below:{\scriptsize $z$}] (z) at (-0.75,-2.2) {};
    \node[v,label=below:{\scriptsize $w$}] (w) at (0.75,-2.2) {};
    \draw[mt] (x)--(p) node[lab,pos=.5]{$S_1$}; \draw (x)--(q) node[lab,pos=.5]{$S_1$}; \draw (p)--(q) node[lab,pos=.5]{$S_1$};
    \draw (x)--(r) node[lab,pos=.5]{$S_2$}; \draw (x)--(u) node[lab,pos=.5]{$S_2$}; \draw[mt] (r)--(u) node[lab,pos=.5]{$S_2$};
    \draw[gone] (x)--(z) node[lab,text=gray,pos=.5]{$S_3$}; \draw[gone] (x)--(w) node[lab,text=gray,pos=.5]{$S_3$}; \draw[mt] (z)--(w) node[lab,pos=.5]{$S_3$};
  \end{tikzpicture}\par\smallskip (c)
  \end{minipage}
  \caption{(a) The instance at one scale. The universe is the set of clusters $x,p,q,r,u,z,w$, and each center defines the set of clusters within distance $5a/8$ of it, namely $S_1=\{x,p,q\}$, $S_2=\{x,r,u\}$, and $S_3=\{x,z,w\}$. The grey center $S_4$ meets only two clusters and is ineligible. (b) The multigraph $H$, with one edge per eligible set containing both endpoints, so $x$ has degree $6$. A maximal matching (bold) satisfies $|\mathcal M|\le\Phi\le 2|\mathcal M|$. (c) With cap $b=2$, $x$ keeps $S_1,S_2$ and forgets $S_3$, so the dashed edges disappear. Only a full vertex can forget an edge, and the lists of unmatched vertices are disjoint, so at most $2n/b$ unmatched vertices are full.}
  \label{fig:capping}
\end{figure}

\paragraph{Cap the incidence lists.} Instead of removing an element that lies in many sets, we keep it and let it forget most of its sets. Fix an arbitrary global order on the sets, say by index. For each element $e$ let its \emph{list} $L(e)$ be the first $b$ sets in this order that contain $e$, or all of them if there are fewer than $b$, where $b$ is a constant fixed below. Define the truncated multigraph $H_b$ on $U$ by putting an edge between $e$ and $e'$ labelled $S$ if and only if $S$ is eligible and $S\in L(e)\cap L(e')$: both endpoints must remember $S$ (\Cref{fig:capping}(c)). Computing $L(e)$ costs no more than the $n$ queries we already spend when visiting $e$, since we scan the sets in order and stop after $b$ hits. The degree of every vertex of $H_b$ is at most $b$ times the maximum set size, regardless of how many sets contain it, so the frequency factor in $\bar d$ has been replaced by the constant $b$ for free. The question is whether a maximal matching of $H_b$ still estimates $\Phi$.

\paragraph{Why capping is safe.}
The lower bound $|\mathcal M_b|\le\Phi$ is unaffected, as every edge of $H_b$ is an edge of~$H$. The danger is in the upper bound. Two vertices left unmatched by a maximal matching $\mathcal M_b$ of $H_b$ might share an eligible set that one of them forgot. Call a vertex \emph{full} if its list was truncated, i.e.\ $|L(e)|=b$. Because every forgotten set belongs to a full vertex, the same argument applies without modification to the unmatched vertices that are not full: no eligible set contains two of them, and each requires its own set in any cover.

It remains to bound the number of full unmatched vertices, and here is the key observation: the lists of unmatched vertices are pairwise disjoint on eligible sets. Indeed, if an eligible set $S$ belonged to the lists of two unmatched vertices $e,e'$, then $ee'$ would be an edge of $H_b$ with both endpoints unmatched, contradicting maximality. Consequently the lists of all unmatched vertices together contain at most $2n$ entries, since each eligible set appears in at most one of them and each ineligible set, having at most two elements, in at most two. As a full vertex contributes $b$ of these entries, there are at most $2n/b$ full unmatched vertices. Writing $I$ for the set of unmatched vertices we get $\SC\ge|I|-2n/b$, and therefore
\[
  \tfrac12\Phi-\tfrac{n}{b} \;\le\; |\mathcal M_b| \;\le\; \Phi .
\]
Choosing $b= 100n/(\varepsilon|U|)$, which is a constant when $|U|=\Theta(n)$, puts the loss $n/b$ inside the additive error $\varepsilon|U|$.

\paragraph{The new balance.} With the frequency factor gone, only the first sparsification step remains. If we remove sets with more than $P$ elements, at cost $\widetilde{O}(n\cdot n/P)$, the degree of $H_b$ is at most $bP$ and one local query costs $\widetilde{O}(bP\cdot n)$. Balancing $n^2/P$ against $bPn$ gives $P=\sqrt n$ and a total of $\widetilde{O}(n^{3/2})$.

\subsubsection{Why the Improved Set Cover Estimator Alone Is Not Enough}
\paragraph{Challenge (2): the improvement does not plug into the framework of~\cite{CKT23,MRTV25}.}

Recall that the universe $U$ at scale $a$ is the set of clusters at this scale, and that the estimator must estimate $\Phi$ up to an additive error of $\varepsilon|U|$. The budget is proportional to $|U|$ rather than to $n$ because the MST pays about $a|U|$ at scale $a$, so an error of $\varepsilon|U|$ in $\Phi$ costs about $\varepsilon a|U|$ in the discount, which sums to $O(\varepsilon L)$ over the scales. So far we have implicitly assumed $|U|=\Theta(n)$, which is what made the cap $b=100n/(\varepsilon|U|)$ a constant, but at a given scale $|U|$ can be much smaller. If $|U|$ is very small, no estimator is needed, since we can compute the whole instance by brute force, i.e., query the distance from a few representative terminals of each cluster to every center, which costs $\widetilde{O}(n)$ queries per cluster, and then compute $\Phi$ exactly. Within a budget of $\widetilde{O}(n^{3/2})$ this handles $|U|\le\sqrt n$, so the estimator must handle every $|U|$ between $\sqrt n$ and $n$.

Take $|U|=\sqrt n$. Fitting the capping loss $2n/b$ into $\varepsilon|U|$ forces $b\approx n/(\varepsilon|U|)=\sqrt n/\varepsilon$, which is no longer a constant, and keeping the exploration cost $bPn$ at $n^{3/2}$ then requires the large-set threshold to shrink to $P\approx|U|/\sqrt n$, so that $|U|$ cancels in $bP\approx\sqrt n/\varepsilon$. Both parameters must thus be defined relative to $|U|$. This is fine in principle, as sets with more than $|U|/\sqrt n$ elements are detected by testing every center against $\widetilde{O}(\sqrt n)$ uniformly random elements of $U$, but it requires sampling uniformly random clusters cheaply, which the representation of~\cite{CKT23,MRTV25} does not allow. There a cluster is represented by a greedily built net of representative terminals, and the only way to get a uniformly random cluster is to pick a random terminal and run a breadth-first search over the terminals to find its cluster, which costs $\widetilde{O}(k)$ queries per attempt while only a $|U|/k$ fraction of the attempts succeed. For this reason, \cite{CKT23,MRTV25} sample terminals instead, which is free, and call a set large if it contains more than $P$ terminals. The number of large sets is then $k/P$, and since all of them are taken into the cover, the error budget forces $P\gtrsim k/(\varepsilon|U|)$. Together with $b\gtrsim n/(\varepsilon|U|)$ this gives $bP\gtrsim nk/(\varepsilon^2|U|^2)$, in which $|U|$ does not cancel, and balancing $bPn$ against the brute-force cost $n|U|$ brings us back to $|U|\approx n^{2/3}$ and $\widetilde{O}(n^{5/3})$. The problem is a mismatch of units, as the capping loss is counted in clusters while set sizes are counted in terminals. To benefit from capping we therefore need a representation of the universe in which sampling a uniformly random element, estimating $|U|$, and measuring set sizes are all cheap and expressed in the same unit.

\subsubsection{Measuring Everything in Sampled Points}
\paragraph{A random sample of terminals as the unit.}
To overcome this issue, instead of constructing a greedy net, we represent the terminals at scale $a$ using a random subset $N$. Specifically, we include each terminal $u$ in $N$ independently with probability roughly inversely proportional to the number of terminals in a small ball around $u$ (up to a logarithmic factor). With high probability, every terminal is within distance $\rho$ of a point of $N$, where $\rho$ is a small constant fraction of $a$. Moreover, a ball with $d$ terminals contributes only about $\log n$ points to $N$ in expectation, since each of them is kept with probability about $\log n/d$, and the terminals can be covered by $\widetilde{O}(L/a)$ such balls, as any $\rho$-separated set of terminals spans an MST with that many edges of length at least $\rho$ and weight at most $2L$. Hence $|N| \le \widetilde{O}(L/a)$. Crucially, we do not construct $N$ explicitly. Whenever the algorithm first visits a terminal $u$, it estimates the local density around $u$ using a few random distance queries and decides on the fly whether $u \in N$. This allows us to sample a uniformly random point of $N$ (and estimate $|N|$) in $\widetilde{O}(k)$ queries by picking random terminals until one belongs to $N$. We then define the set cover instance at scale $a$ over the connected components of $N$, where a center's set consists of the components with a point of $N$ within distance $5a/8$ of it. By setting both the large-set threshold $P \approx |N|/\sqrt{n}$ and the cap $b \approx n/(\varepsilon|N|)$ in terms of $|N|$, the dependency on $|N|$ cancels in $bP \approx \sqrt{n}/\varepsilon$ at every scale. The only trade-off is an additive error of $\varepsilon|N|$ instead of $\varepsilon|U|$, which contributes at most $\varepsilon a|N| \le \widetilde{O}(\varepsilon  L)$ to the discount per scale and is easily absorbed by setting $\varepsilon = 1/\operatorname{polylog}(n)$.

\paragraph{Challenge (3): detecting large sets using a single reusable sample.}
Because $N$ is defined implicitly and sampling a single point of $N$ takes $\widetilde{O}(k)$ queries, we cannot draw a fresh sample of $\widetilde{O}(\sqrt{n})$ points for each of the $n$ centers during the sparsification step as in~\cite{MRTV25}, which would cost $\widetilde{O}(n^{5/2})$ queries. Instead, we draw a single sample of $\widetilde{O}(\sqrt{n})$ points from $N$ at the start of the scale in $\widetilde{O}(k\sqrt{n}) = \widetilde{O}(n^{3/2})$ queries and reuse it across all $n$ centers, removing a sampled point whenever a selected center is within distance $5a/8$ of it. Although reusing a sample across adaptive steps creates dependencies in general, here each sampled point is deleted as soon as it contributes to selecting a center, and a stability argument shows that this reusable sample still accurately detects all large sets. Afterward, the local matching exploration proceeds as in~\cite{MRTV25} in $\widetilde{O}(bPn) = \widetilde{O}(n^{3/2})$ queries.

Finally, working with the components of $N$ rather than the true terminal clusters introduces a few additional technicalities in certifying the discounts and adapting the structural theorem of~\cite{CKT23}, which we handle in the formal analysis to obtain a $(2-\Omega(1))$-approximation in $\widetilde{O}(n^{3/2})$ queries and time.

\subsection{Technical Overview of the Lower Bound}\label{sec:overview-lb}

In this section, we give an overview of the lower bound construction in \Cref{thm:lower}. By
Yao's principle, it suffices to construct two distributions $\Dyes$ and
$\Dno$ over instances such that, with high probability, the optimal cost
under $\Dno$ is more than $2-\eta$ times the optimal cost under $\Dyes$, and
to show that every deterministic algorithm with few queries sees almost the
same distribution of answers under both. We first describe the instances and
explain why their costs differ. We then give a heuristic calculation that explains the exponent $b_s=(3s+7)/(2s+5)$ which approaches  $3/2$. Finally, we describe the obstacles to turning this calculation
into a proof and how we overcome them, step by step. Throughout, $\eta$ and
$s$ are fixed, $d$ is a large constant, and we ignore constant factors.

\subsubsection{The Hard Instances}

\begin{figure}[t]
\centering
\begin{tikzpicture}[x=1cm,y=1cm,
  sp/.style={rectangle,draw=black!85,fill=black!75,minimum size=5pt,inner sep=0pt},
  spw/.style={rectangle,draw=warm!70!black,fill=warm,minimum size=5.5pt,inner sep=0pt},
  ord/.style={rectangle,draw=black!50,fill=black!25,minimum size=3.5pt,inner sep=0pt},
  ctr/.style={circle,draw=black!60,fill=black!10,minimum size=5.5pt,inner sep=0pt},
  hub/.style={circle,draw=accent!60!black,fill=accent,minimum size=6.5pt,inner sep=0pt},
  fl/.style={draw=black!45,line width=0.45pt},
  sl/.style={draw=accent,line width=0.8pt},
  lab/.style={font=\scriptsize},
  ttl/.style={font=\small\bfseries}]
\foreach \cx/\cy in {-2.4/2.0, -1.6/1.2, -2.4/0.4}{
  \foreach \dx/\dy in {0/0, 0.12/0.05, 0.05/0.13, 0.17/0.16, 0.1/0.24}{
    \node[ord] at (\cx+\dx,\cy+\dy) {};
  }
}
\node[lab,align=center] at (-2.0,-0.35) {ordinary\\clusters};
\node[ttl] at (0.1,2.85) {(a) YES};
\foreach \i in {0,...,9}{ \node[sp] (a\i) at (0.6*\i,2.2) {}; }
\foreach \j in {0,...,5}{ \coordinate (ca\j) at (0.45+0.9*\j,0.3); }
\foreach \i in {0,2,5,7}{ \draw[sl] (ca0) -- (a\i); }
\foreach \i in {1,4,6,9}{ \draw[sl] (ca1) -- (a\i); }
\draw[fl] (ca2) -- (a3); \draw[fl] (ca2) -- (a8);
\draw[fl] (ca3) -- (a0); \draw[fl] (ca3) -- (a6);
\draw[fl] (ca4) -- (a2);
\node[hub] at (ca0) {}; \node[hub] at (ca1) {};
\foreach \j in {2,...,5}{ \node[ctr] at (ca\j) {}; }
\node[lab] at (0.9,-0.1) {stars};
\node[lab] at (3.6,-0.1) {fillers};
\node[lab] at (2.7,2.55) {special terminals};
\begin{scope}[shift={(6.6,0)}]
\node[ttl] at (0.1,2.85) {(b) NO};
\foreach \i in {0,2,5,6,7}{ \node[sp] (b\i) at (0.6*\i,2.2) {}; }
\foreach \i in {1,3,4,8,9}{ \node[spw] (b\i) at (0.6*\i,2.2) {}; }
\foreach \j in {0,...,5}{ \coordinate (cb\j) at (0.45+0.9*\j,0.3); }
\foreach \i in {1,3,4,8}{ \draw[sl] (cb0) -- (b\i); }
\foreach \i in {3,4,8,9}{ \draw[sl] (cb1) -- (b\i); }
\draw[fl] (cb2) -- (b3); \draw[fl] (cb2) -- (b8);
\draw[fl] (cb3) -- (b0); \draw[fl] (cb3) -- (b6);
\draw[fl] (cb4) -- (b2);
\node[hub] at (cb0) {}; \node[hub] at (cb1) {};
\foreach \j in {2,...,5}{ \node[ctr] at (cb\j) {}; }
\node[lab] at (0.9,-0.1) {stars};
\node[lab] at (3.6,-0.1) {fillers};
\end{scope}
\end{tikzpicture}
\caption{The construction, schematically with $d=4$. Squares are terminals,
circles are centers, and lines represent unit distances. Ordinary clusters
are the same in both worlds. Fillers have at most two neighbors and give no
saving. In \yes{} the stars collectively cover almost all special terminals.
In \no{} they are confined to the small set of positive-weight terminals
(orange). The weights, not the geometry or the filler probabilities, distinguish
the two distributions.}
\label{fig:instance}
\end{figure}
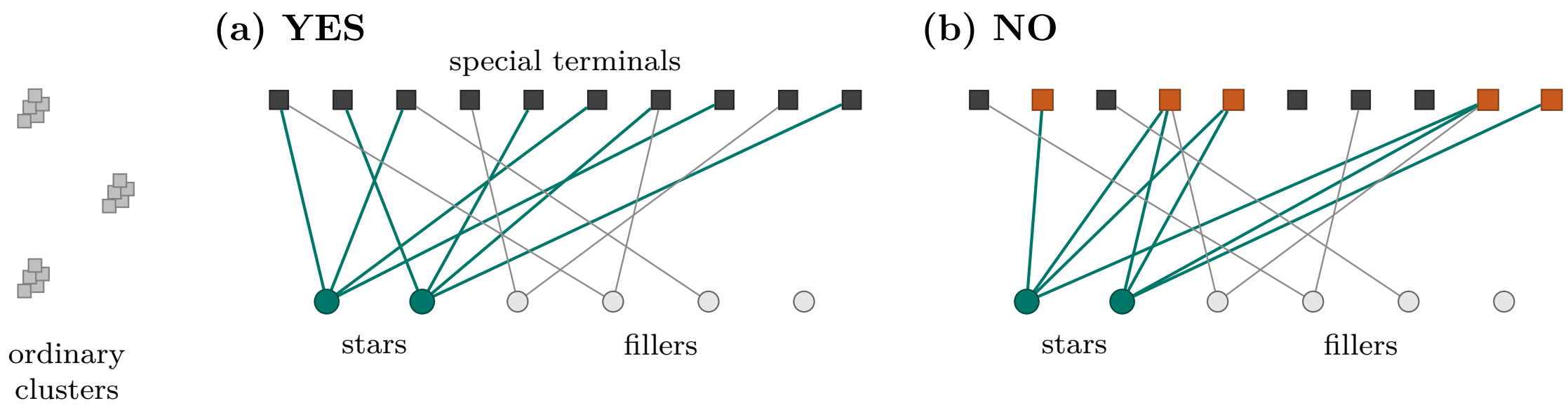

\paragraph{Clusters, special terminals, and centers.}
As in~\cite[Section~7]{CKT23}, the instances hide special terminals among
many more ordinary ones (\cref{fig:instance}). There are $A$ \emph{ordinary
clusters}, each consisting of $R$ terminals at pairwise distance
$\lambda\approx0$, so there are $K=AR$ ordinary terminals. There are $m=dA$
\emph{special terminals}, each at distance $2$ from all other terminals.
Finally, there are $mR$ nonterminal \emph{centers}. Every center $c$ has a
random \emph{neighbor set} $\nb(c)$ of special terminals. It is at distance
$1$ from its neighbors and at distance $2$ from all other points. A uniformly
random bijection decides which terminal plays which role, so the algorithm
knows neither which terminals are special nor how the ordinary terminals are
grouped into clusters. The centers dominate the number of points, so
$n=\Theta(mR)$, while a random terminal is special with probability only
about $m/K=d/R$. The ordinary clusters serve as camouflage. Each of them can
be connected internally at negligible cost, so together they add only about
$2A=2m/d$ to the cost of a Steiner tree, but they make the special terminals
rare and hence expensive to find.

\paragraph{Structured vs. random centers.}
Earlier constructions~\cite{CKT23,MTV26} arrange the centers in public blocks
and hide fixed gadgets in them. In~\cite[Section~7]{CKT23}, a center adjacent
to two special terminals already reveals a star. The construction
of~\cite{MTV26} makes the view of any two terminals uninformative, so that
detecting a star requires three of its terminals, which raises the exponent
from $6/5$ to $7/5$. Its generalization to $\alpha+1$ terminals gives the
exponent $(3\alpha+1)/(2\alpha+1)$, but only for estimation within factor
$\alpha/(\alpha-1)$. We take a different route and replace the structure by
randomness. As we explain next, this decouples the hiding from the cost gap:
we obtain the exponent $(3\alpha+1)/(2\alpha+1)$ with $\alpha=s+2$ for every
factor $2-\eta$.

\paragraph{Stars and the cost gap.}
In our construction, most centers are \emph{fillers}: their neighbor set is empty, a single
special terminal, or a pair of special terminals. Fillers are useless for a
Steiner tree, since connecting two terminals through a center costs $1+1=2$,
the same as connecting them directly. A rare center is a \emph{star}: its
neighbor set consists of $d$ special terminals, which it connects at cost $d$
instead of $2(d-1)$. Each center is a star with probability $\Theta(1/R)$, so
that a special terminal lies in a constant number of stars on average, while
the number of stars is only a small fraction of $m$ because $d$ is large. If
the stars cover almost all special terminals, we can attach every covered
special terminal to a star by a unit edge and then connect the few stars and
the clusters by edges of weight $2$, so $\OPT\approx m$. If the stars touch
only a $\delta$ fraction of the special terminals, for a small constant
$\delta>0$, then almost every special terminal still needs an edge of weight
$2$, so $\OPT\approx2m$. Taking $\delta$ and $1/d$ small compared with $\eta$
makes the ratio larger than $2-\eta$.

\paragraph{The only difference is in random weights.}
The two distributions share all of the above and differ only in where the
stars are placed. Every special terminal $u$ receives an independent random
weight $W_u\ge0$ with mean $1$. Given the weights, each center independently
becomes a star on a $d$-set $H$ with probability proportional to
$\prod_{u\in H}W_u$, while the probabilities of singleton and pair
neighbor sets do not depend on the weights, and the empty neighbor set takes
the remaining probability. In $\Dyes$, every weight is at least $1/2$, so the
stars are spread over all special terminals and cover almost all of them. In
$\Dno$, all but a $\delta$ fraction of the weights are $0$, so all stars lie
among the few terminals of positive weight. As the mean weight is $1$ in both
cases, the expected number of stars is the same. In fact, equal means
already make disjoint stars look the same. Suppose for a moment that $s=1$
and that every weight equals $1$ in $\Dyes$. Given the weights, a fixed
collection of stars occurs with probability proportional to
$\prod_uW_u^{k_u}$, where $k_u$ is the number of these stars that contain
$u$. If the stars are pairwise disjoint, then every $k_u\le1$, and since the
weights are independent with mean $1$, this product has expectation $1$ in
both worlds. If two of the stars share a terminal $u$, then $k_u=2$, and
$\Ex[W_u^2]$ is $1$ in $\Dyes$ but at least $1/\delta$ in $\Dno$, where the
stars crowd onto few terminals. 

We therefore choose
the two weight distributions to have the same first $s$ moments, a standard setup in statistical lower bounds~\cite{WuY19}. For every fixed $s$ this is
compatible with the support conditions above: suitable distributions arise from the
alternating signs of the coefficients in a classical interpolation identity
(\cref{lem:vanishing,lem:weights}). The product form of the star probabilities is what
makes moment matching effective. Roughly speaking, an observation in which
every special terminal belongs to at most $s$ of the observed stars has the
same probability in both worlds, so to tell them apart, the algorithm has to
see some special terminal in $s+1$ stars. Note that the cost gap depends only
on the supports of the two weight distributions, so increasing $s$ improves the hiding
without weakening the gap.

\subsubsection{A Heuristic Count: Where the Exponent Comes From}

We now estimate how many queries an algorithm needs in order to see a special
terminal in $s+1$ stars. This calculation is not a proof, but it determines
the choice of parameters.

\paragraph{Special terminals are expensive to find.}
Since the roles are random, an algorithm can single out a special terminal in
essentially two ways. First, it can query a center against a terminal and
receive the answer $1$. A center has at most two filler neighbors, hidden
among about $K$ terminals, so a query against a terminal about which nothing
is known yet is answered $1$ with probability $O(1/K)$. Second, it can query
a terminal against other terminals until it finds a cluster mate. This takes
about $K/R=A$ queries per terminal, and only a $d/R$ fraction of the
terminals have no cluster mate. Either way, the algorithm pays about $K$
queries for each special terminal that it finds, so with $q\approx KR$
queries it finds about $R$ of them. Let $Q$ be the set of special terminals
found.

\paragraph{Stars hide behind pairs.}
Neither one nor two positive answers at a center reveal a star. A center has
a pair filler with probability $1/4$ but is a star only with probability
$\Theta(1/R)$, so two positive answers at a center are almost always
explained by a pair filler. To recognize a star, the algorithm needs a third
positive answer at the same center. Testing the center against the terminals
in $Q$ provides one only if three members of the star lie in $Q$. Testing it
against terminals that have not been found yet gives a positive answer with
probability $O(1/K)$ per query, and only if the center is a star in the first
place. So this succeeds with probability only $O(1/(KR))$ per query, and the
algorithm needs about $KR$ queries to find even a single star in this way.
Hence, with fewer queries, the algorithm essentially sees only the stars
\emph{visible} from $Q$, namely those with at least three members in $Q$.

\paragraph{Counting visible stars.}  
Every special terminal lies in about $R$ pair fillers, and the other member
of such a pair is a uniformly random special terminal. Thus about
$R|Q|^2/m\approx R^3/m$ centers have two neighbors in $Q$. Each of them is a
star with probability only $O(1/R)$, and a star has a third neighbor in $Q$
with probability $O(|Q|/m)$, so about $R^3/m^2$ stars are visible. Moreover,
a fixed special terminal $u$ is a neighbor of only about $R$ centers. Even if
all of them have a second neighbor in $Q$, the terminal $u$ lies in only
about $R\cdot\frac1R\cdot\frac Rm=\frac Rm$ visible stars. This can
indeed happen, since an adaptive algorithm can put the other members of all
about $R$ pair fillers of $u$ into $Q$. It does not contradict the total
count: $Q$ contains only about $R^3/m$ filler pairs, so only about $R^2/m$
terminals can have all their partners in $Q$ at once, and
$\frac{R^2}m\cdot\frac Rm=\frac{R^3}{m^2}$. Each further
visible star through the same terminal therefore costs a factor of about
$R/m$, and the probability that some special terminal lies in $s+1$ visible
stars is about
\[
  \frac{R^3}{m^2}\Bigl(\frac Rm\Bigr)^s=\frac{R^{s+3}}{m^{s+2}} .
\]

\paragraph{Balancing.}
The two worlds look alike as long as this probability is small, that is, as
long as $R$ is at most a small constant times $m^{(s+2)/(s+3)}$. Since
$n=\Theta(mR)$, this allows $m\approx n^{(s+3)/(2s+5)}$ and
$R\approx n^{(s+2)/(2s+5)}$, and the number of queries is then
$q\approx KR=\Theta(nR)=\Theta(n^{(3s+7)/(2s+5)})$, which is the bound of
\Cref{thm:lower}. As $s$ grows, both $R$ and $m$ approach $\sqrt n$, and the
bound approaches $n^{3/2}$. Heuristically, this is also where constructions
of this type must stop: once $R$ is comparable with $m$, an algorithm with
about $KR\approx n^{3/2}$ queries finds a constant fraction of all special
terminals, and many stars become visible. This is consistent with
\Cref{thm:upper}.

\subsubsection{Hiding the Stars: Keys and Exposure}

The heuristic count makes three simplifying assumptions: that answers carry
no information about the weights unless they reveal a star, that an
algorithm cannot find special terminals and stars faster than described, and
that it suffices to count the stars that are actually seen. The first and the
last are false as stated, the second needs a proof, and all three are
complicated by adaptivity. We now describe how we deal with these issues.

\paragraph{Challenge (1): every answer leaks a little about the weights.}
In the metric, the weights affect every query between a center and a special
terminal, not only those that reveal a star. For example, the first query
$\{c,u\}$ at a center $c$ is answered $1$ slightly more often when $W_u$ is
large, because a star containing $u$ is then more likely. The bias is only
$O(1/(Rm))$ on a probability of order $1/m$, but the algorithm makes many
such queries, and we have to show that the biases do not add up. To this end,
we compare the metric with an auxiliary random process, the \emph{two-key
experiment}, which answers the queries of the algorithm adaptively
(\cref{sec:gated}). In it, every center first draws a filler without using
the weights. If the filler is a pair, its two members are the \emph{keys} of
the center, and the center may also hold a hidden \emph{potential star}: a
$d$-set that contains both keys and is drawn using the weights
(\cref{fig:keys}). The center answers like its filler until both keys have
been queried at it, and only after this \emph{activation} can its potential
star affect the answers. Thus every answer at a center before its activation
is independent of the weights, and a star stays locked until both of its keys
have been found.

Replacing the metric by the two-key experiment changes the distribution of
the transcript by $O(\sqrt{q/(mR^2)}+q/(mR^2))$ in total variation distance.
Before activation, the two processes answer each query with probabilities
that differ by $O(1/(Rm))$, around a base probability of order $1/m$. So
each query contributes $O(1/(mR^2))$ to the relative entropy between the
transcript distributions, and Pinsker's inequality turns the sum over the $q$
queries into the first term. After activation, the two processes differ only
in the probability that a center with two positive answers holds a star.
This probability is $O(1/R)$ in both, and the two values differ by
$O(1/R^2)$. Since only $O(q/m)$ centers are activated in expectation, this
gives the second term. For $q\le c_0KR$ with a small constant $c_0$, both
terms are small, since $KR=\Theta(mR^2)$. This is where the pair fillers are
essential: they make two positive answers at a center an almost certain sign
of a pair, so the first two contacts with a star look like contacts with a
pair. The singleton fillers keep the probability of a positive answer of
order $1/m$ even after many negative answers at the same center, which the
relative entropy bound needs.

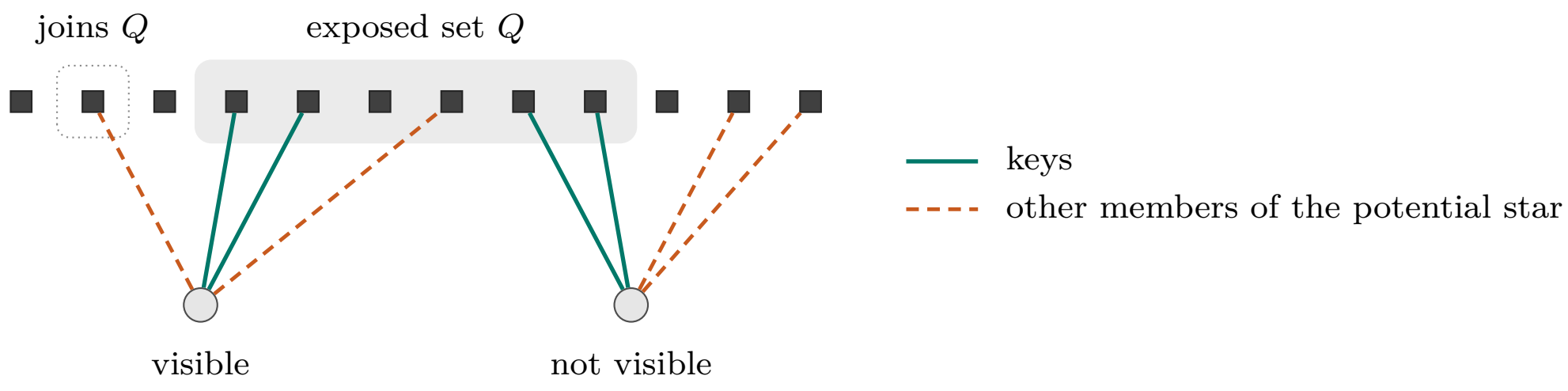
\begin{figure}[t]
\centering
\begin{tikzpicture}[x=1cm,y=1cm,
  sp/.style={rectangle,draw=black!85,fill=black!75,minimum size=5pt,inner sep=0pt},
  ctr/.style={circle,draw=black!70,fill=black!10,minimum size=8pt,inner sep=0pt},
  fl/.style={draw=accent,line width=1pt},
  st/.style={draw=warm,line width=0.9pt,dash pattern=on 3pt off 2pt},
  lab/.style={font=\scriptsize}]
\fill[black!8,rounded corners=4pt] (1.45,1.35) rectangle (5.15,2.05);
\node[lab,anchor=south] at (3.3,2.05) {exposed set $Q$};
\draw[black!45,densely dotted,rounded corners=3pt] (0.3,1.4) rectangle (0.9,2.0);
\node[lab,anchor=south] at (0.6,2.05) {joins $Q$};
\foreach \i in {0,...,11}{ \node[sp] (t\i) at (0.6*\i,1.7) {}; }
\node[ctr] (c1) at (1.5,0) {};
\node[ctr] (c2) at (5.1,0) {};
\draw[fl] (c1) -- (t3); \draw[fl] (c1) -- (t4);
\draw[st] (c1) -- (t6); \draw[st] (c1) -- (t1);
\draw[fl] (c2) -- (t7); \draw[fl] (c2) -- (t8);
\draw[st] (c2) -- (t10); \draw[st] (c2) -- (t11);
\node[lab,below=3pt] at (c1.south) {visible};
\node[lab,below=3pt] at (c2.south) {not visible};
\draw[fl] (7.4,1.2) -- (8.0,1.2);
\node[lab,anchor=west] at (8.1,1.2) {keys};
\draw[st] (7.4,0.8) -- (8.0,0.8);
\node[lab,anchor=west] at (8.1,0.8) {other members of the potential star};
\end{tikzpicture}
\caption{Visibility in the two-key experiment, with $d=4$ and with the ordinary terminals omitted. Squares are special terminals, circles
are centers, and the shaded block is the exposed set $Q$. Solid lines join
each center to its keys, and dashed lines join it to the other members of its
potential star. The left star is visible, since both keys and a third member
lie in $Q$. Closure then adds its last member (dotted) to $Q$. The right star
is not visible, since only its keys lie in $Q$. It can affect an answer only
through a lucky query against one of its unexposed members, or after these
members are exposed for other reasons.}
\label{fig:exposure}
\end{figure}

\paragraph{Challenge (2): bounding what an adaptive algorithm can expose.}
Next, we make the first two steps of the heuristic count rigorous for the
two-key experiment (\cref{sec:reveal}). For the analysis only, we maintain a
set $Q$ of \emph{exposed} special terminals. A special terminal is
\emph{noticed} and added to $Q$ when it first receives a positive answer,
when it has been involved in about $A$ queries, or when one of a few similar
counters becomes large. Moreover, whenever a center has both keys and a third
member of its potential star in $Q$, we add the entire potential star to $Q$. This \emph{closure} only gives the algorithm more information. A query is
\emph{lucky} if it asks a center about an unexposed member of its potential
star. We stop the analysis at the first lucky query, or when one of a few
other unlikely events happens, for example when more than $R$ terminals have
been noticed. Before this stop, no special terminal outside $Q$ has been
involved in a query that met a filler or a potential star. Consequently, the
answers depend only on the bijection, the fillers, and the potential stars
\emph{visible} from $Q$, namely those with both keys and a third member in
$Q$ (\cref{fig:exposure}).

To show that the stop is unlikely, we use an \emph{exchange argument},
similar to those in~\cite{CKT23,MTV26}. Consider a special terminal $u$
outside $Q$. Every earlier query involving $u$ was answered exactly as it
would have been for most ordinary terminals, so swapping the labels of $u$
and of such an ordinary terminal yields another configuration that is
consistent with the transcript so far. Since the bijection is uniform, such
swaps preserve probabilities, and counting them shows that, given the
transcript, a terminal outside $Q$ is special with probability $O(m/K)$ and
belongs to the filler of a given center with probability $O(1/K)$. This
makes the cost of finding special terminals rigorous: $q$ queries notice
$O(q/K)$ special terminals in expectation, so for $q\le c_0KR$, more than
$R$ terminals are noticed only with small probability. A lucky query also
needs a center that holds a potential star. Here we use that a potential star
can be erased without changing the transcript, as long as its center has
given at most two positive answers. Hence conditioning on the transcript can
rule out some potential stars, but never their absence, so a center holds a
potential star with probability $O(1/R)$ even given the transcript, and each
query is lucky with probability $O(1/(KR))$. Finally, the exchange argument
needs a large supply of ordinary terminals to swap with. This supply lasts
unless the algorithm finds the cluster mates of a constant fraction of all
ordinary terminals, which takes about $AK\gg KR$ queries.

\subsubsection{Hiding the Weights: Moment Matching with Bounded Reuse}

\paragraph{Challenge (3): the absence of a star is evidence too.}
By the above, until the first stop the transcript can be computed from the
weight-independent data, namely the bijection and the fillers, together with
the replies to \emph{star requests}, each of which reveals all potential
stars visible from the current set $Q$ (\cref{sec:coupling}). Here $Q$ only
grows and has size at most $2R$. It therefore suffices to show that the
replies have almost the same distribution in the two worlds. Given the
weights, the probability of a sequence of replies is a polynomial in the
weights, and since the weights are independent with matching moments, its
average is the same in both worlds if every weight appears in it with degree
at most $s$. A reply, however, reveals not only the visible stars but also
the absence of all other visible stars, and absence carries information too.
For instance, a terminal of weight $0$ never lies in a star, so each center
that has this terminal as a key and holds no visible star slightly favors
$\Dno$. In the two-key experiment, the probability of such an absence involves
the weights of all candidate supports at the center, and these factors
accumulate over many centers. So the degree of a weight is not controlled by
the number of stars that are actually revealed.

We therefore generate the potential stars differently in the analysis. For
every center with a pair filler and every $d$-set $H$ that contains its keys,
a \emph{proposal} is made independently with a small probability that does
not depend on the weights, and a proposal is retained with probability
$p_H(W)=\prod_{u\in H}(W_u/M)$, where $M$ bounds the weights. The proposal
probability is chosen so that each support is retained with exactly the same
probability as in the two-key experiment. This \emph{proposal model}
differs from the two-key experiment only when a center has two or more retained supports, which happens with probability
$O(1/R^2)$ per center. As we will see, only $O(R^3/m)$ centers have both
keys in $Q$, so this changes the distribution of the replies by only
$O(R/m)$. In the proposal model, a star request reveals all proposals in the
requested region together with their retention bits, rejected proposals
included. The probability of such a record is a weight-independent factor
times a product of factors $p_H(W)$ for retained proposals and $1-p_H(W)$
for rejected ones. The coin with probability $p_H(W)$ is flipped only
when $H$ is proposed, so once the proposals are revealed, the supports that
were not proposed do not depend on $W$. So the degree of $W_u$ is at most the number of revealed
proposals containing $u$, and expanding the products shows that every record
in which no terminal lies in more than $s$ revealed proposals has exactly the
same probability in the two worlds. Thus the two worlds can be coupled until
some terminal lies in $s+1$ revealed proposals.

\paragraph{Challenge (4): bounding reuse when $Q$ is adaptive.}
It remains to show, following the heuristic count, that some terminal lies
in $s+1$ revealed proposals only with small probability. The difficulty is
that $Q$ depends on the transcript, and an adaptive algorithm could try to
steer $Q$ toward sets that contain unusually many filler pairs. We handle
this with a bound that holds uniformly: with high probability, simultaneously
for all sets $Q$ of special terminals, the number of pair fillers with both
members in $Q$ is $O(R|Q|^2/m+|Q|\log m)$, and every special terminal lies in
$O(R)$ pair fillers. For a fixed set of size $h$, the first bound fails with
probability $m^{-\Omega(h)}$ by a Chernoff bound, which beats the at most
$m^h$ sets of that size in a union bound. Hence, up to lower-order terms, every star request, which has
$|Q|\le2R$, concerns a region of proposals of total proposal probability
$O(R^3/m^2)$, in which the proposals containing any fixed terminal have total
probability $O(R/m)$, as in the heuristic count.

Now let $N_u$ be the number of revealed proposals containing $u$. Whether a
proposal is made is independent of everything revealed before it, so each
proposal containing $u$ is revealed with probability equal to its proposal
probability. If $N_u=k$ at that moment, it increases the number of
$(s+1)$-subsets of revealed proposals containing $u$ by $\binom ks$. As the
proposals containing $u$ have total probability $O(R/m)$, this gives
$\Ex\bigl[\binom{N_u}{j+1}\bigr]\le O(R/m)\cdot\Ex\bigl[\binom{N_u}j\bigr]$
for every $j$, and hence
$\Ex\bigl[\binom{N_u}{s+1}\bigr]\le\bigl(O(R/m)\bigr)^s\,\Ex[N_u]$. Summing
over $u$, some terminal lies in $s+1$ revealed proposals with probability
$O(R^3/m^2\cdot(R/m)^s)=O(R^{s+3}/m^{s+2})$, as predicted. The same bound on
the expected number of revealed proposals shows that closure, which adds at
most $d$ terminals per revealed star, adds more than $R$ terminals to $Q$
only with probability $O(R^2/m^2)$. So the exposed set indeed stays of size
at most $2R$.

\paragraph{Putting things together.}
Collecting the errors, for every deterministic algorithm with $q$ queries,
the total variation distance between its transcripts under $\Dyes$ and
$\Dno$ is at most
\[
  O\Bigl(\sqrt{\frac q{mR^2}}+\frac q{KR}+\frac q{AK}
        +\frac{R^{s+3}}{m^{s+2}}+\frac Rm+\frac{\log m}R\Bigr)+m^{-10}.
\]
The first term comes from the two-key experiment; the second from noticing
too many special terminals and from lucky queries; the third from finding too
many clusters; the fourth from the reuse of a weight; and the remaining ones from
the proposal model, closure, and the uniform sparsity bound. Concretely, we
take $A=t^{s+3}$ and $R=\gamma_0t^{s+2}$, so that $R$ is a small constant
times $m^{(s+2)/(s+3)}$, and $q=c_0KR$ for a small constant $c_0$. Then
$n=\Theta(t^{2s+5})$ and the distance is at most $1/10$, so by Le Cam's
two-point method, no algorithm with at most
$c_0KR=\Theta(n^{(3s+7)/(2s+5)})$ queries can estimate $\OPT$ within a
factor $2-\eta$ with probability $2/3$. Padding with extra nonterminals
extends the bound to every $n$, and Markov's inequality extends it to
expected query complexity.

\subsection{Organization}
\Cref{sec:prelim} collects notation and standard facts. \Cref{sec:structure,sec:algorithm} prove \cref{thm:upper}: \cref{sec:structure} proves the structural theorem, and \cref{sec:algorithm} presents and analyzes the $\softO(n^{3/2})$ estimation algorithm. The lower bound is independent of these sections.
\Cref{sec:lb-dist,sec:gated,sec:analysis} prove \cref{thm:lower}:
\cref{sec:lb-dist} constructs the two hard distributions and proves the gap
in their optimal costs, \cref{sec:gated} replaces the metric by a simpler
random process called the two-key experiment, and \cref{sec:analysis}
analyzes this process and completes the proof. We conclude in \cref{sec:conclusion}.

\section{Preliminaries}\label{sec:prelim}

\paragraph{Instances and conventions.}
An instance is a triple $(V,T,w)$ with $|V|=n$ and $|T|=k$. We write $L$ for
the terminal MST weight and $\OPT$ for the Steiner tree cost.
% Requiring a tree
% instead of a connected subgraph loses nothing. Algorithms run on a RAM with
% unit-cost arithmetic, comparisons, and uniform random sampling.
% The notation
% $\softO(\cdot)$ hides fixed powers of $\log(n+2)$, and $\log$ is the natural
% logarithm.
The notation $\softO(\cdot)$ hides fixed powers of $\log(n)$.

\paragraph{Threshold graphs and contraction benefit.}
For $X\subseteq V$ and $a>0$ let $G_a(X)$ be the graph on $X$ whose edges are
the pairs at distance less than $a$, let $\comp(G)$ be the set of connected
components of a graph $G$, and let $\kappa_T(a)=|\comp(G_a(T))|$. Kruskal's
algorithm gives
\begin{equation}\label{eq:kruskal}
  L=\int_0^\infty\bigl(\kappa_T(a)-1\bigr)\,da .
\end{equation}
For a family $\mathcal F$ of subsets of $T$, the \emph{contraction benefit}
$B_T(\mathcal F)$ is the decrease of the MST weight when every set of
$\mathcal F$ is identified at zero cost; for one set $Y$ we write $B_T(Y)$.
For $Y\subseteq T$ let $c(Y)$ be the least weight of a tree spanning $Y$ and
at most one additional point of $V$. For $|Y|\le4$ it is computed from the
distances between $Y$ and $V$ in $O(n)$ time.

\begin{lemma}[Rank monotonicity and lifting]\label{lem:rank}
For every threshold $a$, the decrease of the number of components of $G_a(T)$
caused by a set of zero links, given a fixed set of earlier zero links, does
not increase when $a$ grows or when more earlier links are added. The benefit
is the integral of this decrease over $a$. Moreover, if networks of total
weight $C_{\mathcal F}$ connect each set of $\mathcal F$, then
$L-\OPT\ge B_T(\mathcal F)-C_{\mathcal F}$.
\end{lemma}

\begin{proof}
Apply \eqref{eq:kruskal} before and after adding the zero links. At each
threshold their contribution is their rank in the graphic matroid of the
complete graph on $T$ contracted by the edges of $G_a(T)$ and by the earlier
links; contracting more edges cannot increase it. For the last claim lift an
MST of the contracted metric to the original points and add the networks: this
connects $T$ at weight at most $L-B_T(\mathcal F)+C_{\mathcal F}$.
\end{proof}

\subsection{Probability tools}\label{sec:prob}

All distributions that we compare are on finite sets. Let $P$ and $Q$ be
distributions on a finite set $\Omega$. Their \emph{total variation
distance} is
\[
  \dtv(P,Q)=\max_{E\subseteq\Omega}\bigl|P(E)-Q(E)\bigr|
  =\frac12\sum_{\omega\in\Omega}\bigl|P(\omega)-Q(\omega)\bigr| ,
\]
and it satisfies the triangle inequality. If $Q(\omega)>0$ whenever
$P(\omega)>0$, their \emph{relative entropy} is
\[
  \KL(P\Vert Q)=\sum_{\omega:\,P(\omega)>0}P(\omega)\log\frac{P(\omega)}{Q(\omega)} .
\]
For $x\in[0,1]$, $\Ber(x)$ is the distribution on $\{0,1\}$ that gives $1$
probability $x$. For random objects $T$ and $Z$, we write $\mathcal L(T)$ for
the distribution of $T$, and $\mathcal L(T\mid Z=z)$ for its conditional
distribution given $Z=z$.

\begin{lemma}[Standard facts]\label{lem:prob}
Let $P$ and $Q$ be distributions on a finite set.
\begin{enumerate}[label=(\alph*)]
  \item \emph{Coupling.} If random objects $X$ and $Y$ on one probability
        space have distributions $P$ and $Q$, then
        $\dtv(P,Q)\le\Prob[X\ne Y]$. Moreover, some such pair has
        $\Prob[X\ne Y]=\dtv(P,Q)$.
  \item \emph{Projection.} If $X\sim P$ and $Y\sim Q$, then
        $\dtv\bigl(\mathcal L(\phi(X)),\mathcal L(\phi(Y))\bigr)\le\dtv(P,Q)$
        for every map $\phi$.
  \item \emph{Mixtures.} If random objects $Z$ and $Z'$ have the same
        distribution, then for all random objects $T$ and $T'$,
        \begin{align*}
          \dtv\bigl(\mathcal L(T),\mathcal L(T')\bigr)
          &{}\le\sum_z\Prob[Z=z]\,
             \dtv\bigl(\mathcal L(T\mid Z=z),\,\mathcal L(T'\mid Z'=z)\bigr)\\
          &\le\max_z\dtv\bigl(\mathcal L(T\mid Z=z),\,\mathcal L(T'\mid Z'=z)\bigr),
        \end{align*}
        the sum and the maximum over the values $z$ with $\Prob[Z=z]>0$.
  \item \emph{Chain rule.} Let $P$ and $Q$ be distributions of sequences
        $z=(z_1,\dots,z_q)$ such that $Q(z)>0$ whenever $P(z)>0$. For
        $t\le q$, let $P(\cdot\mid z_{<t})$ and $Q(\cdot\mid z_{<t})$ be the
        conditional distributions of the $t$th entry given that the first
        $t-1$ entries are $z_1,\dots,z_{t-1}$. Then
        \[
          \KL(P\Vert Q)=\sum_{t=1}^q\Ex_{z\sim P}\Bigl[
            \KL\bigl(P(\cdot\mid z_{<t})\Vert Q(\cdot\mid z_{<t})\bigr)\Bigr].
        \]
  \item \emph{Pinsker's inequality.} $\dtv(P,Q)\le\sqrt{\KL(P\Vert Q)/2}$.
  \item \emph{Bernoulli bound.} For $x\in[0,1]$ and $y\in(0,1)$,
        $\KL\bigl(\Ber(x)\Vert\Ber(y)\bigr)\le(x-y)^2/(y(1-y))$.
\end{enumerate}
\end{lemma}

\begin{proof}
(a) For every event $E$,
$P(E)-Q(E)=\Prob[X\in E]-\Prob[Y\in E]\le\Prob[X\in E,\ Y\notin E]
\le\Prob[X\ne Y]$. For the second part, let
$r(\omega)=\min\{P(\omega),Q(\omega)\}$ and
$\rho=\sum_\omega r(\omega)=1-\dtv(P,Q)$. With probability $\rho$, draw
$X=Y$ from the distribution $r/\rho$; otherwise, draw $X$ from
$(P-r)/(1-\rho)$ and $Y$ from $(Q-r)/(1-\rho)$. The last two distributions
have disjoint supports, so $X\ne Y$ exactly in the second case.

(b) Every event about $\phi(X)$ is an event about $X$.

(c) For every event $E$,
\[
  \Prob[T\in E]-\Prob[T'\in E]
  =\sum_z\Prob[Z=z]\Bigl(\Prob[T\in E\mid Z=z]-\Prob[T'\in E\mid Z'=z]\Bigr),
\]
and each bracket is at most $\dtv(\mathcal L(T\mid Z=z),\mathcal L(T'\mid Z'=z))$.

(d) The hypothesis gives $Q(z_1,\dots,z_t)>0$ whenever
$P(z_1,\dots,z_t)>0$, so all the conditional distributions are defined.
Write $P(z)$ and $Q(z)$ as products of the conditional probabilities of the
entries, take logarithms, and in the $t$th term sum out
$z_{t+1},\dots,z_q$.

(e) See \cite[Lemma~11.6.1]{CT06}.

(f) By $\log z\le z-1$,
\[
  \KL\bigl(\Ber(x)\Vert\Ber(y)\bigr)
  \le x\Bigl(\frac xy-1\Bigr)+(1-x)\Bigl(\frac{1-x}{1-y}-1\Bigr)
  =\frac{x^2}y+\frac{(1-x)^2}{1-y}-1=\frac{(x-y)^2}{y(1-y)} .\qedhere
\]
\end{proof}

\begin{proposition}[Le Cam~\cite{LeCam73}]\label{prop:lecam}
If a fair coin decides whether an observation is drawn from $P$ or from $Q$,
every rule guessing the coin from the observation is correct with probability
at most $(1+\dtv(P,Q))/2$. This holds for transcripts of adaptive algorithms.
\end{proposition}

We use Pinsker's inequality with the chain rule for relative entropy, and the
bound
\begin{equation*}
  \dtv(P,Q)\le\sqrt{\KL(P\Vert Q)/2},\qquad
  \KL(\Ber(x)\Vert\Ber(y))\le\frac{(x-y)^2}{y(1-y)}\quad(0<y<1),
\end{equation*}
which follows from $\log z\le z-1$. We use Chernoff, Hoeffding, and
bounded-differences inequalities in their standard exponential forms; in
particular a sum $X$ of independent Bernoulli variables with mean $\mu$
satisfies $\Ex e^X\le e^{(e-1)\mu}$. Finally, a nonnegative supermartingale
$(M_j)$ with $M_0=1$ satisfies $\Prob[\sup_jM_j\ge1/\beta]\le\beta$.

\section{The Structural Theorem}\label{sec:structure}

Throughout \cref{sec:structure,sec:algorithm}
we fix the constants
\[
  e_0=10^{-4},\quad \chi=2-e_0,\quad p=2\chi-2,\quad b_0=p-1,\quad
  \omega=\tfrac{101}{100},\quad \alpha_0=\delta_0=\tfrac1{100},\quad
  \theta_0=\tfrac1{20},\quad \sigma_0=\tfrac1{1000}.
\]
Thus $b_0=1-2e_0$, $p=2-2e_0$, and $b_0<1<p<\chi<2$.

\paragraph{Savings.}
Consider a finite universe $U$, a list of \emph{rows} (subsets of $U$), and a
fixed set of rows declared \emph{eligible}. The \emph{saving} of this system is
\[
  \max_{\mathcal F}\Bigl(\Bigl|\bigcup_{R\in\mathcal F}R\Bigr|-|\mathcal F|\Bigr),
\]
the maximum being over sublists $\mathcal F$ of eligible rows, including the
empty one. Equivalently, it is $|U|$ minus the least number of eligible rows
and singletons covering $U$. Removing elements from $U$ while keeping the
eligibility of every row fixed decreases the saving by at most the number of
removed elements.

\paragraph{Ideal statistics.}
Fix $a_*>0$ and $u\in[0,1)$ and put $a_i=a_*\omega^{i+u}$ for $i\in\mathbb Z$.
Let $\mathcal Q_i=\comp(G_{a_i}(T))$ be the \emph{exact components} at scale
$i$. The \emph{row} of $v\in V$ at scale $i$ consists of the components
$C\in\mathcal Q_i$ containing a terminal within distance $3a_i/5$ of $v$, and it
is eligible if it contains at least three components. Let $Y_i$ be the saving
of $\mathcal Q_i$ with the rows of all points of $V$. For $C\in\mathcal Q_i$ let
\[
  A_i(C)=\max_{Y\subseteq C,\ 2\le|Y|\le4}\bigl(B_T(Y)-c(Y)\bigr)_+ ,
\]
and define the two scale-weighted statistics
\begin{equation}\label{eq:ideal}
  \Gamma=\sum_ia_iY_i,\qquad
  \Lambda=\sum_ia_i\,\bigl|\{C\in\mathcal Q_i:\ A_i(C)\ge\sigma_0a_i\}\bigr| .
\end{equation}

\begin{theorem}[Structural theorem]\label{thm:structure}
For every $a_*>0$ and $u\in[0,1)$,
\begin{equation}\label{eq:structural-charge}
  L\ \le\ \Bigl(2-\frac{89e_0}{90}\Bigr)\OPT+\frac{1001}{4000}\,\Gamma+e_0\Lambda .
\end{equation}
In particular,
\begin{equation}\label{eq:structural-negative}
  \Gamma\le e_0L/100\quad\text{and}\quad\Lambda\le L/10
  \qquad\Longrightarrow\qquad L\le(2-e_0/2)\,\OPT .
\end{equation}
Both statements remain true if the sums in \eqref{eq:ideal} are restricted to
the indices $i$ with $d_{\min}/2<a_i<2\Delta_T$, where $d_{\min}$ and $\Delta_T$
are the smallest and the largest distance between two distinct terminals.
\end{theorem}

\begin{figure}[t]
\centering
\begin{tikzpicture}[x=1cm,y=1cm,
  term/.style={rectangle,draw=black!85,fill=black!75,minimum size=5.5pt,inner sep=0pt},
  stp/.style={circle,draw=black!75,fill=white,minimum size=6.5pt,inner sep=0pt},
  lab/.style={font=\scriptsize},
  ttl/.style={font=\small\bfseries},
  old/.style={draw=black!60,line width=0.6pt},
  gone/.style={draw=black!40,densely dashed,line width=0.6pt},
  new/.style={draw=accent,line width=1.1pt}]
% (a) a cherry
\node[ttl] at (-1.1,3.1) {(a)};
\draw[old] (0,3.0) -- (0,2.2);
\node[stp] (va) at (0,2.2) {};
\node[stp] (ya) at (0,1.2) {};
\node[term] (xa) at (-0.8,0.2) {};
\node[term] (xpa) at (0.9,0.2) {};
\draw[old] (va) -- (ya);
\draw[old] (ya) -- (xa);
\draw[old] (ya) -- (xpa);
\node[lab] at (0.18,1.7) {$z$};
\node[lab] at (-0.55,0.8) {$s$};
\node[lab] at (0.62,0.8) {$s'$};
\node[lab] at (-0.3,2.25) {$v$};
\node[lab] at (-0.3,1.25) {$y$};
\node[lab] at (-0.8,-0.15) {$x$};
\node[lab] at (0.9,-0.15) {$x'$};
% (b) promoting a fringe
\node[ttl] at (2.6,3.1) {(b)};
\node[stp] (pb) at (4.2,3.0) {};
\node[stp] (vb) at (4.2,2.2) {};
\node[stp] (yb) at (3.5,1.2) {};
\node[term] (xb) at (3.0,0.2) {};
\node[term] (xpb) at (4.0,0.2) {};
\node[term] (x2b) at (5.1,1.0) {};
\draw[old] (pb) -- (vb);
\draw[old] (vb) -- (yb);
\draw[old] (yb) -- (xb);
\draw[gone] (yb) -- (xpb);
\draw[old] (vb) -- (x2b);
\draw[new] (xb) -- (xpb);
\node[lab] at (4.4,2.6) {$z$};
\node[lab] at (3.9,2.25) {$v$};
\node[lab] at (3.2,1.25) {$y$};
\node[lab] at (3.0,-0.15) {$x$};
\node[lab] at (4.0,-0.15) {$x'$};
\node[lab] at (5.1,0.65) {$x_2$};
% (c) a short-core star
\node[ttl] at (6.3,3.1) {(c)};
\node[stp] (c1) at (7.4,2.2) {};
\node[stp] (c2) at (8.2,2.45) {};
\node[stp] (c3) at (9.0,2.2) {};
\draw[draw=black!75,line width=1.4pt] (c1) -- (c2) -- (c3);
\draw[old] (c2) -- (8.2,3.1);
\node[term] (t1) at (7.1,0.6) {};
\node[term] (t2) at (8.2,0.6) {};
\node[term] (t3) at (9.4,0.6) {};
\draw[old] (c1) -- (t1);
\draw[gone] (c2) -- (t2);
\draw[gone] (c3) -- (t3);
\draw[new] (t2) to[bend left=35] (t1);
\draw[new] (t3) to[bend left=35] (t1);
\node[lab] at (7.1,0.25) {$x_1$};
\node[lab] at (8.45,0.75) {$x_2$};
\node[lab] at (9.65,0.75) {$x_3$};
\node[lab] at (8.2,2.85) {core};
\end{tikzpicture}
\caption{Steps in the proof of \cref{thm:structure}; squares are terminals and
circles are nonterminals. (a)~A cherry attached at $v$, with trunk $z$, arms
$s\le s'$, primary port $x$, length $r=z+s$, and cost $r+s'$. (b)~Promoting a
fringe: in every cherry the longer arm (dashed) is replaced by the sibling
edge (green), and the cherry center is then suppressed. (c)~A short-core star:
three limbs attached to a nonterminal core of weight at most $\theta_0$ times
the shortest limb length; the shortest limb is kept, and the other two
(dashed) are replaced by terminal edges to its primary port.}
\label{fig:elimination}
\end{figure}
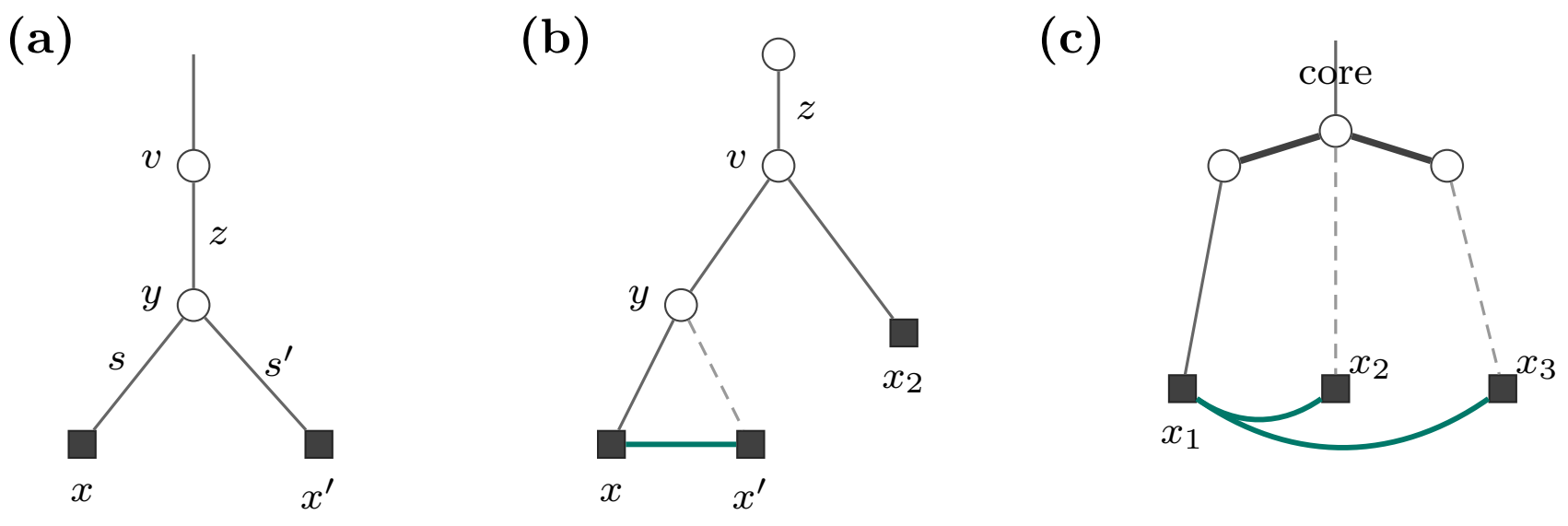

The second statement follows from the first and $L\le2\OPT$, since
$\frac{89}{90}-\frac{1001}{200000}-\frac15>\frac12$. To prove the first we
transform an optimal Steiner tree into a spanning tree on $T$ by local steps,
which are illustrated in \cref{fig:elimination}. The process is a tool of the
analysis and is not executed by any algorithm.

\subsection{Reducing an optimal tree}

Let $K$ be an optimal Steiner tree. \emph{Cleanup} repeatedly deletes
nonterminal leaves and suppresses nonterminals of degree two, replacing a path
$x\,y\,z$ through such a $y$ by the edge $xz$; by the triangle inequality it
does not increase the weight, and afterwards every nonterminal has degree at
least three. Let $E_T(K)$ be the edges of $K$ joining two terminals and
$E_S(K)$ the remaining edges. Edges of $E_T(K)$ are never removed, and the
components of $(T,E_T(K))$ are called \emph{permanent groups}; they only merge
during the process. We track the potential
\[
  \mathcal V(K)=w(E_T(K))+\chi\,w(E_S(K)),
\]
which is initially at most $\chi\OPT$; cleanup does not increase it.

\paragraph{Limbs.}
Let $v$ be a nonterminal. A \emph{direct limb} at $v$ is a tree edge $vx$ with
$x\in T$; its length and cost are $r=W=w(v,x)$, and $x$ is its primary port. A
\emph{cherry} at $v$ is a nonterminal $y$ of degree three adjacent to $v$ and to
two terminals $x,x'$, where $s=w(y,x)\le s'=w(y,x')$; its trunk is $z=w(v,y)$,
its length $r=z+s$, its cost $W=r+s'$, its primary port $x$, and its ports $x$
and $x'$. A limb is \emph{attached} at $v$, and its \emph{region} is the vertex
set of the component of $K$ minus the limb's edge at $v$ that does not
contain $v$. A \emph{fan} is a nonterminal $v$ of degree three whose neighbors
are a nonterminal $v'$, called its parent, and two limbs attached at $v$; its
region is the component of $K-vv'$ containing $v$.

\paragraph{Good reconnections.}
The following steps are performed whenever the added terminal edges weigh at
most $\chi$ times the removed edges; each is followed by cleanup.
\begin{enumerate}[label=(G\arabic*),leftmargin=2.6em]
  \item \emph{One edge:} remove $e\in E_S(K)$ and add a terminal edge joining
        the two components of $K-e$.
  \item \emph{Cherry:} remove the three edges at a cherry center $y$, add
        $xx'$, and add a terminal edge from the region of the cherry to its
        complement.
  \item \emph{Fan:} remove a fan $v$ together with the edges at $v$ and at its
        cherry centers; add the sibling edges of its cherries and the edge
        between its two primary ports, of total weight $I$, and one terminal
        edge from the region of the fan to its complement.
\end{enumerate}
Each step keeps a tree, does not increase $\mathcal V$ (removed edges have
coefficient $\chi$, added ones coefficient $1$), and strictly decreases
$|E_S(K)|$.

\begin{lemma}[Cut barriers]\label{lem:barriers}
Suppose that no good reconnection applies.
\begin{enumerate}[label=(\roman*)]
  \item If $x,x'\in T$ and $e\in E_S(K)$ lies on the tree path between them,
        then $w(x,x')>\chi\,w(e)$. Consequently no component of
        $G_a(T)\cup E_T(K)$ has terminals on both sides of an edge
        $e\in E_S(K)$ with $\chi w(e)\ge a$.
  \item In every cherry $s'<s/(\chi-1)$, and every limb has
        $W\le(1+\frac1{\chi-1})\,r$.
  \item Every terminal pair with exactly one terminal in the region of a limb
        of length $r$ is at distance greater than
        $\chi z+(\chi-1)(s+s')\ge pr$, where $z=r$ and $s=s'=0$ for a direct
        limb. Hence, for $a\le pr$, no component of $G_a(T)\cup E_T(K)$ meets
        both the region and its complement.
\end{enumerate}
\end{lemma}

\begin{proof}
(i) Otherwise (G1) applies to $e$ and $xx'$. A path of $G_a(T)\cup E_T(K)$
between the two sides of $e$ has a step crossing the cut of $e$; it is not a
tree edge, so it is a pair at distance less than $a\le\chi w(e)$.
(ii) Apply (i) to $x,x'$ and $e=yx'$: $s+s'\ge w(x,x')>\chi s'$. Then
$W=r+s'\le r+s/(\chi-1)$.
(iii) For a direct limb this is (i) with $e=vx$. For a cherry, a crossing pair
$f$ with $w(f)\le\chi W-w(x,x')$ would allow (G2). Hence
$w(f)>\chi(z+s+s')-(s+s')=\chi z+(\chi-1)(s+s')\ge2(\chi-1)(z+s)=pr$, using
$\chi\le2$ and $s\le s'$. Terminal edges of $K$ do not cross tree cuts.
\end{proof}

\subsection{Recorded events and their charges}

When no good reconnection applies, the process performs one of the following
events, preferring a star, and then cleans up and exhausts good reconnections
again.

\paragraph{Short-core stars.}
A \emph{short-core star} consists of three limbs with pairwise disjoint regions,
each attached outside the regions of the other two, whose attachment points are
joined by a subtree of nonterminals (the \emph{core}) of weight at most
$\theta_0\ell$, where $\ell$ is the smallest of the three lengths. The event
keeps a limb of length $\ell$ and deletes the edges of the other two limbs; for
each deleted cherry it adds the sibling edge $xx'$, and for each deleted limb it
adds an edge from its primary port to the kept primary port. The \emph{charge}
is the total cost of the two deleted limbs. The event records its \emph{center},
the attachment point of the kept limb, and the index $i$ with
$a_i\le p\ell<\omega a_i$.

\begin{lemma}[Stars]\label{lem:stars}
A star increases $\mathcal V$ by at most $(\frac{21}{10}-\chi)$ times its charge
and permanently joins its three primary ports. The total charge $W_S$ of all
stars satisfies $W_S\le\frac52\Gamma$.
\end{lemma}

\begin{proof}
Let $x_1,x_2,x_3$ be the primary ports, $r_1,r_2,r_3$ the lengths, and $h$ the
kept limb, so $r_h=\ell$. For $j\ne h$, \cref{lem:barriers}(iii) and the tree
path give $pr_j<w(x_j,x_h)\le r_j+\theta_0\ell+\ell$, so
$r_j<(1+\theta_0)\ell/b_0$. Every primary port is within $r_j+\theta_0\ell$ of
the recorded center $v_h$, and since $a_i>p\ell/\omega$,
\[
  \frac{w(v_h,x_j)}{a_i}<\frac\omega p\Bigl(\frac{1+\theta_0}{b_0}+\theta_0\Bigr)
  <\frac35 .
\]
As $a_i\le p\ell\le pr_j$ for all $j$, \cref{lem:barriers}(iii) places the
three primary ports in different components of $G_{a_i}(T)\cup E_T(K)$, hence in
different exact components, and the row of $v_h$ is eligible.

Fix $i$ and consider the bipartite incidence graph $\Psi$ between the stars
recorded at scale $i$ and the exact components $\mathcal Q_i$, each star being
adjacent to the components of its primary ports. When a star is recorded, all
earlier stars have their ports joined by permanent edges, so every connected
component of $\Psi$ lies inside one component of $G_{a_i}(T)\cup E_T(K)$, and the
new star meets three different ones. Hence $\Psi$ is a forest. If it has $f$
stars and $c$ connected components, it touches $u=2f+c$ exact components; the
at most $f$ rows of the recorded centers are eligible and cover them, so
$Y_i\ge u-f\ge f$.

By \cref{lem:barriers}(ii) a deleted limb costs at most
$(1+\frac1{\chi-1})(1+\theta_0)\ell/b_0$, so the charge is below
$4.21\,\ell<\frac52a_i$, and $W_S\le\sum_i\frac52a_iY_i=\frac52\Gamma$.
For a deleted limb $j$ the added edges weigh at most
$(s_j+s'_j)+(r_j+\theta_0\ell+\ell)\le2W_j+\theta_0\ell$, using
$\ell\le r_j$ and $s_j\le s'_j$ (for a direct limb $s_j=s'_j=0$). Since the
charge is at least $2\ell$, the added weight is at most $(2+\theta_0)$ times the
charge, and the removed edges have coefficient $\chi$. The added edges keep the
tree connected and join the ports.
\end{proof}

\paragraph{Promotions.}
Root $K$ at a terminal. A fan is a \emph{fringe} if its region contains no
nonterminal other than the fan and its cherry centers (terminal subtrees below
its ports may be arbitrary). Let its limbs have lengths $r_1,r_2$, put
$R_f=r_1+r_2$, and let $z$ be the weight of its parent edge. If it has at least
one cherry, its \emph{promotion} replaces in each cherry the longer arm $yx'$ by
the sibling edge $xx'$ and suppresses $y$. The charge is the sum of the longer
arms $s'_j$ of its cherries. The promotion is \emph{small} if $s_j<\alpha_0r_j$
for every cherry, and \emph{regular} if it is not small and $z\ge\delta_0R_f$.
Other fringes are not promoted.

\begin{lemma}[Small promotions]\label{lem:small}
A promotion increases $\mathcal V$ by at most $(2-\chi)$ times its charge. The
total charge $W_B$ of small promotions satisfies
$W_B\le\frac{\alpha_0}{(1-\alpha_0)(\chi-1)}\OPT<\OPT/90$.
\end{lemma}

\begin{proof}
The new edge costs at most $s+s'\le2s'$. If $s_j<\alpha_0(z_j+s_j)$ then
$s_j<\alpha_0z_j/(1-\alpha_0)$, and \cref{lem:barriers}(ii) gives
$s'_j<\alpha_0z_j/((1-\alpha_0)(\chi-1))$. A small promotion destroys the
nonterminal--nonterminal edges $vy_j$ of its cherries. The total weight of
nonterminal--nonterminal edges never increases: such edges arise only by
suppressing a nonterminal between two nonterminals, which replaces two of them
by a lighter one. It is initially at most $\OPT$.
\end{proof}

\begin{lemma}[Regular promotions]\label{lem:regular}
Every regular promotion can be assigned a pair $(i,C)$ with $C\in\mathcal Q_i$
and $A_i(C)\ge\sigma_0a_i$, such that its charge is less than $a_i$ and no
terminal set $C$ is assigned twice. Hence the total charge $W_R$ of regular
promotions satisfies $W_R\le\Lambda$.
\end{lemma}

\begin{proof}
By \cref{lem:barriers}(iii) applied to each limb and the other primary port,
$b_0r_1<r_2$ and $b_0r_2<r_1$; with $\ell=\min\{r_1,r_2\}$ this gives
$R_f<2\ell/b_0$. Let $a=a_i$ be the least grid scale above $R_f/(\chi-1)$, so
$a\le\omega R_f/(\chi-1)<2.03\,\ell$.

\emph{All ports lie in one exact component $C$ of $G_a(T)$.} Ports of different
limbs are at distance at most $R_f+\sum_j(s'_j-s_j)<R_f/(\chi-1)$, because
$s'_j-s_j<(\frac1{\chi-1}-1)s_j$. For a cherry $j$, \cref{lem:barriers}(iii)
applied to $x_{3-j}$ gives $\chi z_j+(\chi-1)(s_j+s'_j)<w(x_j,x_{3-j})\le R_f$,
so $w(x_j,x'_j)\le s_j+s'_j<R_f/(\chi-1)$.

\emph{$C$ lies in the region of the fringe.} As (G3) does not apply, every
terminal pair leaving the region is at distance greater than
$\chi(z+W_1+W_2)-I$, where $I\le R_f+\sum_j(s_j+s'_j)$. Since
$(\chi-1)s'_j-s_j\ge-(2-\chi)r_j$, this is at least
$\chi z+(2\chi-3)R_f\ge(\chi\delta_0+b_0)R_f>\omega R_f/(\chi-1)\ge a$.

\emph{Local evidence.} Some cherry $j$ has $s_j\ge\alpha_0r_j$; let
$Y=\{x_j,x'_j,x_{3-j}\}\subseteq C$. Treat the permanent edges as earlier zero
links. For $t\le p\ell$, \cref{lem:barriers}(iii) separates $x_{3-j}$ from the
other two, and for $t\le\chi s_j$, \cref{lem:barriers}(i) for the edge
$yx_j$ separates $x_j$ from $x'_j$. By \cref{lem:rank},
$B_T(Y)\ge p\ell+\min\{\chi s_j,p\ell\}$. Connecting $Y$ through $y$ costs at
most $s_j+s'_j+z_j+r_{3-j}=R_f+s'_j$. Using $s'_j<s_j/(\chi-1)$, $s_j<\ell/b_0$,
and $R_f<2\ell/b_0$ in both cases of the minimum,
\[
  B_T(Y)-c(Y)\ge\Bigl(\chi-\frac1{\chi-1}\Bigr)s_j
  -\Bigl(\frac2{b_0}-p+\frac\chi{b_0}-p\Bigr)\ell
  >0.9997\,\alpha_0\ell-0.0012\,\ell>0.008\,\ell>\sigma_0a .
\]
Thus $A_i(C)\ge\sigma_0a_i$, and the charge is
$\sum_js'_j<R_f/(\chi-1)<a_i$.

\emph{No repetition.} Before the promotion the terminals of the region form at
most four permanent groups, and afterwards at most two. Every later regular
promotion has three ports in three different permanent groups (they are
separated by the edges $yx_j$ and by the limb regions), all in its assigned
component. Since groups only merge, the terminal set $C$ cannot be assigned
again. Finally each promotion raises $\mathcal V$ by at most $(2-\chi)$ times
its charge, as in \cref{lem:small}.
\end{proof}

\subsection{The process terminates}

\begin{lemma}\label{lem:termination}
If $E_S(K)\ne\varnothing$ and no good reconnection applies, then a short-core
star, a small promotion, or a regular promotion exists.
\end{lemma}

\begin{proof}
Contract every permanent group to a vertex and root the tree at a group. A
\emph{core} is a component of the subgraph induced by the nonterminals; choose a
core $\mathcal K$ below which no other core lies. Every vertex of $\mathcal K$
has one parent edge, its other neighbors are its children, and its children
outside $\mathcal K$ are groups whose subtrees contain no nonterminal; edges to
them are direct limbs. The height of a vertex of $\mathcal K$ is the number of
nonterminals on a longest downward path in $\mathcal K$.

Three limbs attached at one vertex form a short-core star with a core of weight
zero; this includes the parent edge if it leads to a group. Assume there is no
star. Then a vertex of height one has exactly two group children and a
nonterminal parent, so it is a cherry at its parent. A vertex of height two has
exactly two children, both limbs and at least one a cherry, and a nonterminal
parent: it is a fan, and by the choice of $\mathcal K$ a fringe. A root of
height at most two would have a group parent, so $\mathcal K$ has a vertex of
height three.

Suppose no fringe can be promoted, so each fringe has a non-small cherry and a
parent edge shorter than $\delta_0R_f<d_0\ell$, where $d_0=2\delta_0/b_0<0.02001$
and $\ell$ is its shortest limb length. Let $v'$ have height three and a
fringe child $v$ with shortest limb length $\ell_1$; as $v'$ has degree at least
three, it has another child. If that child is a limb of length $r$, then
\cref{lem:barriers}(iii) for the shortest limb of $v$ gives
$r>(b_0-d_0)\ell_1$, and two limbs of $v$ together with this limb form a star
whose core, the edge $vv'$, weighs less than $d_0\ell_1<\theta_0\min\{r,\ell_1\}$.
Otherwise it is a fringe with shortest limb length $\ell_2$; we may assume
$\ell_1\le\ell_2$. Then
$(b_0-d_0)\ell_2<(1+d_0)\ell_1$, so $\ell_2<1.042\,\ell_1$, the two parent edges
weigh less than $d_0(\ell_1+\ell_2)<0.041\,\ell_1<\theta_0\ell_1$, and two limbs
of the first fringe with one limb of the second form a star.
\end{proof}

\begin{proof}[Proof of \cref{thm:structure}]
Every event strictly decreases $|E_S(K)|$, so by \cref{lem:termination} the
process ends with a spanning tree of $T$, whose weight is at least $L$ and
equals its potential. Good reconnections and cleanup do not increase
$\mathcal V$, so by \cref{lem:stars,lem:small,lem:regular},
\[
  L\le\chi\OPT+\bigl(\tfrac{21}{10}-\chi\bigr)W_S+(2-\chi)(W_B+W_R)
  \le(2-e_0)\OPT+\tfrac{1001}{4000}\Gamma+e_0\bigl(\tfrac{\OPT}{90}+\Lambda\bigr),
\]
which is \eqref{eq:structural-charge}. Recorded star scales satisfy
$\frac56d_{\min}\le a_i\le\Delta_T$, because an eligible row contains two ports
within $\frac65a_i$ of each other and the ports of a star are separated at
scale $a_i$; a regular scale exceeds $d_{\min}$, and its component is a proper
subset of $T$, so it is below $\Delta_T$. Hence only indices in the stated range
are used.
\end{proof}

\section{The \texorpdfstring{$\softO(n^{3/2})$}{three-halves} Estimation Algorithm}\label{sec:algorithm}

We now prove \cref{thm:upper}. After normalization (\cref{sec:certificates}),
the estimator computes at every scale a lower estimate of the saving on a
random cover, and a lower estimate of the band statistic, and returns an upper
estimate of $L$ minus the larger of the two discounts of
\cref{lem:saving-sound,lem:band-sound}. Throughout, $\beta$ denotes a failure
parameter that is inverse-polylogarithmic in $n$, chosen in \cref{sec:mutual}.

\subsection{Normalization and sound statistics}\label{sec:certificates}

\paragraph{Flooring.}
If $k\le1$ we return $0$, and if $k=2$ the terminal distance. Otherwise we query
one terminal against $T$ and let $R_{\mathrm{met}}$ be the largest answer, so
$R_{\mathrm{met}}\le\Delta_T\le2R_{\mathrm{met}}$ and $R_{\mathrm{met}}\le L_0$, where
$L_0$ and $\OPT_0$ denote the original costs. For a target improvement
$\eta_{\mathrm w}\in(0,1)$ put $\zeta=\eta_{\mathrm w}/100$ and
$\lambda_{\mathrm f}=\zeta R_{\mathrm{met}}/k$, and replace every distance between
distinct points by $\max\{w(x,y),\lambda_{\mathrm f}\}$. This is a metric, one
query of it costs one original query, and terminal distances now lie in
$[\lambda_{\mathrm f},2R_{\mathrm{met}}]$. A cleaned optimal tree has at most $2k-3$
edges, so the new optimum satisfies
$\OPT_0\le\OPT\le\OPT_0+2k\lambda_{\mathrm f}\le(1+4\zeta)\OPT_0$. Hence an
estimate in $[\OPT,(2-\eta_{\mathrm w})\OPT]$ for the floored metric lies in
$[\OPT_0,(2-\eta_{\mathrm w}/2)\OPT_0]$. From now on we work with the floored
metric and write $L,\OPT$ for its costs.

\paragraph{MST estimation.}
By the estimator of Czumaj and Sohler~\cite{CS09}, for a fixed accuracy $\nu>0$
we obtain with probability $1-\beta'$ values with
\begin{equation}\label{eq:mst-interval}
  L/(1+\nu)\le\underline L\le L\le\overline L\le(1+\nu)L
\end{equation}
using $\softO(k\,\nu^{-O(1)}\log(1/\beta'))$ queries and time.

\paragraph{The grid.}
Fix $\delta=10^{-10}$ and $\tau=\sigma_0/100$. Put $q_0=\lceil\log_\omega2\rceil+1$,
$I=\lceil\log_\omega(4R_{\mathrm{met}}/\lambda_{\mathrm f})\rceil+q_0+3$, and
\begin{equation}\label{eq:grid}
  a_i=\lambda_{\mathrm f}\,\Theta\,\omega^{\,i-q_0},\qquad 0\le i<I,
\end{equation}
where $\Theta$ has density $1/(\Theta\log\omega)$ on $[1,\omega)$, so that the
shift is log-uniform (propose $\Theta$ uniformly and accept it with
probability $1/\Theta$). The grid starts below $\lambda_{\mathrm f}/2\le d_{\min}/2$,
ends above $4R_{\mathrm{met}}\ge2\Delta_T$, and its largest scale is below
$5R_{\mathrm{met}}\le5L$. Hence $I=O(\log(k+2))$, $\sum_ia_i\le10L/(\omega-1)$,
and \cref{thm:structure} applies to the finite sums over this grid.

\subsubsection*{The saving statistic on a cover}

For each scale let $N_i\subseteq T$ be a $\rho_i$-cover of $T$, that is, every
terminal is within distance $\rho_i$ of $N_i$, where $0<\rho_i\le\delta a_i$.
Let $\mathcal C_i=\comp(G_{a_i}(N_i))$. The row of $v\in V$ consists of the
components containing a point of $N_i$ within distance $5a_i/8$ of $v$; it is
eligible if it contains at least three components, and this eligibility is
never recomputed after deletions. Let $\Phi_i$ be the saving of $\mathcal C_i$
with these rows. Fix a map $\pi_i:T\to N_i$ that moves points by at most
$\rho_i$ and fixes $N_i$, and let $\mathcal P_i$ be the partition of $T$ into
the preimages of the components of $\mathcal C_i$. Writing $\preceq$ for
refinement,
\begin{equation}\label{eq:sandwich}
  \comp\bigl(G_{a_i-2\rho_i}(T)\bigr)\ \preceq\ \mathcal P_i\ \preceq\ \mathcal Q_i ,
\end{equation}
because a pair at distance less than $a_i-2\rho_i$ is mapped to a pair at
distance less than $a_i$, while edges of $G_{a_i}(N_i)$ and the attachments
$x\pi_i(x)$ are edges of $G_{a_i}(T)$.

\begin{lemma}[The saving is sound]\label{lem:saving-sound}
Let $Q_S$ be a fixed integer with $\frac54\omega^{-Q_S}<1-2\delta$. Then
$\sum_ia_i\Phi_i\le32Q_S\,(L-\OPT)$.
\end{lemma}

\begin{proof}
Fix a scale, write $a=a_i$ and $\rho=\rho_i$, and take a list of eligible rows
attaining $\Phi=\Phi_i$. In its incidence graph between rows and components,
every connected component has saving at least two, since otherwise a single
eligible row would do better. With $u$ components, $z$ rows, and $c$ connected
components, a spanning forest has $f=u+z-c$ edges, and
$f-2z=\Phi-c\ge\Phi/2$. Deleting a row of forest degree at most two does not
decrease $f-2z$; after all such deletions $f\ge3z$. For each forest edge pick a
retained point of the component within $5a/8$ of the row's center. The
resulting stars cost at most $5af/8$, and their identifications have rank
$f-z$ in $\mathcal P_i$, hence by \eqref{eq:sandwich} at every threshold below
$a-2\rho$. With $r=f-2z\ge z$ their net benefit is at least
\[
  (a-2\rho)(f-z)-\tfrac58af\ \ge\ a\bigl[(\tfrac38-2\delta)r-(\tfrac14+2\delta)z\bigr]
  \ \ge\ \tfrac1{16}ar\ \ge\ \tfrac1{32}a\Phi .
\]
Now process the scales of one residue class modulo $Q_S$ in increasing order.
An earlier star has diameter at most $\frac54\omega^{-Q_S}a<a-2\rho$, so it lies
inside one component of $G_{a-2\rho}(T)$; therefore, at every threshold
$t\le a-2\rho$, each component of $G_t(T)$ together with all earlier links still
lies in one part of $\mathcal P_i$, and the marginal rank of the current stars is
still $f-z$. Telescoping and \cref{lem:rank} bound the sum over the class by
$32(L-\OPT)$; summing the $Q_S$ classes proves the claim.
\end{proof}

\subsubsection*{The band statistic}

For $C\in\mathcal C_i$ and $Y\subseteq C$, let $B_C(Y)$ be the integral over
$t\in[\tau a_i,\,a_i-2\rho_i]$ of the number of components of $G_t(C)$ meeting
$Y$, minus one, and define the \emph{band score}
\begin{equation}\label{eq:band-score}
  D_i(C)=\max\Bigl\{0,\ \max_{Y\subseteq C,\ 2\le|Y|\le4}
  \bigl[B_C(Y)-c(Y)-6\rho_i\bigr]\Bigr\}.
\end{equation}
If $|C|\le B$, it is computed from the distances between $C$ and $V$ in
$O(nB^4+\poly(B))$ time, and $0\le D_i(C)\le3a_i$. Let $\beta_X(x,y)$ be the
bottleneck distance between $x$ and $y$ in the complete graph on $X$. For
$x,y\in C$ we have $\beta_T\le\beta_C\le\beta_T+2\rho_i$: if
$\beta_T(x,y)<a_i-2\rho_i$, project a bottleneck path to $N_i$; otherwise use that
$C$ is connected in $G_{a_i}(N_i)$. The integrand above is determined by the at
most three merge heights of $Y$ in the bottleneck hierarchy, clipped to the band,
so $B_C(Y)$ exceeds the corresponding quantity for $T$ by at most $6\rho_i$.

\begin{lemma}[The band statistic is sound]\label{lem:band-sound}
Let $Q_Q$ be a fixed integer with $\omega^{-Q_Q}<\tau/2$. For all subfamilies
$\mathcal F_i\subseteq\mathcal C_i$,
$\sum_i\sum_{C\in\mathcal F_i}D_i(C)\le Q_Q\,(L-\OPT)$.
\end{lemma}

\begin{proof}
Fix a residue class modulo $Q_Q$ and process its scales in increasing order,
identifying an attaining set $Y_C$ for each $C$ with $D_i(C)>0$. At a threshold
$t$ in the band, sets in different components of $\mathcal C_i$ lie in different
parts of $\comp(G_t(T))$ by \eqref{eq:sandwich}, so their ranks add. A set from an
earlier scale $a'\le\omega^{-Q_Q}a_i<\tau a_i$ lies in a component of
$G_{a'}(N_{i'})$, hence in one component of $G_t(T)$, so it does not change ranks
in the current band. The marginal benefit of the current sets is therefore at
least $\sum_C(B_C(Y_C)-6\rho_i)$, and \cref{lem:rank} with the one-point networks
of cost $c(Y_C)$ bounds the sum over the class by $L-\OPT$.
\end{proof}

\subsection{Local random-greedy matching on multigraphs}\label{sec:matching}

Let $H$ be a finite loopless multigraph on a vertex set $\mathcal U$ with
$m=|\mathcal U|\ge1$ and with $p$ labeled edge copies, and put $N_H=m+p+2$. A
uniformly random order of the copies (their \emph{ranks}) defines the
random-greedy maximal matching, which scans the copies by increasing rank and
adds a copy whenever both endpoints are free. We use the local oracles
of~\cite{nguyen2008constant,yoshida2009improved,Behnezhad21}. For a copy $e$
with endpoint $u$, the directed oracle $\EO(e,u)$ decides whether $u$ is still
unmatched just before the rank of $e$: it scans the other copies $f$ at $u$ of
smaller rank in increasing order, calls $\EO(f,u_f)$ for their other endpoints
$u_f$, and answers ``no'' at the first positive call. The vertex oracle
$\VO(v)$ scans the copies $e=vu$ at $v$ in increasing order and returns the
first one with $\EO(e,u)=$``yes'', or ``unmatched''. Answers are memoized
within one call of $\VO$. A copy reported by $\VO(v)$ is the matched copy of $v$.

\begin{lemma}[Call bound]\label{lem:calls}
Fix $H$ and a directed copy $(e,u)$, and let $Q(e,u)$ be the number of
invocations of $\EO(e,u)$, cached ones included, summed over separate calls
$\VO(v)$ for all $v\in\mathcal U$. Over a uniformly random order,
$\Ex Q(e,u)=O(\log N_H)$. Consequently, for a uniformly random root independent
of the order, the expected number of oracle invocations is
$O((1+p/m)\log N_H)$, and the expected number of invocations at a fixed vertex
$x$ is $O((1+d_x)\log N_H/m)$, where $d_x$ is its degree.
\end{lemma}

\begin{proof}
The argument is that of~\cite[Section~3]{Behnezhad21}, adapted to parallel
copies (see also~\cite[Section~5]{MRTV25}). Within one call of $\VO$ every
directed state is evaluated once and invokes each lower-ranked state at most
once, so $Q(e,u)\le m(p+1)\le N_H^2$ for every order.

The stack of an invocation is a walk $w_0,w_1,\dots,w_\ell=u$ along copies
$e_1,\dots,e_\ell=e$ of decreasing rank. It is a simple path: at a first repeated
vertex the closing copy has smaller rank than the copy used at the earlier
visit, so it was tested there, and its answer ``no'' certifies a lower-ranked
matched copy at its other endpoint, which would have stopped the later scan
first. Let $r(f)$ be the round in which $f$ is removed by the parallel version
of greedy matching (in each round the copies that are rank minima at both
endpoints among the surviving copies are matched and their neighbors removed).
Along a stack $r$ does not increase, and it decreases strictly every two steps.
Indeed, a matched queried copy recurses only to lower-ranked copies at its head,
which were removed in earlier rounds. An unmatched queried copy $e_j$ is
removed in the round of its matched blocker, which lies at its head because the
scan at its tail reached it; the next queried copy is either this blocker, which
is matched, or a lower-ranked copy at the same vertex, removed in an earlier
round. Parallel greedy matching on $H$ is parallel greedy MIS on the
line graph of $H$, a simple graph on the $p$ copies; padding it with isolated
vertices to $N_H$ vertices, the theorem of Fischer and Noever~\cite{FN18} shows
that, for a suitable constant $C$, all stacks have length at most
$D=C\log N_H$ except with probability $N_H^{-3}$.

For a stack $P=(e_1,\dots,e_\ell)$ ending at $(e,u)$ under order $\pi$, let
$f(\pi,P)$ rotate the ranks along $P$: $e_j$ receives the rank of $e_{j+1}$ for
$j<\ell$, and $e_\ell$ the rank of $e_1$. For a fixed image, the preimage stacks
are nested suffixes of one another. Indeed, let two of them first diverge,
reading backwards from $(e,u)$, at a common copy $e'$ preceded by $f'$ in the
first stack, where $\alpha=\pi_1(e')\le\pi_2(e')$. By simplicity $f'$ is not on
the second stack, so $\pi_2(f')=\alpha$, and the two orders agree on all copies
of rank below $\alpha$. If $f'$ were unmatched under $\pi_2$, its matched blocker
of rank below $\alpha$ would also block the first stack before $f'$ or before
$e'$; so $f'$ is matched under $\pi_2$, and the second stack would have stopped
at $f'$ before reaching $e'$, a contradiction. Since $P$ and its image determine
$\pi$, every image has at most $D$ preimages with short stacks. Counting the
pairs $(\pi,P)$ gives $\Ex Q(e,u)\le D+N_H^2\cdot N_H^{-3}$. The consequences
follow by summing over directed copies and dividing by $m$.
\end{proof}

\subsubsection*{Lazy ranks on padded candidate lists}

In our application the incident copies of a vertex are not listed explicitly.
Instead every vertex $v$ has a fixed indexable list of \emph{candidate slots};
testing a slot returns null or an incident copy with a shared identifier. Every
copy occurs in exactly one slot at each endpoint, and a successful test also
locates the reverse slot. Preparing a vertex and testing a slot cost at most
$C_*$, and the slot-cost bounds at $v$ sum to $W_v$; these bounds are fixed
before the ranks are drawn.

\begin{lemma}[Padded lists]\label{lem:padded}
There is an implementation of the vertex oracle with correct answers and an
event $\mathcal E$ of probability $1-N_*^{-\Omega(1)}$, for a polynomial bound
$N_*$ on the number of slots, such that for a uniformly random root
\begin{equation}\label{eq:padded}
  \Ex\bigl[\ind_{\mathcal E}\operatorname{Cost}\bigr]
  \le\softO\Bigl(C_*\Bigl(1+\frac pm\Bigr)+\frac1m\sum_vW_v\Bigr).
\end{equation}
Degrees need not be known. The bound holds separately for several nonnegative
cost accounts.
\end{lemma}

\begin{proof}
Ranks are independent uniform numbers in $[0,1)$, exposed lazily on the common
breakpoints $0,2^{-P_0},2^{1-P_0},\dots,1$, where $2^{-P_0}$ is below the reciprocal
of $N_*$. Every vertex $v$ maintains a boundary $b_v$, initially $0$, such that
all incident ranks below $b_v$ are known, and the following invariant holds
after each step: every assigned rank is stored at both endpoints, and every
unassigned copy $uv$ has rank uniform on $[\max\{b_u,b_v\},1)$, independently.
To extend $v$ from $[a,b)$ with $a=b_v$, note that every other endpoint has
$b_u\le a$ or $b_u\ge b$, because the breakpoints are common. In the first case
the rank of an unassigned copy lies in $[a,b)$ with probability
$p_{a,b}=(b-a)/(1-a)$, independently; on success it receives a fresh uniform
rank in $[a,b)$, which is stored at both endpoints. In the second case the
rank is already at least $b$. We realize this by Bernoulli$(p_{a,b})$ coins on
all slot indices, skipping failures by geometric jumps, and testing only the
selected slots; coins on null slots and on assigned copies are discarded. After
the step we set $b_v=b$; the invariant follows because distinct unassigned
copies receive independent interval conditions. The oracles above only need the copies at $v$ in rank order below a requested value,
which this procedure provides.

For an opened interval, the selected slots cost $O(p_{a,b}W_v+C_*\log N_*)$
with high probability. The number of incident ranks below $x$ is binomial with
mean $xd_v$, so if the last boundary of $v$ is $b_v=2x$, then
$b_vd_v=O(j_v+\log N_*)$, where $j_v$ is the number of rank-ordered requests at
$v$. Since the interval probabilities sum to at most $2b_v$, the cost at $v$ is
$\softO((W_v/d_v)(j_v+1)+C_*)$ on the joint high-probability event. By
\cref{lem:calls} the expected value of $j_v$ for a random root is
$O((1+d_v)\log N_H/m)$, under the original law of the ranks, and an isolated
vertex is entered only as the root. Summing over $v$ proves \eqref{eq:padded}.
\end{proof}

\subsection{Independent terminal covers and completeness}\label{sec:covers}

At scale $a=a_i$ put $\rho=\delta a$, $h=\lceil\log(k+2)\rceil$, $r_c=\rho/h$, and
$\psi=\lceil e\log(4kI/\beta)\rceil$. For $x\in T$ let $d_x$ be the number of other
terminals within distance $r_c$ of $x$. Every terminal $x$ is \emph{retained}
independently with probability $q_x=1-(d_x/(d_x+1))^\psi$, and one fixed terminal
is always retained; $N=N_i$ is the retained set and $g=g_i=|N_i|$. The bit of $x$
is resolved on demand, without computing $d_x$: we draw uniform terminals and
query their distance to $x$, retaining $x$ if $x$ itself is drawn before $\psi$
other terminals within distance $r_c$, and rejecting it otherwise. The draws for
different pairs of a terminal and a scale use private independent randomness,
and resolved bits are cached.

\begin{lemma}[Covers]\label{lem:cover}
Except with probability $\beta$, simultaneously at all scales, $N_i$ is a
$\rho_i$-cover of $T$ and
\begin{equation}\label{eq:cover-size}
  g_i\le H_c+\frac{4\psi hL}{\delta a_i},\qquad H_c=2+2\psi+\log(2I/\beta).
\end{equation}
Moreover, for every adaptive procedure that does not observe unresolved draws,
the total number $W$ of draws and the number $A$ of retained terminals it has
found satisfy
\begin{equation}\label{eq:factory-work}
  \Prob\bigl[\,W>2k\bigl(A+\log(1/\beta)\bigr)\text{ at some time}\,\bigr]\le\beta .
\end{equation}
\end{lemma}

\begin{proof}
A draw hits its current target with probability $1/k$, given the past, and every
such hit reveals a new retained terminal. Hence
$\exp(W/(2k)-\#\text{hits})$ is a nonnegative supermartingale, and the maximal
inequality gives \eqref{eq:factory-work}.

For the cover property, let $B_j(x)$ be the set of terminals within $j$ hops of
$x$ in the graph joining terminals at distance at most $r_c$. Since
$|B_h(x)|\le k$, some $j<h$ has $|B_j(x)|/|B_{j+1}(x)|\ge k^{-1/h}\ge e^{-1}$.
Every $y\in B_j(x)$ has its closed neighborhood in $B_{j+1}(x)$, so
$\sum_{y\in B_h(x)}1/(d_y+1)\ge e^{-1}$, and the probability that no terminal of
$B_h(x)$ is retained is at most $\exp(-\psi/e)\le\beta/(4kI)$. All of $B_h(x)$ is
within distance $hr_c=\rho$ of $x$. For the size, $q_x\le\psi/(d_x+1)$, and
$\sum_x1/(d_x+1)$ is the expected number of local minima of a random order in
this graph; local minima are pairwise more than $r_c$ apart, so a doubled MST
tour bounds their number by $1+2L/r_c$. Hence $\Ex g\le1+\psi(1+2hL/\rho)$, and
the Bernoulli exponential moment gives $g\le2\Ex g+\log(2I/\beta)$ except with
probability $\beta/(2I)$. Union bounds finish the proof.
\end{proof}

\begin{corollary}\label{cor:factory}
Let $\mathcal E$ be any event and suppose $\Ex[\ind_{\mathcal E}A]\le\mu_A$. For
$\beta_A,\beta_W\in(0,1)$, the event $\mathcal E$ and
$W>2k(\mu_A/\beta_A+\log(1/\beta_W))$ occur together with probability at most
$\beta_A+\beta_W$.
\end{corollary}

\begin{proof}
Apply Markov's inequality to $\ind_{\mathcal E}A$ and \eqref{eq:factory-work}
with $\beta_W$.
\end{proof}

Uniform terminals accepted exactly when retained are independent uniform
samples of $N$, and $s$ of them cost $\softO(k(s+1))$ draws by
\eqref{eq:factory-work}. Stopping after $s=C\xi^{-2}\log(2/\beta)$ acceptances
among $t_s$ proposals and setting $g_\pm=(1\pm\xi/2)ks/t_s$ gives, with
probability $1-\beta$,
\begin{equation}\label{eq:g-interval}
  (1-\xi)g\le g_-\le g\le g_+\le(1+\xi)g .
\end{equation}
On the event of \cref{lem:cover}, using $\sum_ia_i\le10L/(\omega-1)$,
\begin{equation}\label{eq:retained-mass}
  \sum_ia_ig_i\le\mathcal AL,\qquad
  \mathcal A=1+\frac{10H_c}{\omega-1}+\frac{4\psi hI}{\delta},
\end{equation}
and $\mathcal A$ is polylogarithmic in $n$. We use the notation of
\cref{sec:certificates} for these covers, with $\rho_i=\delta a_i$.

\paragraph{Completeness of the saving.}
On the cover event,
\begin{equation}\label{eq:retained-domination}
  Y_i\le\Phi_i .
\end{equation}
Indeed, if a center is within $3a_i/5$ of $x$ in an exact component, then
$\pi_i(x)$ lies in the same exact component and within
$3a_i/5+\rho_i<5a_i/8$ of the center; distinct exact components contain
distinct representative components by \eqref{eq:sandwich}. Thus every eligible
ideal row family yields an eligible family on $\mathcal C_i$ with at least the
same union and the same number of rows.

\paragraph{Completeness of the band statistic.}
Let $B_Q=\lceil400\mathcal A\rceil$ and
$T_Q=\sum_i\sum_{C\in\mathcal C_i,\,|C|\le B_Q}D_i(C)$.

\begin{lemma}[Band completeness]\label{lem:band-complete}
On the cover event, $T_Q\le Q_Q(L-\OPT)$. Moreover, with probability at least
$1-10^{-4}$ over the shift of the grid,
\begin{equation}\label{eq:band-completeness}
  \Lambda\le2T_Q/\sigma_0+L/100 .
\end{equation}
\end{lemma}

\begin{proof}
The first claim is \cref{lem:band-sound}. Call an exact component
\emph{persistent} if it is a part of $\mathcal P_i$. By \eqref{eq:sandwich}, a
component of $\mathcal Q_i$ that is also a component of $G_{a_i-2\rho_i}(T)$ is
persistent, so the number of nonpersistent components is at most
$\kappa_T(a_i-2\rho_i)-\kappa_T(a_i)$, the number of MST edges with weight in
$[a_i-2\rho_i,a_i)$. Under the log-uniform shift an MST edge of weight $w$
contributes expected scale weight
$\frac1{\log\omega}\int_w^{w/(1-2\delta)}da=\frac{2\delta w}{(1-2\delta)\log\omega}$,
so by Markov's inequality the scale-weighted number of nonpersistent components
is at most $L/200$ with probability at least $1-10^{-4}$. Persistent components
with more than $B_Q$ retained points have scale weight at most
$\sum_ia_ig_i/(B_Q+1)\le L/400$ by \eqref{eq:retained-mass}.

Let $C\in\mathcal Q_i$ be any other component with $A_i(C)\ge\sigma_0a_i$, so that
$C\cap N_i$ is a component of $\mathcal C_i$ with at most $B_Q$ points, and project
an attaining set $Y$ by $\pi_i$. The benefit decreases by at most $3\rho_i$, the
one-point connection cost increases by at most $8\rho_i$, restricting to the band
loses at most $3\tau a_i+6\rho_i$, and the penalty is $6\rho_i$. Hence
$D_i(C\cap N_i)\ge A_i(C)-23\rho_i-3\tau a_i\ge\sigma_0a_i/2$; the projected set
has at least two points, since otherwise its benefit would be at most
$3\rho_i<\sigma_0a_i$. Summing proves \eqref{eq:band-completeness}.
\end{proof}

\subsection{Scores of retained components and the estimator}\label{sec:scores}

At each scale we shall compute $Z_i$ with
\begin{equation}\label{eq:saving-contract}
  0\le Z_i\le\Phi_i,\qquad Z_i\ge\Phi_i/2-\eps g_i,\qquad
  \eps=\frac{e_0}{10^5\mathcal A}.
\end{equation}
A breadth-first search in $G_a(N)$ from a retained point either lists its
component, if it has at most $B$ points, or stops after $B+1$ points; each popped
point is compared with all terminals, resolving retention only for those within
distance $a$, so the search costs $O(kB)$ queries and discloses at most $B+1$
retained points. Given two complete components within $5a/8$ of a center $v$,
the original eligibility of the row of $v$ is decided by scanning the terminals
within $5a/8$ of $v$ outside the two components until a retained one is found;
it lies in a third component.

Let $D$ be the union of the balls $A_v=\{x\in T:w(v,x)\le5a/8\}$ of at most $s$
selected centers. Let $\mathcal C_0$ be the components with at most $B$ points
that meet $D$, let $U$ be the other components with at most $B$ points, let
$b_{\mathrm{large}}$ be the number of larger components, and let $\Psi_U$ be the
saving of the system restricted to $U$, with the original eligibility.

\begin{lemma}[Deletion and scores]\label{lem:score}
We have
\begin{equation}\label{eq:deletion}
  \Phi\le b_{\mathrm{large}}+|\mathcal C_0|+\Psi_U,\qquad
  |\mathcal C_0|-2s+\Psi_U\le\Phi .
\end{equation}
Suppose a graph on $U$ has expected random-greedy matching size $\mu$ with
$\Psi_U/2-\Delta\le\mu\le\Psi_U$. Then a bounded score of a uniform retained
point has mean $(|\mathcal C_0|+\mu)/g$, and with $O(\xi^{-2}\log(2/\beta))$
samples one obtains, with probability $1-\beta$, a value $Z$ with $0\le Z\le\Phi$
and
\begin{equation}\label{eq:score-loss}
  Z\ge\Phi/2-b_{\mathrm{large}}/2-\Delta-2s-3\xi g .
\end{equation}
\end{lemma}

\begin{proof}
The first inequality holds because removing elements while keeping eligibility
fixed decreases the saving by at most their number. For the second, add the
eligible selected rows to a family attaining $\Psi_U$. These rows do not meet
$U$, and every component of $\mathcal C_0$ lies in a selected row; an ineligible
selected row leaves at most two of them uncovered. For the score, explore the
component $C$ of a uniform retained point. If it is large the score is $0$; if
$C\in\mathcal C_0$ it is $1/|C|$; if $C\in U$, then with probability $1/|C|$ we
run the matching oracle at $C$ with fresh ranks and score half of the matched
indicator, and otherwise score $0$. Every $C\in U$ is initiated with probability
$1/g$ and the matched indicators sum to $2\mu$ in expectation, which gives the
mean. Let $\widehat\vartheta$ be the empirical mean and
$Z=\max\{0,g_-\max\{0,\widehat\vartheta-\xi\}-2s\}$. On the Hoeffding and
calibration events, $Z\le\max\{0,|\mathcal C_0|+\mu-2s\}\le\Phi$ by
\eqref{eq:deletion}, and $Z\ge|\mathcal C_0|+\mu-2s-3\xi g$, which together with
\eqref{eq:deletion} gives \eqref{eq:score-loss}.
\end{proof}

\paragraph{The band estimate.}
With an independent sample, score a uniform retained point by
$D_i(C)/(3a_i|C|)$ if its component $C$ has at most $B_Q$ points, and $0$
otherwise; the mean is $\sum_{|C|\le B_Q}D_i(C)/(3a_ig_i)$. With
$\xi_Q=\sigma_0/(10^5\mathcal A)$, the estimate
$\widehat T_i=3a_ig_-\max\{0,\widehat\vartheta_i-\xi_Q\}$ loses at most
$9\xi_Qa_ig_i$, so $\widehat T=\sum_i\widehat T_i$ satisfies
\begin{equation}\label{eq:band-estimate}
  0\le\widehat T\le T_Q,\qquad T_Q-\widehat T\le\sigma_0L/1000 .
\end{equation}
Each sample costs $O(nB_Q^4)$ time and $O(B_Q)$ disclosures, so this costs
$\softO(n)$ in total.

\subsubsection*{The estimate and its guarantee}

Let $S=\sum_ia_iZ_i$, let $\overline L$ be as in \eqref{eq:mst-interval}, and return
\begin{equation}\label{eq:output}
  \widehat C=\overline L-\max\Bigl\{\frac S{32Q_S},\ \frac{\widehat T}{Q_Q}\Bigr\}.
\end{equation}

\begin{proposition}\label{prop:estimate}
Let $c_*=\min\{e_0/(32000Q_S),\ \sigma_0/(1000Q_Q),\ e_0/8,\ 1/10\}$ and
$0<\nu\le c_*/10$. On the events of \cref{lem:cover,lem:band-complete}, of
\eqref{eq:mst-interval}, and of \eqref{eq:saving-contract} and
\eqref{eq:band-estimate}, the output satisfies
$\OPT\le\widehat C\le(2-c_*)\OPT$ for the floored metric.
\end{proposition}

\begin{proof}
By \eqref{eq:saving-contract} and \cref{lem:saving-sound},
$S/(32Q_S)\le L-\OPT$, and by \eqref{eq:band-estimate} and
\cref{lem:band-complete}, $\widehat T/Q_Q\le L-\OPT$; since $\overline L\ge L$, we
get $\widehat C\ge\OPT$. By \eqref{eq:retained-domination},
\eqref{eq:saving-contract}, and \eqref{eq:retained-mass},
$\Gamma\le\sum_ia_i(2Z_i+2\eps g_i)\le2S+e_0L/1000$. If $S\le e_0L/1000$ and
$\widehat T\le\sigma_0L/1000$, then $\Gamma\le e_0L/250$ and, by
\cref{lem:band-complete}, $\Lambda\le4L/1000+L/100<L/10$, so
\cref{thm:structure} gives $L\le(2-e_0/2)\OPT$ and
$\widehat C\le(1+\nu)(2-e_0/2)\OPT\le(2-c_*)\OPT$. Otherwise one discount is at
least $c_*L$, and $\widehat C\le(1+\nu-c_*)L\le(2-c_*)\OPT$ since $L\le2\OPT$.
\end{proof}

With $\eta_{\mathrm w}=c_*$, the flooring step turns this into the factor
$2-\eta_0$ with $\eta_0=c_*/2$ for the original metric. It remains to compute
the $Z_i$ within the claimed budget.

\paragraph{Small covers.}
If $g\le100t/\eps$, where $t=\lceil\sqrt n\,\rceil$, we resolve all retention
bits, read the distances within $N$ and between $N$ and all centers, and
greedily accept an eligible row whenever it contains two components not yet
claimed, claiming all its components. If $u$ components are claimed by $f$ rows
then $u\ge2f$, the claimed family certifies saving $u-f\ge u/2$, and every
eligible row contains at most one unclaimed component, so $\Phi\le u$. Thus
$Z=u-f$ satisfies \eqref{eq:saving-contract}, at cost
$\softO((n+k)t\,\poly(1/\eps))$.

\subsection{A reusable sample for deleting large rows}\label{sec:pool}

The following lemma selects few rows that cover most of every large row, using
a single sample whose records are reused across all rows. It is related to the
sparse vector technique~\cite{KMS21} and to the transfer from stability to
generalization in adaptive data analysis~\cite{DworkFHPRR15,BassilyNSSSU16}.

\begin{lemma}[Reusable sample]\label{lem:pool}
Let $R_1,\dots,R_{N_r}$ be fixed rows in a universe of size $g$, let $t\ge1$, and
let $0<\beta<1/4$. A procedure that knows $t$ but not $g$ uses
$N_s=O(t\log(t/\beta)\log((N_r+1)t/\beta))$ independent uniform samples of the
universe and $O(N_rN_s)$ membership tests, and selects at most $t$ rows such that,
except with probability $\beta$, every selected row contains at least $g/t$
elements outside the previously selected rows, and in the end every row
contains at most $4g/t$ elements outside the selected rows.
\end{lemma}

\begin{proof}
Keep the sample records as \emph{live} records. Process the rows in order; if
row $R_j$ contains $c$ live records, accept it with probability
\begin{equation}\label{eq:pool-logistic}
  p(c)=\frac1{1+\exp((H-c)/b)},\qquad H=2N_s/t,
\end{equation}
and on acceptance delete its records. Stop after $t$ acceptances. Only the
acceptance bits are used later.

\emph{Stability.} Compare samples that differ by one added record. Along a common
transcript the record matters only up to the first accepted row containing it.
At a row containing it, with $q_b=e^{1/b}-1$, the acceptance probabilities
$p'$ and $p$ with and without the record satisfy $p'=(1+q_b)p/(1+q_bp)$, so
$p'/p\le e^{1/b}$ and $(1-p)/(1-p')=1+q_bp$. In one direction there is at most
one acceptance factor; in the other the rejection factors multiply to at most
$\exp(q_b\sum p)$, where the sum of the predictable hazards $p$ exceeds
$\log(1/\delta')$ only with probability $\delta'$ because
$\exp(\sum p)\prod(1-p)$ stopped at the first relevant acceptance is a
supermartingale. Hence the transcript is $(\eps_b,\delta')$-stable with
$\eps_b=\max\{1/b,\,q_b\log(1/\delta')\}$. A replacement is a removal followed
by an addition, so under replacement the transcript is
$(2\eps_b,(1+e^{\eps_b})\delta')$-stable.

\emph{Transfer.} Let $X_1,\dots,X_{N_s}$ be i.i.d.\ with law $\nu$, let the transcript
$\mathsf T$ be $(\eps,\delta_1)$-stable under replacement, and let
$\nu_{\mathsf T}=N_s^{-1}\sum_l\mathcal L(X_l\mid\mathsf T)$. Except on transcripts of
probability $2\delta_1/\zeta$, simultaneously for all $f:\mathcal U\to[0,1]$,
\begin{equation}\label{eq:posterior-transfer}
  \nu_{\mathsf T}(f)\le e^\eps\nu(f)+\zeta,\qquad \nu(f)\le e^\eps\nu_{\mathsf T}(f)+\zeta .
\end{equation}
To see this, replace a uniformly random coordinate $X_I$ by a fresh draw. The
joint law of $X_I$ and the new transcript is $\nu\otimes\mathcal L(\mathsf T)$, and
stability compares it with the joint law of $X_I$ and the old transcript in both
directions with additive error $\delta_1$; Markov's inequality over transcripts
gives \eqref{eq:posterior-transfer}.

\emph{Margins.} Put $\varrho=1/t$, $\eps=\log(4/3)$, $\zeta=\varrho/16$,
$\delta_1=\beta\varrho/128$, and $\beta_0=\beta\varrho/64$, and choose
$b=\Theta(\log(t/\beta))$ so that the stability parameters hold. If
$N_s\ge Ctb\log((N_r+1)t/\beta)$, then except with probability $\beta_0$ every
accepted row has empirical fresh fraction at least $7\varrho/4$ and every rejected
row at most $9\varrho/4$, since a decision beyond these margins has probability at
most $\exp(-N_s/(4tb))$. Except on transcripts of probability $\beta_0/\zeta$ this
event has posterior probability at least $1-\zeta$. The fresh part of a row is
determined by the row and the earlier acceptance bits, so its posterior mean is
at least $(1-\zeta)7\varrho/4$ on acceptance and at most $9\varrho/4+\zeta$ on
rejection. By \eqref{eq:posterior-transfer}, the true fresh fractions are at
least $\frac{303}{256}\varrho>\varrho$ and at most $\frac{151}{48}\varrho<4\varrho$,
respectively. The total exceptional probability is at most
$(\beta_0+2\delta_1)/\zeta=\beta/2$. Accepted fresh sets are disjoint, so after
$t$ acceptances the universe is covered; otherwise all rows were tested, and
later deletions only shrink them.
\end{proof}

The rule \eqref{eq:pool-logistic} can be evaluated exactly with integer
arithmetic by choosing $b=1/(t\log z_*)$ for $z_*=1+1/q_*$ with an integer
$q_*=\Theta(t\log(2t/\beta))$, so that $p(c)=1/(1+z_*^{2N_s-tc})$. The sample
records must not influence the order of the rows or any other test; in our
application all decisions are taken before the selected balls are scanned.

\subsection{Mutual incidence capping and proof of \Cref{thm:upper}}\label{sec:mutual}

Fix a scale with $g>50t/\eps$, and let $\xi=\eps/100$ and $B=\lceil10/\eps\rceil$.
Apply \cref{lem:pool} to the rows $A_v\cap N$ of all centers, with uniform
retained samples, and let $D$ be the union of the selected balls $A_v$ on $T$.
This costs $\softO(nt)$ queries for the pool and $O(kt)$ for the balls, and on
the pool event
\begin{equation}\label{eq:pool-residual}
  |A_v\cap(N\setminus D)|\le P:=4g/t\qquad\text{for every center }v .
\end{equation}
Define $\mathcal C_0$ and $U$ as in \cref{sec:scores}, with $s\le t$. For
$C\in U$ let $L(C)$ be the first $b=\lceil100n/(\eps g_-)\rceil$ centers, in a
fixed public order, within distance $5a/8$ of $C$ (all of them if there are
fewer); it is found in $O(nB)$ time. The multigraph on $U$ has a \emph{port edge}
$(v,\min\{x,x'\},\max\{x,x'\})$ between $C$ and $C'$ whenever $v$ belongs to
$L(C)\cap L(C')$, the row of $v$ is eligible, and $x\in C$, $x'\in C'$ are
retained points within $5a/8$ of $v$.

\begin{lemma}[Mutual capping]\label{lem:mutual-cap}
Every independent set $I\subseteq U$ of this multigraph satisfies
$\Psi_U\le|U|-|I|+2n/b$. Consequently every maximal matching, and hence the
expected random-greedy matching size $\mu$, satisfies
\begin{equation}\label{eq:mutual-matching}
  \Psi_U/2-n/b\le\mu\le\Psi_U .
\end{equation}
\end{lemma}

\begin{proof}
An eligible center lies in the lists of at most one member of $I$, since
otherwise those two members would be adjacent, and an ineligible center is
within $5a/8$ of at most two components. Hence
$\sum_{C\in I}|L(C)|\le2n$, and at most $2n/b$ members of $I$ have a full list.
Each other member kept all its incident centers, so an eligible row covers at
most one of them, and every cover of $U$ by eligible rows and singletons has at
least $|I|-2n/b$ elements. For a maximal matching of size $z$, the unmatched
vertices form an independent set, so $\Psi_U\le2z+2n/b$; conversely, the
centers of the matched edges cover $2z$ components with at most $z$ eligible
rows, so $z\le\Psi_U$.
\end{proof}

The candidate slots of $C\in U$ are the triples $(v,x,y)$ with $v\in L(C)$,
$x\in C$ within $5a/8$ of $v$, and $y\in T$; there are at most $kbB$ of them. A
slot is tested by checking $y\notin D$ and $w(v,y)\le5a/8$, then resolving the
retention of $y$, exploring its component $C'$, rejecting it if it is large,
meets $D$, or equals $C$, preparing $L(C')$, checking $v\in L(C')$, and testing
the eligibility of $v$. This recognizes exactly the port edges, and the reverse
slot is $(v,y,x)$. By \eqref{eq:pool-residual}, at most $bBP$ slots of $C$ reach
a retained $y$; each costs $O(nB)$ time and $O(B)$ disclosures, and the others
cost $O(1)$ and disclose nothing. Thus, in the notation of
\cref{lem:padded}, for the account of ordinary work and for the account of
disclosures respectively,
\begin{equation}\label{eq:fast-slots}
  d_C\le bBP,\qquad
  W_C=O(kbB+nB^2bP),\ C_*=O(nB);\qquad
  W'_C=O(B^2bP),\ C'_*=O(B).
\end{equation}
Using the score of \cref{lem:score} with $\Delta=n/b$ from
\eqref{eq:mutual-matching}, and $b_{\mathrm{large}}\le g/(B+1)$,
\[
  Z\ge\Phi/2-\tfrac{\eps g}{20}-\tfrac{\eps g}{100}-\tfrac{\eps g}{25}-\tfrac{3\eps g}{100}
  \ge\Phi/2-\eps g,
\]
so \eqref{eq:saving-contract} holds. Since $g>50t/\eps$ we have
$bP=O(n/(\eps t))$ and $kb=O(n^2/t)$, so by \cref{lem:padded} a matching call
from a uniform root costs $\softO((n+n^2/t)\poly(1/\eps))$ time and
$\softO((1+n/t)\poly(1/\eps))$ disclosures in expectation on its
high-probability event. Only $O(\xi^{-2}\log(1/\beta))$ calls are made per
scale. Together with the pool, the masks, and the explorations, the work at
one scale is
\begin{equation}\label{eq:fast-work}
  \softO\bigl((nt+n^2/t)\poly(1/\eps)\bigr),
\end{equation}
and the expected number $\mu_A$ of disclosures on the joint good event is
$\softO((t+n/t)\poly(1/\eps))$. By \cref{cor:factory} the number of draws is
$\softO(k(t+n/t)\poly(1/\eps))$ except with small probability.

\begin{proof}[Proof of \cref{thm:upper}]
Run the estimator~\eqref{eq:output} of \cref{sec:scores} with $t=\lceil\sqrt n\,\rceil$, using the
small-cover routine when $g_+\le100t/\eps$ and the routine of this subsection
otherwise. Choose the per-scale failure probabilities inverse-polylogarithmic
so that the cover, calibration, pool, score, MST, and rank-implementation
failures sum to less than $1/12$; the persistence failure of
\cref{lem:band-complete} is below $10^{-4}$. Bound the expected work and
disclosures of the uncapped algorithm on these events, and impose a global cap
of a large constant times these expectations; by Markov's inequality and
\cref{cor:factory} the cap is hit with probability less than $1/12$, and a cap
hit is treated as failure. The total failure probability is below $1/3$, and
on success \cref{prop:estimate} and flooring give the factor $2-\eta_0$. As
$\eps$, $\mathcal A$, $B$, and $B_Q$ are polylogarithmic and there are $O(\log n)$
scales, \eqref{eq:fast-work} gives $\softO(n^{3/2})$ queries and time.
\end{proof}

\section{The Lower-Bound Construction}\label{sec:lb-dist}

In the rest of the paper, we prove our lower bound, \cref{thm:lower}.
Recall that it states that for every $\eta\in(0,1)$ and every integer
$s\ge1$, estimating $\OPT$ within a factor $2-\eta$ requires
$\Omega(n^{b_s})$ queries, where $b_s=(3s+7)/(2s+5)$ tends to $3/2$ as $s$
grows. As outlined in \cref{sec:overview-lb}, we construct two distributions
$\Dyes$ and $\Dno$ over instances such that, with high probability, the
optimal cost under $\Dno$ is more than $2-\eta$ times the optimal cost under
$\Dyes$, and we show that every algorithm with few queries sees almost the
same distribution of answers under both. This section constructs the two
distributions and proves the cost gap. \Cref{sec:gated} replaces the metric
by a simpler random process, the two-key experiment, and
\cref{sec:analysis} analyzes this process and proves \cref{thm:lower}.
This part is independent of the upper bound, and its parameters are defined
afresh.

\medskip
Throughout, fix $\eta\in(0,1)$ and an integer $s\ge1$. Here
$2-\eta$ is the approximation factor, and $s$ is the number of moments that
the two weight distributions will share. Larger $s$ gives a stronger lower
bound. All constants below may depend on $\eta,s$, but not on the input
size or the algorithm.

Each instance has just three types of points: ordinary terminals in tight
clusters, special terminals hidden among them, and nonterminal centers.
The centers are sampled independently given weights on the special
terminals. There is no need to partition the special terminals into groups,
to order centers within rows, or to select a hidden block.

\subsection{Clusters, special terminals, and centers}

Choose constants in the following order:
\begin{equation}\label{eq:lb-constants}
  \delta=\frac{\eta}{100},\qquad
  H_0=64\log\frac4\delta,\qquad
  d\ge\max\left\{4,\frac{2+8H_0}{\delta}\right\},
\end{equation}
where $d$ is an integer. Let $t$ be a growing integer, and set
\begin{equation}\label{eq:lb-sizes}
  A=t^{s+3},\qquad R=\lfloor\gamma_0t^{s+2}\rfloor,\qquad
  K=AR,\qquad m=dA,\qquad \lambda=K^{-3}.
\end{equation}
The small constant $\gamma_0>0$ will be fixed in \cref{sec:lb-proof}.
% Here $A,R,K$ denote, respectively, the number of ordinary clusters, the
% size of a cluster, and the number of ordinary terminals. Finally, $m$ is the number
% of special terminals.

The point set $V$ contains three types of points.
The terminal set $T$ consists of the $K$ \emph{ordinary terminals}, grouped
into $A$ \emph{ordinary clusters} of $R$ terminals each, and the $m$
\emph{special terminals}. The remaining $mR$ points are nonterminals, which
we call \emph{centers}. Since $mR=dAR=dK$,
\begin{equation}\label{eq:lb-n}
  n=|V|=K+m+mR=(d+1)K+m,\qquad |T|=K+m=\Theta(n).
\end{equation}
The point set $V$ consists of $K+m$ terminals and $mR$ centers. The algorithm knows $V$ and $T$, so it can tell the centers from the terminals, but it does not know which terminals are special or which ordinary terminals share a cluster. To hide these roles, take a uniformly random
bijection $\pi$ from $T$ to the positions
\[
  [m]\ \sqcup\ ([A]\times[R]).
\]
Write $S=\pi^{-1}([m])$ for the special terminals and
$G_a=\pi^{-1}(\{a\}\times[R])$ for ordinary cluster $a$. We identify a
special terminal with its position in $[m]$ when no confusion can arise.

% The public point set consists of $K+m$ terminals and $mR$ centers. Thus
% \begin{equation}\label{eq:lb-n}
%   n=(d+1)K+m,\qquad |T|=K+m=\Theta(n).
% \end{equation}
% To hide the terminal roles, take a uniformly random bijection $\pi$ from
% $T$ to the positions
% \[
%   [m]\ \sqcup\ ([A]\times[R]).
% \]
% Write $S=\pi^{-1}([m])$ for the special terminals and
% $G_a=\pi^{-1}(\{a\}\times[R])$ for ordinary cluster $a$. We identify a
% special terminal with its position in $[m]$ when no confusion can arise.

\paragraph{The weights.}
We use two finitely supported distributions $\nu_{\mathsf y},\nu_{\mathsf n}$ on a
fixed interval $[0,M]$ satisfying
\begin{equation}\label{eq:moments}
  \Ex_{\nu_{\mathsf y}}[W^j]=\Ex_{\nu_{\mathsf n}}[W^j]
  \quad(0\le j\le s),\qquad \Ex[W]=1,
\end{equation}
and
\begin{equation}\label{eq:weight-supports}
  \Prob_{\nu_{\mathsf y}}[W\ge1/2]=1,\qquad
  \Prob_{\nu_{\mathsf n}}[W>0]\le\delta .
\end{equation}
Here $W$ denotes a single weight drawn from the distribution in the
subscript, and the condition $\Ex[W]=1$ holds for both distributions.
\Cref{lem:weights} below constructs such distributions. Independently of $\pi$,
give the special positions independent weights $W_1,\dots,W_m$, using
$\nu_{\mathsf y}$ in \yes{} and $\nu_{\mathsf n}$ in \no.

\paragraph{The neighbor sets.}
Given the weights, sample independently for each center $c$ a neighbor set
$\nb(c)\subseteq[m]$. For a $d$-set $H$, put
\[
  a_W(H)=\frac{H_0}{dR\binom md}\prod_{u\in H}W_u,
  \qquad \beta=\frac1{4\binom m2}.
\]
The distribution of $\nb(c)$ is
\begin{equation}\label{eq:static-law}
\begin{array}{c|c}
  \text{neighbor set} & \text{probability}\\ \hline
  \varnothing & \displaystyle\frac12-\sum_{|H|=d}a_W(H)\\[3pt]
  \{u\} & \displaystyle\frac1{4m}\quad\text{for each }u\\[3pt]
  \{u,v\} & \displaystyle\beta\quad\text{for each pair }\{u,v\}\\[3pt]
  H,\ |H|=d & a_W(H)\quad\text{for each }d\text{-set }H .
\end{array}
\end{equation}
A center with $d$ neighbors is a \emph{star}, and the others are \emph{fillers}.
The probabilities are valid for all weights once $R$ is sufficiently large,
since $\sum_Ha_W(H)\le H_0M^d/(dR)\le1/2$.

The singleton and pair probabilities do not depend on the weights.
The pair fillers hide the first two contacts with a star. Singleton fillers
keep this background present even after many negative answers. The empty
set absorbs the remaining mass. In particular, we do \emph{not} normalize
star probabilities by a random sum of weights: their product form will be
essential for moment matching.

\paragraph{The metric.}
For distinct points $x,y$, define
\begin{equation}\label{eq:lb-metric}
  w(x,y)=
  \begin{cases}
    \lambda, & x,y\in G_a\text{ for some ordinary cluster }G_a,\\
    1, & x=c\text{ is a center},\ y\in S,\ \pi(y)\in\nb(c),\text{ or conversely},\\
    2, & \text{otherwise},
  \end{cases}
\end{equation}
and put $w(x,x)=0$. Let $\Dyes,\Dno$ be the resulting distributions.
Their only difference is the weight distribution. All distances are fixed before
the algorithm starts.

\begin{lemma}\label{lem:metric}
Every instance in the support of $\Dyes$ or $\Dno$ is a metric.
\end{lemma}

\begin{proof}
Only the triangle inequality needs checking. As every distance is at most
$2$, a violation needs two distances whose sum is less than $2$. Two
$\lambda$-edges lie within one ordinary cluster, where the third distance is
also $\lambda$. A $\lambda$-edge and a unit edge cannot share an endpoint:
the former has only ordinary endpoints, and the latter has none.
\end{proof}

\subsection{The gap in the optimal costs}

Contracting the ordinary clusters costs $\lambda(K-A)$ and leaves
$\kappa=A+m$ \emph{terminal pieces}: one for each cluster and one for each
special terminal. For a set $X$ of centers, let $H_X$ be the bipartite graph
on these pieces and $X$ whose edges are the unit distances, and let $r(X)$
be the size of a maximum forest in $H_X$. The quantity $r(X)-2|X|$ is the
\emph{profit} from using $X$.

\begin{lemma}[Profit identity]\label{lem:profit}
For every constructed instance,
\begin{equation}\label{eq:exact-cost}
  \OPT=\lambda(K-A)+2(\kappa-1)
       -\max_X\bigl(r(X)-2|X|\bigr).
\end{equation}
The maximum may be restricted to sets of stars.
\end{lemma}

\begin{proof}
For fixed $X$, take the $\lambda$-spanning trees within clusters and a
maximum forest of $H_X$. The remaining
$\kappa+|X|-r(X)$ components can be joined by edges of weight $2$:
an edge of weight $1$ between different components would contradict the
maximality of the forest. This gives a tree of weight
$\lambda(K-A)+2(\kappa-1)-(r(X)-2|X|)$.

Conversely, a feasible tree using $X$ has $K+m+|X|-1$ edges. At most $K-A$
are $\lambda$-edges, and at most $r(X)$ are unit edges. Writing all other
edges as weight $2$ gives the same expression as a lower bound. Finally,
deleting a filler from a maximum forest removes at most two edges and saves
the penalty $2$ for that center. It cannot decrease the profit.
\end{proof}

\begin{lemma}[Cost gap]\label{lem:gap}
Define
\[
  U_m=\lambda(K-A)+(1+\eta/50)m-2,\qquad
  V_m=\lambda(K-A)+(2-\eta/50)m-2.
\]
With probability $1-e^{-\Omega(m)}$, a \yes{} instance has $\OPT\le U_m$. With
the same probability, a \no{} instance has $\OPT\ge V_m$. Moreover,
\begin{equation}\label{eq:cost-gap}
  V_m-(2-\eta)U_m
  =\frac{47\eta+\eta^2}{50}\,m
   +(1-\eta)\bigl(2-\lambda(K-A)\bigr)>0 .
\end{equation}
\end{lemma}

\begin{proof}
\emph{The \no{} case.}
Let $S_+=\{u:W_u>0\}$. Every star lies in $S_+$, and
$|S_+|\le2\delta m$ except with probability $e^{-\Omega(m)}$.
A component of an incidence forest with $v$ special terminals and $x$
centers contributes $v-x-1$ to its profit. Summing over components shows
that stars have total profit at most $|S_+|$. By \cref{lem:profit},
\[
  \OPT\ge\lambda(K-A)+2(A+m-1)-2\delta m\ge V_m.
\]

\emph{The \yes{} case.}
There are enough stars to cover almost all special terminals, but few
enough that their center penalties are negligible. To see this, for
$1\le h\le m$ let
\[
  E_h=\binom mh^{-1}\sum_{|H|=h}\prod_{v\in H}W_v
\]
be the average weight product over all $h$-subsets of $[m]$, and for
$u\in[m]$ let
\[
  E_{d-1}^{(-u)}=\binom{m-1}{d-1}^{-1}
  \sum_{\substack{|H|=d-1\\u\notin H}}\prod_{v\in H}W_v
\]
be the same average over the $(d-1)$-subsets that avoid $u$.
With probability $1-e^{-\Omega(m)}$,
$E_d$ and all $E_{d-1}^{(-u)}$ lie in $[1/2,2]$. Indeed, let
$\overline W$ be the mean of all weights, and $\overline W^{(-u)}$ the mean
of the weights other than $u$. By Hoeffding's inequality,
$|\overline W-1|\le1/(8d)$ except with probability $e^{-\Omega(m)}$, and
deterministically $|\overline W^{(-u)}-\overline W|\le M/(m-1)$ for every
$u$. An average of products over the $h$-subsets of $n$ numbers in $[0,M]$
differs from the $h$-th power of their mean by at most $\binom h2M^h/n$,
since a uniformly random $h$-tuple of indices repeats an index with
probability at most $\binom h2/n$. So, for large $m$, $E_d$ and every
$E_{d-1}^{(-u)}$ are within $1/4$ of a power $x^h$ with $h\in\{d-1,d\}$ and
$|x-1|\le1/(4d)$. Such powers lie in $[3/4,e^{1/4}]$, hence the averages lie
in $[1/2,2]$.

Condition on such weights. Each of the $mR$ centers is a star with
probability $\sum_{|H|=d}a_W(H)=H_0E_d/(dR)$, independently. So the number
$Z$ of stars is binomial with mean
$H_0mE_d/d\in[H_0m/(2d),2H_0m/d]$, and by a Chernoff bound, except with
probability $e^{-\Omega(m)}$,
\[
  \frac{H_0m}{4d}\le Z\le\frac{4H_0m}{d}.
\]
Given $Z$, the star supports are independent, and each is a $d$-set $H$
chosen with probability proportional to $\prod_{v\in H}W_v$. The $d$-sets
containing $u$ are the sets $\{u\}\cup H'$ with $|H'|=d-1$ and $u\notin H'$,
and $\binom{m-1}{d-1}/\binom md=d/m$. So each support contains a fixed
terminal $u$ with probability
\[
  \frac dm\,\frac{W_uE_{d-1}^{(-u)}}{E_d}\ge\frac d{8m},
\]
where the inequality uses $W_u\ge1/2$ (\yes{}), $E_{d-1}^{(-u)}\ge1/2$
and $E_d\le2$.
Let $U$ be the number of special terminals that lie in no star support.
Since $Z\ge H_0m/(4d)$, each $u$ is uncovered with probability at most
$(1-d/(8m))^Z\le e^{-H_0/32}$. So $\Ex[U]\le me^{-H_0/32}<\delta m/2$, by the
choice of $H_0$. Given $Z=z$, $U$ is a function of the $z$ independent
supports, and replacing one support changes $U$ by at most $d$. Since
$z\le4H_0m/d$, the bounded-differences inequality gives
\[
  \Prob\Bigl[U\ge\Ex[U]+\frac{\delta m}2\Bigr]
  \le\exp\Bigl(-\frac{2(\delta m/2)^2}{zd^2}\Bigr)
  \le\exp\Bigl(-\frac{\delta^2m}{8H_0d}\Bigr)=e^{-\Omega(m)}.
\]
So at most $\delta m$ terminals are uncovered, except with probability
$e^{-\Omega(m)}$.

Take $X$ to be all stars. For each covered terminal, pick one unit edge
to a star containing it. These edges form a forest in $H_X$: every special
terminal has degree one in it, so it has no cycle. Hence $r(X)\ge(1-\delta)m$.
Since $|X|=Z\le4H_0m/d$, we get $r(X)-2|X|\ge(1-\delta)m-8H_0m/d$.
\Cref{lem:profit}, with $\kappa=A+m$ and $A=m/d$, gives
\[
  \OPT\le\lambda(K-A)+2(A+m-1)-\Bigl((1-\delta)m-\frac{8H_0m}d\Bigr)
  =\lambda(K-A)+\left(1+\delta+\frac{2+8H_0}{d}\right)m-2.
\]
By \eqref{eq:lb-constants}, $d\ge(2+8H_0)/\delta$, so
$(2+8H_0)/d\le\delta$, and the coefficient of $m$ is at most
$1+2\delta=1+\eta/50$. Hence $\OPT\le U_m$.
Finally, we verify \eqref{eq:cost-gap}. Write $\ell=\lambda(K-A)$. Then
\[
  (2-\eta)U_m=(2-\eta)\ell+(2-\eta)\Bigl(1+\frac\eta{50}\Bigr)m-2(2-\eta),
  \qquad
  (2-\eta)\Bigl(1+\frac\eta{50}\Bigr)=2-\frac{48\eta}{50}-\frac{\eta^2}{50}.
\]
Subtracting this from $V_m=\ell+(2-\eta/50)m-2$, term by term,
\[
  V_m-(2-\eta)U_m=\frac{47\eta+\eta^2}{50}\,m+(1-\eta)(2-\ell).
\]
Both terms are positive. For the first, $\eta>0$ and $m>0$. For the second,
$1-\eta>0$, and $\ell<\lambda K=K^{-2}\le1$, so $2-\ell>1$.
\end{proof}

\subsection{Constructing the two weight distributions}

The cost argument needed only the support conditions
\eqref{eq:weight-supports} and the common mean $1$. The indistinguishability
argument will use the additional matching moments. These requirements are
compatible for every fixed $s$. We use the following classical identity
from Lagrange interpolation. It says that divided differences of order $n$
vanish on polynomials of degree less than $n$. We include its short proof for
completeness.

\begin{lemma}[A classical interpolation identity]\label{lem:vanishing}
Let $x_0<x_1<\dots<x_n$ be real numbers, and let
$w_i=\prod_{j\ne i}(x_i-x_j)^{-1}$ for $0\le i\le n$. Then
\[
  \sum_{i=0}^nw_i\,x_i^{\,r}=0\qquad(0\le r\le n-1).
\]
\end{lemma}

\begin{proof}
For values $y_0,\dots,y_n$, the Lagrange interpolation polynomial
$p(x)=\sum_iy_iL_i(x)$, with $L_i(x)=\prod_{j\ne i}\frac{x-x_j}{x_i-x_j}$, has degree at most $n$, and
$p(x_k)=y_k$ for every $k$, since $L_i(x_k)$ is $1$ for $k=i$ and $0$ for
$k\ne i$. It is the only such polynomial: the difference of two of them has
degree at most $n$ and vanishes at the $n+1$ points $x_k$, so it is zero. The
numerator of $L_i$ is a product of $n$ factors $x-x_j$, so the coefficient of
$x^n$ in $L_i$ is $w_i$, and in $p$ it is $\sum_iw_iy_i$. Now take
$y_i=x_i^{\,r}$ with $r\le n-1$. By uniqueness, $p(x)=x^r$, whose coefficient
of $x^n$ is $0$.
\end{proof}

\begin{lemma}[Moment-matched weights]\label{lem:weights}
For every integer $s\ge1$ and $\delta>0$, there are finitely supported distributions
on $[0,M]$, for some $M=M(s,\delta)$, satisfying
\eqref{eq:moments} and \eqref{eq:weight-supports}.
\end{lemma}

\begin{proof}
Take an integer $B\ge3$ and the $s+2$ nodes
\[
  x_0=0<x_1=1<x_2=B<x_3=B^2<\dots<x_{s+1}=B^s,
\]
that is, $x_{j+1}=B^j$ for $1\le j\le s$, and let
$w_i=\prod_{j\ne i}(x_i-x_j)^{-1}$. By \cref{lem:vanishing} with $n=s+1$,
$\sum_iw_ix_i^{\,r}=0$ for $0\le r\le s$. In $\prod_{j\ne i}(x_i-x_j)$,
exactly the $s+1-i$ factors with $j>i$ are negative, so the signs of the
$w_i$ alternate.

\emph{The two distributions.} Let $a_i=|w_i|$. Since the signs alternate,
the identity says
\[
  \sum_{i\text{ even}}a_i\,x_i^{\,r}=\sum_{i\text{ odd}}a_i\,x_i^{\,r}
  \qquad(0\le r\le s).
\]
For $r=0$, call the common value $T_0$. The \no{} distribution puts mass
$a_i/T_0$ on $x_i$ for each even $i$, and the \yes{} distribution puts mass
$a_i/T_0$ on $x_i$ for each odd $i$. Both have total mass $1$, and their
moments of every order $r\le s$ agree. Let $\mu_0$ be their common mean.
Dividing every node by $\mu_0$ multiplies the $r$-th moment of both
distributions by $\mu_0^{-r}$, so the moments still agree, and the mean
becomes $1$. This gives \eqref{eq:moments}.

\emph{Sizes of the masses.} The distances from $x_0=0$ to the other nodes are
$1,B,\dots,B^s$, so $a_0=B^{-s(s+1)/2}$. The distances from $x_1=1$ are $1$
and $B^h-1=B^h(1-B^{-h})$ for $1\le h\le s$, so
\[
  \frac{a_1}{a_0}=\prod_{h=1}^s(1-B^{-h})^{-1}\in\Bigl[1,\,1+\frac1{B-2}\Bigr],
\]
using $\prod_h(1-B^{-h})\ge1-\sum_{h\ge1}B^{-h}=1-\frac1{B-1}$. For
$1\le j\le s$, the distances from $x_{j+1}=B^j$ are $B^j$ (to $0$),
$B^j(1-B^{-j})$ (to $1$), $B^j(1-B^{h-j})$ (to $B^h$ with $h<j$), and
$B^h(1-B^{j-h})$ (to $B^h$ with $h>j$). Their powers of $B$ multiply to
$B^{s(s+1)/2+j(j+1)/2}$, and the remaining $s$ factors are each at least
$1/2$. Hence
\[
  \frac{a_{j+1}}{a_0}\le2^s\,B^{-j(j+1)/2}\qquad(1\le j\le s).
\]

\emph{The supports.} The even indices other than $0$ are $i=j+1$ with
$j\ge1$, so $1\le T_0/a_0\le1+s2^s/B$. On the \yes{} side, the odd indices
other than $1$ are $i=j+1$ with $j\ge2$, and for them
$a_{j+1}x_{j+1}/a_0\le2^sB^{-j(j-1)/2}\le2^s/B$. Hence
\[
  \mu_0=\frac{\sum_{i\text{ odd}}a_ix_i}{T_0}
  =\frac{a_1/a_0+O_s(B^{-1})}{1+O_s(B^{-1})}=1+O_s(B^{-1}).
\]
Now take $B$ large, depending on $s$ and $\delta$. The smallest \yes{} weight
is $x_1/\mu_0=1/\mu_0\ge1/2$. Rescaling does not move mass off $0$, so the
\no{} mass away from zero is
$\sum_{i\ge2\text{ even}}a_i/T_0\le s2^s/B\le\delta$. Finally, the largest
node is $B^s$, so both distributions live on $[0,M]$ with $M=B^s/\mu_0$.
This gives \eqref{eq:weight-supports}.
\end{proof}

Moment matching is a standard tool in statistical lower bounds~\cite{WuY19}.
Here its role is to hide which terminals can participate in stars, while
leaving the fraction of such terminals very different in the two worlds.

\section{Replacing the Metric by a Surrogate}\label{sec:gated}

\Cref{sec:gated,sec:analysis} show that an algorithm
making few queries cannot tell $\Dyes$ from $\Dno$: what it sees has almost
the same distribution in both cases. The two input distributions differ only
in the weights, and the weights matter only through the stars. In the metric,
however, the weights have a small effect on every query between a center and a
special terminal. For example, the first such query $\{c,u\}$ at a center is
answered $1$ slightly more often when $W_u$ is large, because a star
containing $u$ is then more likely. In this section we remove these small
effects. We replace each metric distribution $\mathcal D_\bullet$ by a simpler
random process $\mathcal G_\bullet$ that answers the algorithm's queries
adaptively. This process is called the \emph{two-key experiment}. In it, the
answers at a center do not depend on the weights until that center has
answered $1$ twice. \Cref{lem:gated} shows that this replacement changes what
the algorithm sees only slightly. \Cref{sec:analysis} then compares
$\mathcal G_{\mathsf y}$ with $\mathcal G_{\mathsf n}$.

\paragraph{Transcripts.}
Fix a deterministic algorithm $\mathcal A$ making at most $q$ queries,
with no repeated pair. A query is a pair $\{x,y\}$ of points, and its
answer is $w(x,y)$. The \emph{transcript} of a run of $\mathcal A$ is the
sequence $\bigl((p_1,z_1),\dots,(p_k,z_k)\bigr)$ of its queries $p_i$ and
their answers $z_i$, in the order in which they are asked. A \emph{prefix} is
the sequence of its first $i$ entries, for some $0\le i\le k$. We write
$\tau_j$ for the prefix of the first $j-1$ entries, the prefix \emph{before
query $j$}. The next pair asked by $\mathcal A$ is determined by the
preceding answers, and its output is determined by its transcript. For a
distribution $\mathcal D$, write $\trans(\mathcal A,\mathcal D)$ for the
distribution of the transcript when the input is drawn from
$\mathcal D$.

\paragraph{Experiments.}
An \emph{experiment} is a randomized procedure that answers the queries
of $\mathcal A$ one at a time. Unlike an input, it is not a fixed metric: its
answers may depend on the order in which the queries are asked. We run
$\mathcal A$ against an experiment $\mathcal G$ exactly as against an input. This
defines the transcript and its distribution
$\trans(\mathcal A,\mathcal G)$. Experiments are used only in the analysis, and
the hard inputs remain the metrics of \cref{sec:lb-dist}.

\subsection{The two-key experiment}

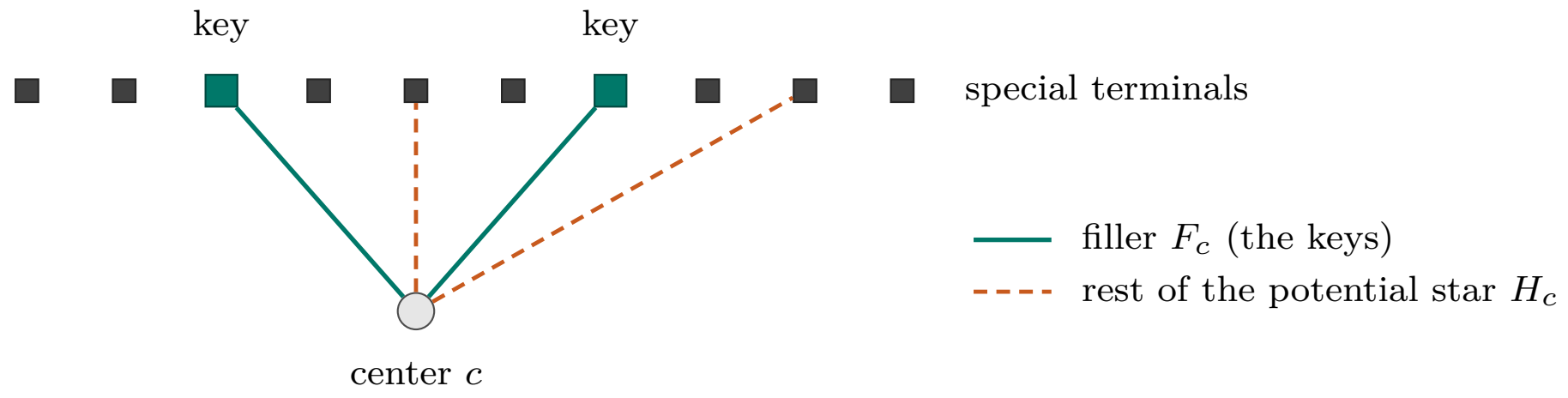
\begin{figure}[ht]
\centering
\begin{tikzpicture}[x=1cm,y=1cm,
  sp/.style={rectangle,draw=black!85,fill=black!75,minimum size=5pt,inner sep=0pt},
  key/.style={rectangle,draw=accent!60!black,fill=accent,minimum size=7pt,inner sep=0pt},
  ctr/.style={circle,draw=black!70,fill=black!10,minimum size=8pt,inner sep=0pt},
  fl/.style={draw=accent,line width=1pt},
  st/.style={draw=warm,line width=0.9pt,dash pattern=on 3pt off 2pt},
  lab/.style={font=\scriptsize}]
\foreach \i in {0,1,3,4,5,7,8,9}{ \node[sp] (t\i) at (0.75*\i,1.7) {}; }
\node[key] (t2) at (1.5,1.7) {};
\node[key] (t6) at (4.5,1.7) {};
\node[ctr] (c) at (3.0,0) {};
\node[lab,below=3pt] at (c.south) {center $c$};
\draw[fl] (c) -- (t2); \draw[fl] (c) -- (t6);
\draw[st] (c) -- (t4); \draw[st] (c) -- (t8);
\node[lab,above=3pt] at (t2.north) {key};
\node[lab,above=3pt] at (t6.north) {key};
\node[lab,anchor=west] at (7.1,1.7) {special terminals};
\draw[fl] (7.3,0.55) -- (7.9,0.55);
\node[lab,anchor=west] at (8.0,0.55) {filler $F_c$ (the keys)};
\draw[st] (7.3,0.15) -- (7.9,0.15);
\node[lab,anchor=west] at (8.0,0.15) {rest of the potential star $H_c$};
\end{tikzpicture}
\caption{A center $c$ in the two-key experiment, with $d=4$. Its filler is
the pair of keys (solid), and its potential star $H_c$ consists of the keys
and $d-2$ further special terminals (dashed). Until both keys have been
queried at $c$, the center answers $1$ exactly for the keys. Afterwards it
also answers $1$ for the other members of $H_c$, unless $H_c$ contains a
terminal that was answered $2$ at $c$ earlier.}
\label{fig:keys}
\end{figure}

In the two-key experiment, each center $c$ draws a \emph{filler} $F_c$, a
random set of at most two special terminals, without using the weights. If
$F_c$ is a pair, its two members are the \emph{keys} of $c$, and $c$ may also
hold a hidden \emph{potential star} $H_c$: a $d$-set that contains both keys
and is drawn using the weights. The center answers like its filler until both
keys have been queried at $c$. Only then can its star affect the answers
(\cref{fig:keys}). Thus each star stays locked until both of its keys have been queried at its
center. Formally,
the experiment is defined as follows.

Keep the same random bijection and weights. Independently for every center
$c$, sample a weight-independent filler $F_c$ with
\[
  \Prob[F_c=\varnothing]=\tfrac12,\qquad
  \Prob[F_c=\{u\}]=\frac1{4m},\qquad
  \Prob[F_c=\{u,v\}]=\beta.
\]
On singletons and pairs these are the probabilities of
\eqref{eq:static-law}. Unlike \eqref{eq:static-law}, the filler puts
probability $\frac12$ on $\varnothing$ instead of $\frac12-\sum_Ha_W(H)$, and
it does not depend on the weights.
If $F_c=e$ is a pair, its two members are the \emph{keys} of $c$, and
we sample a \emph{potential star} $H_c$ by
\begin{equation}\label{eq:potential-star}
  \Prob[H_c=H\mid F_c=e,W]=\frac{a_W(H)}{\beta}
  \quad\text{for every }d\text{-set }H\supseteq e,
\end{equation}
and leave it absent with the remaining probability. For other fillers
$H_c$ is always absent. We represent absence by $H_c=\varnothing$.
Potential stars are independent given the weights and fillers.

Until both keys have been
queried at $c$, answer a query $\{c,u\}$ by $1$ exactly when $u\in F_c$.
The query of the second key is still answered by this rule (so it
is answered $1$) and then \emph{activates} $c$. Let $B_c$ be the special
terminals queried at $c$ before activation that are not keys. They were
all answered $2$. After activation, answer by $1$
exactly when
\[
  u\in F_c
  \quad\text{or}\quad
  \bigl(H_c\text{ exists},\ H_c\cap B_c=\varnothing,\ u\in H_c\bigr).
\]
A potential star meeting $B_c$ is \emph{discarded}: it never affects
answers, but we keep it as a hidden object. This avoids contradicting an
earlier negative answer. All queries other than center--special-terminal
queries use \eqref{eq:lb-metric}. Denote the two experiments by
$\mathcal G_{\mathsf y},\mathcal G_{\mathsf n}$. In particular, the
answers at $c$ depend on the weights only after $c$ is activated, that is,
after $c$ has answered $1$ to both of its keys.

The following bound both validates \eqref{eq:potential-star} for large $R$
and records its useful sparsity. For any prescribed set $P$ of $p\le d$
special positions,
\begin{equation}\label{eq:prescribed}
  \sum_{H\supseteq P}a_W(H)
  \le\frac{H_0M^d}{dR}\,
      \frac{\binom{m-p}{d-p}}{\binom md}
  {}\le\frac{H_0M^d\,d^{\,p-1}}{Rm^p}
  =O\left(\frac1{Rm^p}\right).
\end{equation}
Indeed, every weight is at most $M$, so each term is at most
$H_0M^d/(dR\binom md)$, there are $\binom{m-p}{d-p}$ sets $H\supseteq P$, and
$\binom{m-p}{d-p}/\binom md=\prod_{i<p}\frac{d-i}{m-i}\le(d/m)^p$.
In particular, for every pair $e$ and every $u\notin e$,
\begin{equation}\label{eq:mark-bounds}
  \Prob[H_c\text{ exists}\mid F_c=e,W]\le\frac{B_0}{R},\qquad
  \Prob[u\in H_c\mid F_c=e,W]\le\frac{B_1}{Rm},
\end{equation}
where $B_0=2(d-1)H_0M^d$ and $B_1=2(d-2)B_0$. To see this, divide
\eqref{eq:prescribed} by $\beta=1/(2m(m-1))$: for $P=e$ it gives $B_0/R$, and
for $P=e\cup\{u\}$ it gives $(d-2)B_0/(R(m-2))\le B_1/(Rm)$, since $m\ge4$.
These estimates hold for \emph{every} realization of the weights.
We henceforth take $R\ge2B_0$. Then the probabilities in
\eqref{eq:potential-star} add up to at most $B_0/R\le1/2$, so
\eqref{eq:potential-star} is a valid distribution, and $H_c$ is absent with
probability at least $1/2$.

\subsection{Comparison with the metric}\label{sec:gated-compare}

\begin{lemma}[Comparison with the metric]\label{lem:gated}
For either $\bullet\in\{\mathsf y,\mathsf n\}$,
\begin{equation}\label{eq:gated-error}
  \dtv\bigl(\trans(\mathcal A,\mathcal D_\bullet),
            \trans(\mathcal A,\mathcal G_\bullet)\bigr)
  \le C\left(\sqrt{\frac q{mR^2}}+\frac q{mR^2}\right).
\end{equation}
\end{lemma}

\setcounter{claim}{0}%
The rest of this subsection proves \cref{lem:gated}. Fix $\bullet$, and let
$C_0=dH_0M^d$, so that $B_0\le2C_0$. We assume $m\ge8$ and $R\ge4C_0$ (so
that $R\ge2B_0$). Both hold for all large $t$, since $C_0$ depends only on
$\eta$ and $s$.

\paragraph{Plan.}
We use a third experiment, the \emph{hybrid} $\mathsf{Hyb}$, defined below.
At each center it answers like $\mathcal G_\bullet$ until the center has
answered $1$ twice, and like $\mathcal D_\bullet$ afterwards. For fixed $\pi$
and $W$ (see the reductions below), we show that
\begin{align}
  \dtv\bigl(\trans(\mathcal A,\mathcal D_\bullet),
            \trans(\mathcal A,\mathsf{Hyb})\bigr)
  &\le32C_0\sqrt{\frac q{mR^2}},\label{eq:gated-hyb}\\
  \dtv\bigl(\trans(\mathcal A,\mathsf{Hyb}),
            \trans(\mathcal A,\mathcal G_\bullet)\bigr)
  &\le\frac{16C_0^2\,q}{mR^2}.\label{eq:hyb-exp}
\end{align}
The triangle inequality then gives \eqref{eq:gated-error}. The two bounds
need different arguments. Before a
center's second answer $1$, $\mathcal D_\bullet$ and $\mathsf{Hyb}$ answer
each query with slightly different probabilities, and we add up these small
differences using relative entropy. The hybrid and $\mathcal G_\bullet$
differ only in the set from which a center answers after its second answer
$1$. These sets differ with probability $O(1/R^2)$ at each center, and in
expectation only $O(q/m)$ centers receive a second answer $1$. Here we use a
coupling.

\paragraph{Reductions.}
First, we may assume that every transcript has exactly $q$ entries: if
$\mathcal A$ stops after $k<q$ queries, we append $q-k$ copies of a symbol
$\bot$. The padded transcript and the original one determine each other, so
by \cref{lem:prob}(b) this changes none of the distances in
\eqref{eq:gated-error}--\eqref{eq:hyb-exp}.

Second, we may fix $\pi$ and $W$. The pair $(\pi,W)$ has the same
distribution under $\mathcal D_\bullet$ and $\mathcal G_\bullet$, so by
\cref{lem:prob}(c) it suffices to prove \eqref{eq:gated-error} conditionally
on every value of $(\pi,W)$, with a bound that does not depend on this value.
From now on, $\pi$ and $W$ are fixed, all probabilities are conditional on
them, and the hybrid is defined in this setting.

Third, by \eqref{eq:lb-metric}, the fixed $\pi$ determines the answer to every
query between two terminals ($\lambda$ or $2$), between two centers ($2$), and
between a center and an ordinary terminal ($2$), in $\mathcal D_\bullet$,
$\mathsf{Hyb}$ and $\mathcal G_\bullet$ alike. Only a query $\{c,u\}$ between
a center $c$ and a special terminal $u$ has a random answer, $1$ or $2$.

\paragraph{Answer rules and local histories.}
Since $\mathcal A$ is deterministic, the query that follows a prefix $\tau$
is determined by $\tau$, and we call it the \emph{next query}. The \emph{answer
rule} of an experiment gives, for every prefix $\tau$ of positive
probability, the probability of each answer to the next query given $\tau$.
The probability of a transcript is the product of the probabilities of its
answers given the preceding prefixes, so the answer rule determines the
transcript distribution.

For a prefix $\tau$ and a center $c$, the \emph{local history} of $c$ in
$\tau$ is the pair $(P,B)$, where $P$ is the set of special terminals $u$
such that $\{c,u\}$ was answered $1$ in $\tau$, and $B$ is the set of those
answered $2$. If $|P|\ge2$, let $e$ be the set of the first two terminals
answered $1$ at $c$, and $B^\circ$ the set of terminals answered $2$ at $c$
before the second $1$. Thus $e\subseteq P$ and $B^\circ\subseteq B$, and $e$
and $B^\circ$ do not change after the second answer $1$ at $c$.

\paragraph{Three functions.}
We use two distributions on subsets $N\subseteq[m]$:
\begin{itemize}
  \item the distribution $\Phi$ of a filler: $\Phi(\varnothing)=\frac12$,
        $\Phi(\{v\})=\frac1{4m}$ for each $v$, $\Phi(\{v,v'\})=\beta$ for
        each pair, and $\Phi(N)=0$ for every other set $N$;
  \item the distribution $\mu_W$ of a neighbor set, given by
        \eqref{eq:static-law}: equal to $\Phi$ on singletons and pairs,
        $\mu_W(H)=a_W(H)$ for each $d$-set $H$,
        $\mu_W(\varnothing)=\frac12-\sigma$, where
        $\sigma=\sum_{|H|=d}a_W(H)$, and $\mu_W(N)=0$ for every other set
        $N$.
\end{itemize}
For a distribution $\psi$ on subsets of $[m]$ and disjoint
$P,B\subseteq[m]$, let
\[
  \psi(P,B)=\sum_{N\supseteq P,\ N\cap B=\varnothing}\psi(N)
\]
be the probability that a set drawn from $\psi$ contains $P$ and avoids $B$.
If $(P,B)$ is the local history of a center whose neighbor set is drawn from
$\psi$, then $\psi(P,B)$ is the probability that this center gives exactly the
recorded answers. Let
\[
  f(P,B)=\Phi(P,B),\qquad g(P,B)=\mu_W(P,B),\qquad
  s(P,B)=\sum_{\substack{|H|=d,\ H\supseteq P\\ H\cap B=\varnothing}}a_W(H).
\]
The next claim says that $g$ is $f$ plus a small correction.

\begin{claim}\label{cl:fgs}
Let $P,B\subseteq[m]$ be disjoint, let $p=|P|$ and $b=|B|$, and let
$u,v\notin B$ be distinct.
\begin{enumerate}[label=(\roman*)]
  \item $g(P,B)=f(P,B)+s(P,B)-\sigma\cdot\ind[P=\varnothing]$.
  \item $f(\varnothing,B)=\frac12+\frac{m-b}{4m}+\beta\binom{m-b}2
        \in\bigl[\frac12,1\bigr]$,\quad
        $f(\{v\},B)=\frac1{4m}+\beta(m-b-1)\in\bigl[\frac1{4m},\frac3{4m}\bigr]$,
        \quad and $f(\{v,u\},B)=\beta$.
  \item If $p\le2$, then $0\le s(P,B)\le C_0/(Rm^p)$. Moreover,
        $0\le s(\varnothing,B)\le\sigma\le C_0/R\le\frac14$.
\end{enumerate}
\end{claim}

\begin{claimproof}{\cref{cl:fgs}}
(i) The two distributions agree on singletons and pairs. On $\varnothing$,
$\mu_W$ has $\sigma$ less than $\Phi$, and $\varnothing$ contains $P$ only if
$P=\varnothing$. Finally, $\mu_W$ also gives probability to the $d$-sets, and
those that contain $P$ and avoid $B$ have total probability $s(P,B)$.

(ii) A set of positive $\Phi$-probability that avoids $B$ is $\varnothing$,
one of the $m-b$ singletons outside $B$, or one of the $\binom{m-b}2$ pairs
outside $B$. Such a set that contains $v$ is $\{v\}$ or one of the $m-b-1$
pairs $\{v,w\}$ with $w\notin B\cup\{v\}$, and the only such set that
contains both $v$ and $u$ is $\{v,u\}$. The ranges follow from
$\beta(m-1)=\frac1{2m}$ and $\beta\binom m2=\frac14$.

(iii) By \eqref{eq:prescribed},
$s(P,B)\le\sum_{H\supseteq P}a_W(H)\le H_0M^dd^{\,p-1}/(Rm^p)\le C_0/(Rm^p)$
for $p\le2$. Also $s(\varnothing,B)\le s(\varnothing,\varnothing)=\sigma$,
the case $p=0$ gives $\sigma\le C_0/R$, and $C_0/R\le\frac14$ since
$R\ge4C_0$.
\end{claimproof}

\paragraph{The metric and the experiment.}
For a pair $e$ and a set $B^\circ$ disjoint from $e$, let $\mathcal H$ be the
family of $d$-sets $H\supseteq e$ with $H\cap B^\circ=\varnothing$, and let
$a=\sum_{H\in\mathcal H}a_W(H)$. Define two distributions on subsets of
$[m]$, both supported on $\{e\}\cup\mathcal H$:
\[
  \psi^{\mathrm{hyb}}(e)=\frac\beta{\beta+a},\quad
  \psi^{\mathrm{hyb}}(H)=\frac{a_W(H)}{\beta+a};
  \qquad
  \psi^{\mathrm{exp}}(e)=1-\frac a\beta,\quad
  \psi^{\mathrm{exp}}(H)=\frac{a_W(H)}\beta
  \qquad(H\in\mathcal H).
\]
By \eqref{eq:mark-bounds}, $a\le B_0\beta/R\le\beta/2$, so both are
distributions. Both describe the set from which a center $c$ answers after
its second answer $1$, given its answers until then. In $\mathcal D_\bullet$
this set is $\nb(c)$, whose conditional distribution is $\psi^{\mathrm{hyb}}$
(\cref{cl:hyb} below), and $\mathcal G_\bullet$ draws it from
$\psi^{\mathrm{exp}}$ (\cref{cl:rules}).

From now on, let $\tau$ be a prefix of positive probability in the experiment
considered, let $\{c,u\}$ be the next query, with $u$ special, and let
$(P,B)$ be the local history of $c$ in $\tau$. Then $u\notin P\cup B$, since
no pair is queried twice. If $|P|\ge2$, we use $\psi^{\mathrm{hyb}}$ and
$\psi^{\mathrm{exp}}$ with the $e$ and $B^\circ$ of $c$. The next claim writes
the answer rules of $\mathcal D_\bullet$ and $\mathcal G_\bullet$ in the same
form: a ratio of probabilities of two local histories.

\begin{claim}\label{cl:rules}
Given $\tau$, the probability that the answer to $\{c,u\}$ is $1$ is:
\begin{enumerate}[label=(\roman*)]
  \item in $\mathcal D_\bullet$:\quad
        $\displaystyle x=\frac{g(P\cup\{u\},B)}{g(P,B)}$;
  \item in $\mathcal G_\bullet$:\quad
        $\displaystyle x_0=\frac{f(P\cup\{u\},B)}{f(P,B)}$ if $|P|\le1$,
        \quad and\quad
        $\displaystyle\frac{\psi^{\mathrm{exp}}(P\cup\{u\},B)}
                           {\psi^{\mathrm{exp}}(P,B)}$ if $|P|\ge2$.
\end{enumerate}
\end{claim}

\begin{claimproof}{\cref{cl:rules}}
Since $\pi$ and $W$ are fixed, the random data of different centers are
independent: the neighbor sets in $\mathcal D_\bullet$, and the pairs
$(F_{c'},H_{c'})$ in $\mathcal G_\bullet$. As $\mathcal A$ is deterministic
and all other answers are fixed, $\tau$ occurs exactly when the data of every
center produce the answers recorded at that center. Hence, given $\tau$, the
data of $c$ have their prior distribution conditioned on producing the
answers recorded at $c$, and the probability of the answer $1$ depends only on
these answers.

(i) In $\mathcal D_\bullet$, the center $c$ answers $1$ exactly for the
members of $\nb(c)$, which has distribution $\mu_W$. So the probability that
the answer is $1$ is the probability that $\nb(c)$ contains $P\cup\{u\}$ and
avoids $B$, divided by the probability that $\nb(c)$ contains $P$ and avoids
$B$. This is $x$.

(ii) In $\mathcal G_\bullet$, the center $c$ is activated exactly at its
second answer $1$, and up to and including that query, it answers $1$ exactly
for the members of $F_c$, which has distribution $\Phi$. If $|P|\le1$, then
$c$ is not activated, and the query $\{c,u\}$ is also answered by membership
in $F_c$, even if it activates $c$. As in (i), this gives $x_0$.

If $|P|\ge2$, the answers at $c$ before its second $1$ are produced exactly
when $F_c\supseteq e$ and $F_c\cap B^\circ=\varnothing$, that is, when
$F_c=e$ (as $|F_c|\le2$). Then $c$ was activated with $B_c=B^\circ$, and
afterwards it answers $1$ exactly for the members of the set $N_c$ that
equals $H_c$ if $H_c$ exists and avoids $B^\circ$, and equals $e$ otherwise.
Given $F_c=e$, \eqref{eq:potential-star} shows that $N_c$ has distribution
$\psi^{\mathrm{exp}}$. Every set in its support contains $e$ and avoids
$B^\circ$, so the answers after activation are produced exactly when $N_c$
contains $P$ and avoids $B$. Hence the probability that the answer is $1$ is
$\Phi(e)\,\psi^{\mathrm{exp}}(P\cup\{u\},B)$ divided by
$\Phi(e)\,\psi^{\mathrm{exp}}(P,B)$.
\end{claimproof}

\paragraph{The hybrid.}
Let $\mathsf{Hyb}$ be the experiment whose answer rule gives every query with
a fixed answer that answer, and, in the situation above,
\[
  \Prob_{\mathsf{Hyb}}[\text{the answer is }1\mid\tau]=
  \begin{cases}
    x_0=\dfrac{f(P\cup\{u\},B)}{f(P,B)}, & |P|\le1,\\[10pt]
    \dfrac{\psi^{\mathrm{hyb}}(P\cup\{u\},B)}{\psi^{\mathrm{hyb}}(P,B)},
      & |P|\ge2.
  \end{cases}
\]
The next claim shows that, at a center that has answered $1$ at least
twice, $\mathsf{Hyb}$ answers like $\mathcal D_\bullet$.

\begin{claim}\label{cl:hyb}
If $|P|\ge2$, then
$\psi^{\mathrm{hyb}}(P\cup\{u\},B)/\psi^{\mathrm{hyb}}(P,B)=x$.
\end{claim}

\begin{claimproof}{\cref{cl:hyb}}
Under $\mu_W$, the set $e$ has probability $\beta$, each $H\in\mathcal H$ has
probability $a_W(H)$, and every other set that contains $e$ and avoids
$B^\circ$ has probability $0$. These probabilities add up to $\beta+a$, so
$\psi^{\mathrm{hyb}}$ is $\mu_W$ conditioned on containing $e$ and avoiding
$B^\circ$. If $P'\supseteq e$ and $B'\supseteq B^\circ$, then every set that
contains $P'$ and avoids $B'$ also contains $e$ and avoids $B^\circ$, so
$g(P',B')=(\beta+a)\,\psi^{\mathrm{hyb}}(P',B')$. Apply this to
$(P\cup\{u\},B)$ and to $(P,B)$, and divide.
\end{claimproof}

In summary, the probability that a query to a center $c$ is answered $1$ is
given by the following table.
\begin{center}
\renewcommand{\arraystretch}{1.25}
\begin{tabular}{lccc}
  \toprule
  & $\mathcal D_\bullet$ & $\mathsf{Hyb}$ & $\mathcal G_\bullet$\\
  \midrule
  $c$ has answered $1$ at most once & $x$ & $x_0$ & $x_0$\\
  $c$ has answered $1$ at least twice & $x$ & $x$ (\cref{cl:hyb})
    & $\psi^{\mathrm{exp}}$-ratio\\
  \bottomrule
\end{tabular}
\end{center}
Thus $\mathcal D_\bullet$ and $\mathsf{Hyb}$ differ only at queries to
centers that have answered $1$ at most once, and $\mathsf{Hyb}$ and
$\mathcal G_\bullet$ differ only at queries to centers that have answered $1$
at least twice.

\paragraph{Proof of \texorpdfstring{\eqref{eq:gated-hyb}}{(8.5)}.}
The next claim says that before a center's second answer $1$, the metric and
the hybrid answer $1$ with almost the same probability, which is of order
$1/m$.

\begin{claim}\label{cl:close}
If $|P|\le1$, then
\[
  \frac1{4m}\le x_0\le\frac2{m-1}\le\frac12
  \qquad\text{and}\qquad
  |x-x_0|\le\frac{16C_0}{Rm}.
\]
\end{claim}

\begin{claimproof}{\cref{cl:close}}
Let $F=f(P,B)$, $F'=f(P\cup\{u\},B)$, $\Delta=g(P,B)-F$ and
$\Delta'=g(P\cup\{u\},B)-F'$. Then $x_0=F'/F$ and
$x=(F'+\Delta')/(F+\Delta)$, so
\[
  x-x_0=\frac{F(F'+\Delta')-F'(F+\Delta)}{F(F+\Delta)}
       =\frac{\Delta'-x_0\Delta}{F+\Delta},
  \qquad\text{hence}\qquad
  |x-x_0|\le\frac{|\Delta'|+x_0|\Delta|}{F+\Delta}.
\]
We use \cref{cl:fgs} throughout.

\emph{Case $P=\varnothing$.} Here $F\in[\frac12,1]$ and
$F'=f(\{u\},B)\in[\frac1{4m},\frac3{4m}]$, so
$x_0\in[\frac1{4m},\frac3{2m}]$. Moreover,
\[
  \Delta=s(\varnothing,B)-\sigma\in\Bigl[-\frac{C_0}R,\,0\Bigr],\qquad
  \Delta'=s(\{u\},B)\in\Bigl[0,\,\frac{C_0}{Rm}\Bigr].
\]
Hence $F+\Delta\ge\frac12-\frac14=\frac14$, and
\[
  |x-x_0|\le4\Bigl(\frac{C_0}{Rm}+\frac3{2m}\cdot\frac{C_0}R\Bigr)
  =\frac{10\,C_0}{Rm}.
\]

\emph{Case $P=\{v\}$.} Here $F=f(\{v\},B)\in[\frac1{4m},\frac3{4m}]$ and
$F'=\beta=\frac1{2m(m-1)}$, so
$x_0=\beta/F\in[\frac2{3(m-1)},\frac2{m-1}]$. Moreover,
\[
  \Delta=s(\{v\},B)\in\Bigl[0,\,\frac{C_0}{Rm}\Bigr],\qquad
  \Delta'=s(\{v,u\},B)\in\Bigl[0,\,\frac{C_0}{Rm^2}\Bigr].
\]
Hence $F+\Delta\ge F\ge\frac1{4m}$, and, as $x_0\le\frac2{m-1}\le\frac3m$,
\[
  |x-x_0|\le4m\Bigl(\frac{C_0}{Rm^2}+\frac3m\cdot\frac{C_0}{Rm}\Bigr)
  =\frac{16\,C_0}{Rm}.
\]

In both cases $\frac1{4m}\le x_0\le\frac2{m-1}$, since
$\frac2{3(m-1)}\ge\frac1{4m}$ and $\frac3{2m}\le\frac2{m-1}$. Finally,
$\frac2{m-1}\le\frac12$ as $m\ge8$.
\end{claimproof}

Now apply the chain rule, \cref{lem:prob}(d), to the transcript
distributions of $\mathcal D_\bullet$ and $\mathsf{Hyb}$. Its hypothesis
holds: at a query to a center that has answered $1$ at most once,
$\mathsf{Hyb}$ gives both answers positive probability, as $0<x_0<1$, and at
every other step the two answer rules are equal. Consider the term of step
$t$, for a prefix drawn from $\mathcal D_\bullet$.
\begin{itemize}
  \item If the next query has a fixed answer (or is $\bot$), both
        conditional distributions are the same point mass, so the term is
        $0$.
  \item If the next query is to a center that has answered $1$ at least
        twice, the two answer rules are equal by \cref{cl:hyb}, so the term
        is $0$.
  \item If the next query is to a center that has answered $1$ at most
        once, the two conditional distributions of the answer are $\Ber(x)$
        and $\Ber(x_0)$, where the outcome $1$ stands for the answer $1$. By
        \cref{cl:close}, $x_0(1-x_0)\ge\frac1{4m}\cdot\frac12=\frac1{8m}$,
        and by \cref{lem:prob}(f) and \cref{cl:close},
        \[
          \KL\bigl(\Ber(x)\Vert\Ber(x_0)\bigr)
          \le\frac{(x-x_0)^2}{x_0(1-x_0)}
          \le\Bigl(\frac{16C_0}{Rm}\Bigr)^2\cdot8m
          =\frac{2048\,C_0^2}{R^2m}.
        \]
\end{itemize}
So each of the $q$ terms is at most $2048C_0^2/(R^2m)$, and the relative
entropy between the two transcript distributions is at most
$2048C_0^2q/(R^2m)$. By Pinsker's inequality, \cref{lem:prob}(e), the total
variation distance is at most $\sqrt{1024\,C_0^2q/(R^2m)}=32C_0\sqrt{q/(mR^2)}$.
This proves \eqref{eq:gated-hyb}.

\paragraph{Proof of \texorpdfstring{\eqref{eq:hyb-exp}}{(8.6)}.}
The hybrid and $\mathcal G_\bullet$ differ only in the distribution of the
set from which a center answers after its second answer $1$. The next claim
shows that these two distributions are very close.

\begin{claim}\label{cl:psi}
For every pair $e$ and every set $B^\circ$ disjoint from $e$,
\[
  \dtv\bigl(\psi^{\mathrm{hyb}},\psi^{\mathrm{exp}}\bigr)
  =\frac{a^2}{\beta(\beta+a)}\le\Bigl(\frac{B_0}R\Bigr)^2.
\]
\end{claim}

\begin{claimproof}{\cref{cl:psi}}
For $H\in\mathcal H$,
\[
  \psi^{\mathrm{exp}}(H)-\psi^{\mathrm{hyb}}(H)
  =a_W(H)\Bigl(\frac1\beta-\frac1{\beta+a}\Bigr)
  =\frac{a\,a_W(H)}{\beta(\beta+a)}\ge0,
\]
and these differences add up to $a^2/(\beta(\beta+a))$. For $e$,
\[
  \psi^{\mathrm{hyb}}(e)-\psi^{\mathrm{exp}}(e)
  =\frac\beta{\beta+a}-\frac{\beta-a}\beta
  =\frac{\beta^2-(\beta-a)(\beta+a)}{\beta(\beta+a)}
  =\frac{a^2}{\beta(\beta+a)}.
\]
The total variation distance is half the sum of the absolute differences,
which gives the equality. Finally,
$a^2/(\beta(\beta+a))\le(a/\beta)^2$, and $a\le B_0\beta/R$ by
\eqref{eq:mark-bounds}.
\end{claimproof}

To use \cref{cl:psi}, we run $\mathsf{Hyb}$ and $\mathcal G_\bullet$ in a form
in which each center draws, at its second answer $1$, the whole set from
which it answers afterwards. This is the principle of deferred decisions.

\begin{claim}\label{cl:drawn}
Let $\psi$ be $\psi^{\mathrm{hyb}}$ or $\psi^{\mathrm{exp}}$. The following
procedure has the answer rule of $\mathsf{Hyb}$, respectively of
$\mathcal G_\bullet$. A query with a fixed answer gets that answer. A query to
a center that has answered $1$ at most once is answered $1$ with probability
$x_0$. When a center $c$ gives its second answer $1$, it draws a set $N_c$
from $\psi$, with the $e$ and $B^\circ$ of $c$, independently of everything
else. At every later query $\{c,u\}$, it answers $1$ exactly when
$u\in N_c$.
\end{claim}

\begin{claimproof}{\cref{cl:drawn}}
At a query to a center that has answered $1$ at most once, the procedure
follows the rule by construction. Now consider a query $\{c,u\}$ after $c$
has drawn $N_c$, and let $(P,B)$ be the local history of $c$. Every set $N$
in the support of $\psi$ contains $e$ and avoids $B^\circ$, so the answers
given at $c$ since the draw agree with $N_c=N$ exactly when $N\supseteq P$
and $N\cap B=\varnothing$. Since $N_c$ was drawn independently of everything
else, the probability of the prefix together with the event $N_c=N$ is a
factor that does not depend on $N$, times
$\psi(N)\,\ind[N\supseteq P,\ N\cap B=\varnothing]$. So, given the prefix,
$N_c=N$ with probability
$\psi(N)\,\ind[N\supseteq P,\ N\cap B=\varnothing]/\psi(P,B)$, and the answer
is $1$ with probability $\psi(P\cup\{u\},B)/\psi(P,B)$. This is the rule of
$\mathsf{Hyb}$ for $\psi=\psi^{\mathrm{hyb}}$, and the rule of
$\mathcal G_\bullet$ for $\psi=\psi^{\mathrm{exp}}$, by \cref{cl:rules}.
\end{claimproof}

We now run the two experiments together. Let
$U_1,\dots,U_q,V_1,\dots,V_q$ be independent and uniform in $[0,1]$. For all
$e$ and $B^\circ$, \cref{lem:prob}(a) gives a pair of random sets with
distributions $\psi^{\mathrm{hyb}}$ and $\psi^{\mathrm{exp}}$ that differ with
probability $\dtv(\psi^{\mathrm{hyb}},\psi^{\mathrm{exp}})$. Fix a map that
turns a uniform number into such a pair. Run $\mathsf{Hyb}$ and
$\mathcal G_\bullet$ side by side, each in the form of \cref{cl:drawn}. At
step $t$, each run looks only at its own past, and:
\begin{itemize}
  \item a query with a fixed answer gets that answer;
  \item a query to a center that has answered $1$ at most once is answered
        $1$ exactly when $U_t<x_0$, with $x_0$ computed from the run's own
        past;
  \item if this answer gives the center its second $1$, the run feeds $V_t$
        into the map for its own $e$ and $B^\circ$. The hybrid run keeps the
        first set of the resulting pair as $N_c$, and the experiment run keeps
        the second;
  \item a later query to that center is answered by membership in the run's
        own $N_c$.
\end{itemize}
Each run uses $U_t$ and $V_t$ only through its own past, exactly as it would
alone. So, by \cref{cl:drawn}, the hybrid run has the transcript distribution
of $\mathsf{Hyb}$, and the experiment run that of $\mathcal G_\bullet$.

Suppose that the two transcripts agree before step $t$. Then the two runs ask
the same query, have the same local histories, and compute the same $x_0$,
$e$ and $B^\circ$. So they give the same answer to a query with a fixed answer
and to a query to a center that has answered $1$ at most once, since they use
the same $U_t$. At a query to a center $c$ that has answered $1$ at least
twice, they answer by membership in their own sets $N_c$. These two sets were
obtained from the same $V_s$ and the same $e$ and $B^\circ$, at the step $s$
where $c$ got its second answer $1$. If they are equal, the answers agree.

Call a step an \emph{activation} if it gives some center its second answer
$1$. Let $E_s$ be the event that the runs agree before step $s$, that step $s$
is an activation, and that the two sets obtained from $V_s$ differ. By the
previous paragraph, the transcripts differ only if some $E_s$ occurs. The
first two conditions are decided by $U_1,\dots,U_s$ and
$V_1,\dots,V_{s-1}$, and $V_s$ is independent of these. So, given the first
two conditions, the two sets differ with probability at most $(B_0/R)^2$ by
\cref{cl:psi}, and
\[
  \Prob[E_s]\le\Bigl(\frac{B_0}R\Bigr)^2
  \Prob[\text{step $s$ is an activation in the hybrid run}].
\]
Step $s$ can be an activation only if it is a query to a center that has
answered $1$ exactly once, and then it is one with probability
$x_0\le2/(m-1)$ given the past, by \cref{cl:close}. So the expected number of
activations in the hybrid run is at most $2q/(m-1)\le4q/m$. By
\cref{lem:prob}(a) and $B_0\le2C_0$,
\[
  \dtv\bigl(\trans(\mathcal A,\mathsf{Hyb}),
            \trans(\mathcal A,\mathcal G_\bullet)\bigr)
  \le\Prob[\text{the transcripts differ}]
  \le\sum_{s=1}^q\Prob[E_s]
  \le\Bigl(\frac{B_0}R\Bigr)^2\frac{4q}m
  \le\frac{16C_0^2\,q}{mR^2}.
\]
This proves \eqref{eq:hyb-exp}.

\begin{proof}[Proof of \cref{lem:gated}]
For fixed $\pi$ and $W$, \eqref{eq:gated-hyb}, \eqref{eq:hyb-exp} and the
triangle inequality give
\[
  \dtv\bigl(\trans(\mathcal A,\mathcal D_\bullet),
            \trans(\mathcal A,\mathcal G_\bullet)\bigr)
  \le32C_0\sqrt{\frac q{mR^2}}+\frac{16C_0^2\,q}{mR^2}
  \le C\left(\sqrt{\frac q{mR^2}}+\frac q{mR^2}\right)
\]
with $C=\max\{32C_0,16C_0^2\}$. This bound does not depend on $\pi$ and $W$,
so by the reductions it holds for random $\pi$ and $W$ and for the original
transcripts.
\end{proof}

\subsection{Erasing an unseen star}\label{sec:erase}

From now on, an answer $1$ to a query between a center and a special
terminal is called \emph{positive}. The next lemma concerns
$\mathcal G_\bullet$ alone. Its hidden data are $\pi$, $W$, the fillers, and
the potential stars. Together with $\mathcal A$, they determine the
transcript. Fix a center $c$, and let $Z$ be all hidden data other than
$H_c$. \emph{Deleting} the potential star of $c$ means making $H_c$ absent
while keeping $Z$.

\begin{lemma}[Erasing an unseen star]\label{lem:erase}
Let $\tau$ be a prefix in which $c$ has at most two positive answers.
\begin{enumerate}[label=(\alph*)]
  \item If the transcript begins with $\tau$, then it still begins with
        $\tau$ after the potential star of $c$ is deleted.
  \item For every value $z$ of $Z$ such that the event that $Z=z$ and
        the transcript begins with $\tau$ has positive probability,
        \[
          \Prob[H_c\text{ is present}\mid Z=z,\ \text{the transcript begins
          with }\tau]\le\frac{B_0}R .
        \]
\end{enumerate}
\end{lemma}

\begin{proof}
(a) Compare the runs with and without $H_c$, and suppose that they agree
before some query of $\tau$. Then $\mathcal A$ asks the same query in both.
Only the answer to a query $\{c,u\}$ with $u$ special can depend on $H_c$. Up
to and including its activating query, $c$ answers by membership in $F_c$, and
its activation time and $B_c$ depend only on $F_c$ and on the queries at $c$,
which agree. After activation, the star can only change an answer $2$ into
$1$, for some $u\notin F_c$. Activation happens only after both members of
$F_c$ were answered $1$ at $c$, so such an answer would be a third positive
answer at $c$ in $\tau$, which is excluded. Hence the two answers agree, and
by induction the two runs agree on all of $\tau$.

(b) If $F_c$ is not a pair (this is part of $z$), then $H_c$ is absent. So
let $F_c=e$. Potential stars are independent given the weights and the
fillers, so given $Z=z$, the star $H_c$ still has the distribution
\eqref{eq:potential-star}, and it is present with probability $p\le B_0/R$ by
\eqref{eq:mark-bounds}. Given $Z=z$, the value of $H_c$ determines whether
the transcript begins with $\tau$. Some value does, so by (a) absence does.
Let $y\le p$ be the probability, given $Z=z$, that $H_c$ is present and the
transcript begins with $\tau$. Then the probability in (b) is
\[
  \frac{y}{(1-p)+y}\le\frac{p}{(1-p)+p}=p\le\frac{B_0}R ,
\]
as $y\mapsto y/((1-p)+y)$ is increasing. In words, conditioning on $\tau$
may rule out some potential stars, but never their absence.
\end{proof}

Since (b) holds for every value of $Z$, the same bound holds given that
the transcript begins with $\tau$ together with any part of the other hidden
data.

\section{Lower-Bound Analysis}\label{sec:analysis}

We now compare the two-key experiments $\mathcal G_{\mathsf y}$ and
$\mathcal G_{\mathsf n}$ and prove \cref{thm:lower}. \Cref{sec:reveal} shows
that few queries single out only a few special terminals. \Cref{sec:coupling}
shows that the potential stars visible from these terminals reveal little
about the weights. \Cref{sec:lb-proof} combines these bounds with
\cref{lem:gated}.

\subsection{What few queries can reveal}\label{sec:reveal}

We work in either two-key experiment. The aim is to show that
few queries can single out only a few special terminals. Several
components of this subsection, such as the bookkeeping of noticed terminals and
the exchange arguments of \cref{lem:cluster,lem:special}, are similar to
those in earlier works~\cite{CKT23,MTV26}. The new ingredients handle potential stars that
have not yet affected any answer.

\subsubsection{Noticed and exposed terminals}\label{sec:bookkeeping}

Put $\theta=1/16$ and $D=\lfloor\theta A\rfloor$. Recall that $\tau_j$ is the
prefix before query $j$, and let $\deg_j(u)$ count earlier queries
involving $u$. An ordinary terminal is \emph{marked} once it receives a
$\lambda$-answer. For an ordinary cluster $G$, let $e_j(G)$ be the number
of terminals outside $G$ queried against its members before query $j$.
For a special terminal $u$, define
\begin{align*}
  b^F_j(u)&=\#\{x\text{ ordinary}: \{x,c\}\text{ was queried before }j
                  \text{ for some }c\text{ with }u\in F_c\},\\
  b^H_j(u)&=\#\{x\text{ ordinary}: \{x,c\}\text{ was queried before }j
                  \text{ for some }c\text{ with }u\in H_c\}.
\end{align*}
The second counter includes discarded potential stars. These counters
measure the ordinary labels that cannot safely exchange roles with $u$.
Two deterministic budgets will be useful:
\begin{equation}\label{eq:counter-budgets}
  \sum_G e_{q+1}(G)\le2q,\qquad
  \sum_{u\in S}b^F_{q+1}(u)\le2q.
\end{equation}
The latter uses $|F_c|\le2$.

\begin{definition}[Noticing and closure]\label{def:exposed}
Maintain sets $\mathcal N\subseteq Q\subseteq S$, initially empty.
A terminal $u\in S\setminus Q$ is \emph{noticed} and added to both sets
when it first receives answer $1$, its query count reaches $D$, or its
$b^F$ counter reaches $\theta K$.
After adding newly noticed terminals in a fixed order, close $Q$ under
the rule
\begin{equation}\label{eq:closure}
  F_c\subseteq Q,\quad |F_c|=2,\quad |H_c\cap Q|\ge3
  \qquad\Longrightarrow\qquad H_c\subseteq Q.
\end{equation}
Repeat the rule until it makes no further addition. A terminal added by
closure is \emph{exposed}, but is not subsequently counted as a new
noticed terminal.

A query $\{c,u\}$ is \emph{lucky} if, just before it, $u\in S\setminus Q$
and $u\in H_c$. This definition includes discarded stars and does not
require the answer to be $1$.
\end{definition}

The sets are analytical objects, not information given to $\mathcal A$.
Closure deliberately exposes an entire potential star as soon as three
of its members, including its keys, are available. This only gives
the analysis more information.

Stop the analysis at the first query $j_*$ where any of the following
occurs:
\begin{enumerate}[label=(\roman*),leftmargin=2.6em]
  \item at least $\theta K$ ordinary terminals have been marked
        (\emph{cluster stop});
  \item some $u\in S\setminus Q$ has $b^H(u)\ge\theta K$
        (\emph{star-counter stop});
  \item the query is lucky;
  \item $|\mathcal N|>R$ (\emph{notice overflow});
  \item $|Q|>2R$ (\emph{closure overflow}).
\end{enumerate}
Luck is tested against the pre-query set $Q$, before updating any counters
or sets, and it has priority if several stops occur together. For the other
stops use a fixed tie order in which closure overflow comes last. Set $j_*=q+1$ if none occurs.

Call a special terminal \emph{quiet} if no earlier query $\{c,u\}$ had
$u\in F_c\cup H_c$.

\begin{lemma}[Before the first stop]\label{lem:structure-lb}
At the start of every query $j\le j_*$:
every special terminal outside $Q$ is quiet, both keys of every active
center lie in $Q$, and every center with at least three positive answers
has its potential star entirely in $Q$.
\end{lemma}

\begin{proof}
A query to a filler member returns $1$ and notices it unless it is already
in $Q$. A query to a potential-star member outside $Q$ is lucky. Thus
terminals still outside $Q$ are quiet. Activation requires two positive
key answers, so both keys are in $Q$. A third positive answer
requires a non-key star member. Without a lucky query that member was
already in $Q$, and \eqref{eq:closure} then adds the whole star.
\end{proof}

\subsubsection{The exchange argument}

Condition on a prefix $\tau_j$, and let $\Omega_j$ be its consistent
configurations: the values of the hidden data for which the transcript
begins with $\tau_j$. To simplify notation, we slightly abuse notation and
also write $\tau_j$ for this event. Thus
$\Prob[\tau_j]=\Prob[\Omega_j]$, and $\Prob[E\mid\tau_j]$ is the probability
of $E$ given that the transcript begins with $\tau_j$. We use the following elementary counting principle, which
also applies to nonuniform configurations. If every configuration in an
event $E\subseteq\Omega_j$ has at least $F$ probability-preserving moves
into $\Omega_j$, and each target receives at most $B$ such moves, then
\begin{equation}\label{eq:exchange-principle}
  \Prob[E\mid\tau_j]\le B/F.
\end{equation}
Indeed, summing the probability mass of all moves gives
$F\Prob[E]\le B\Prob[\Omega_j]$.

Every move below transposes two terminal labels in $\pi$, keeping all data
attached to positions fixed. The labels exchange their roles, clusters,
weights, and memberships in fillers and potential stars. Since $\pi$ is
uniform and independent of that data, each move preserves probability.
Only consistency with the prefix and the number of inverse moves need
checking.

\begin{lemma}[Clusters are hard to find]\label{lem:cluster}
The probability of a cluster stop is $O(q/(AK))$.
\end{lemma}

\begin{proof}
Count newly marked terminals up to and including the query that causes
the cluster stop. Terminals already having query count at least $D$
contribute at most $2q/D=O(q/A)$ marks. Clusters with
$e_j(G)\ge\theta K$ contribute at most
$2qR/(\theta K)=O(q/A)$ marks, by \eqref{eq:counter-budgets}.

For every other new mark, fix the queried pair $\{u,v\}$ and consider
configurations where $u$ is unmarked, $\deg_j(u)<D$, and $u,v$ lie in a
cluster $G$ with $e_j(G)<\theta K$. Transpose $u$ with an ordinary terminal
$x\notin G$, excluding $x$ if it is marked, if it was queried against
$G$, or if its cluster contains a terminal previously queried against $u$.
Before the cluster stop at most
$2\theta K+DR+R\le K/2$ labels are excluded, for large $A$.
No old $\lambda$-answer is lost, no new one is created, and all
center answers involving either ordinary terminal remain $2$.
Thus at least $K/2$ moves preserve the prefix. In a target, $x$ lies in
the cluster of $v$, so there are at most $R$ inverse moves.
By \eqref{eq:exchange-principle}, the conditional probability of such
a mark is at most $2R/K=2/A$.

Summing over the $2q$ query endpoints gives $O(q/A)$ expected marks.
A cluster stop needs $\theta K$ marks, so Markov's inequality proves
the claim.
\end{proof}

For a terminal $u$, let $E_j(u)$ be the event that no cluster stop has
occurred, $u$ is special and quiet, and
\[
  \deg_j(u)<D,\qquad b^F_j(u)<\theta K,\qquad b^H_j(u)<\theta K.
\]
Before the first stop this event holds for every special terminal
outside $Q$.

\begin{lemma}[Special terminals are hard to recognize]\label{lem:special}
For every terminal $u$, center $c$, and prefix,
\begin{equation}\label{eq:exchange}
  \Prob[E_j(u)\mid\tau_j]\le\frac{4m}{3K},\qquad
  \Prob[E_j(u),\,u\in F_c\mid\tau_j]\le\frac8{3K}.
\end{equation}
The same statements hold under the distribution in which the potential star
of one fixed center is deleted and all other data keep their original
distribution.
\end{lemma}

\begin{proof}
Transpose $u$ with an ordinary label $x$. Exclude marked $x$, labels
whose cluster contains a terminal queried against $u$, and labels counted
by $b^F_j(u)$ or $b^H_j(u)$. At most
$3\theta K+DR\le K/4$ labels are excluded, leaving $3K/4$ moves.

All these moves preserve the prefix. Terminal--terminal answers remain
$2$, by the first two exclusions. By quietness, earlier center queries
at $u$ met neither a filler nor a potential star. By the counter
exclusions the same remains true of earlier center queries at $x$
after it takes $u$'s position. Consequently neither positive answers nor
activation or discard decisions change.

In a target, the moved label $x$ is special, giving at most $m$ inverse
moves. For the second event it also belongs to $F_c$, giving at most two.
Apply \eqref{eq:exchange-principle}. Deleting one fixed potential star
commutes with the same transpositions, so the argument also applies
when the star is deleted.
\end{proof}

\subsubsection{Most stops are unlikely}

\begin{lemma}[Other stops are unlikely]\label{lem:screening}
The probability that the first stop is anything other than closure
overflow is at most
\begin{equation}\label{eq:screening}
  C\left(\frac q{AK}+\frac q{KR}\right).
\end{equation}
\end{lemma}

\begin{proof}
The cluster term is \cref{lem:cluster}. We bound the other three types
of stop along prefixes with no earlier stop. All conditional estimates
below condition only on the observable prefix $\tau_j$. Having
no earlier stop is part of the event being bounded, not additional
conditioning on hidden information.

\emph{Notices.}
On a query that is not lucky, a new answer-$1$ notice must come from
$u\in F_c$, so \cref{lem:special} bounds its probability by $O(1/K)$.
There are at most $2q/D$ crossings of query count $D$, and the crossing
times and labels are determined by the transcript before the query.
Each yields a new special notice with probability $O(m/K)$, contributing
$O(qm/(DK))=O(q/K)$ in expectation. Finally,
\eqref{eq:counter-budgets} permits at most $2q/(\theta K)$ notices from
the $b^F$ threshold. Only terminals outside $Q$ are counted, so vertices
already added by closure do not enter this accounting.
The expected number of such direct notices is $O(q/K)$.
Markov's inequality at $R$ bounds a first notice overflow by $O(q/(KR))$.

\emph{Star counters.}
An increment of $b^H(u)$ for $u\notin Q$ comes from an ordinary-terminal
query at a center with a potential star. By \cref{lem:structure-lb},
that center has at most two positive answers: otherwise its entire
potential star is already in $Q$. By \cref{lem:erase}, the conditional
probability that such a center has a potential star is $O(1/R)$.
One query increments at most $d$ star counters. The expected total
number of relevant increments, including a possible stopping query, is
therefore $O(q/R)$. Reaching $\theta K$ gives probability $O(q/(KR))$.

\emph{Lucky queries.}
Let $L_j$ be the event that query $j$ is lucky and no stop occurred before
it. Fix a prefix $\tau=\tau_j$ of positive probability and its next pair
$\{c,u\}$. Only a pair between a center $c$ and a terminal $u$ can be lucky.
We show that $\Prob[L_j\mid\tau]=O(1/(KR))$.

Write a configuration as $(Z,h)$, where $h$ is the value of $H_c$
($h=\varnothing$ if $H_c$ is absent) and $Z$ consists of all other hidden
data. The star $H_c$ is drawn by \eqref{eq:potential-star}, which uses only
$F_c$ and $W$, and these are part of $Z$. So
$\Prob[Z,h]=\Prob[Z]\,\rho_Z(h)$ with $\rho_Z(h)=\Prob[H_c=h\mid F_c,W]$,
and by \eqref{eq:mark-bounds},
\begin{equation}\label{eq:rho-bounds}
  \sum_{h\ne\varnothing}\rho_Z(h)\le\frac{B_0}R,
  \qquad
  \sum_{h\ni u}\rho_Z(h)\le\frac{B_1}{Rm}\quad\text{if }u\notin F_c ,
\end{equation}
and both sums are $0$ if $F_c$ is not a pair. Let $P_0^c$ be the distribution of
$(Z,\varnothing)$ when $(Z,H_c)$ is drawn as usual. It deletes the potential
star of $c$ and keeps all other data, so it gives probability $\Prob[Z]$ to
$(Z,\varnothing)$. This is deletion, not conditioning on absence. The
experiment itself produces $(Z,\varnothing)$ with probability
$\Prob[Z]\,\rho_Z(\varnothing)\ge(1-B_0/R)\Prob[Z]$. Keeping only the
configurations of $\tau$ in which $H_c$ is absent, we get
\begin{equation}\label{eq:delete-star}
  \Prob[\tau]
  \ge\sum_{Z:\,(Z,\varnothing)\in\tau}\Prob[Z]\,\rho_Z(\varnothing)
  \ge\Bigl(1-\frac{B_0}R\Bigr)\sum_{Z:\,(Z,\varnothing)\in\tau}\Prob[Z]
  =\Bigl(1-\frac{B_0}R\Bigr)P_0^c[\tau].
\end{equation}

Now let $(Z,h)\in\tau\cap L_j$. Luck means that $u\in h$ and
$u\in S\setminus Q$. Since no stop occurred before query $j$, the event
$E_j(u)$ holds, and $c$ has at most two positive answers: otherwise
\cref{lem:structure-lb} would give $H_c\subseteq Q$, contradicting
$u\in H_c\setminus Q$. Hence, by \cref{lem:erase}(a), $(Z,\varnothing)$ is
still in $\tau$. It is also in $E_j(u)$, because deleting $H_c$ only removes
memberships in a potential star. Whether $u\in F_c$ is decided by $Z$.
Grouping the configurations $(Z,h)$ by $Z$ and using \eqref{eq:rho-bounds},
we get
\[
  \Prob[\tau,\,L_j,\,u\in F_c]
  \le\sum_{\substack{Z:\,(Z,\varnothing)\in\tau\cap E_j(u)\\ u\in F_c}}
     \Prob[Z]\sum_{h\ne\varnothing}\rho_Z(h)
  \le\frac{B_0}{R}\,P_0^c[\tau,\,E_j(u),\,u\in F_c],
\]
and in the same way, with the second bound in \eqref{eq:rho-bounds},
\[
  \Prob[\tau,\,L_j,\,u\notin F_c]\le\frac{B_1}{Rm}\,P_0^c[\tau,\,E_j(u)].
\]
By \cref{lem:special} under $P_0^c$, the right-hand sides are at most
$\frac{B_0}R\cdot\frac8{3K}\,P_0^c[\tau]$ and
$\frac{B_1}{Rm}\cdot\frac{4m}{3K}\,P_0^c[\tau]$. Adding them and using
\eqref{eq:delete-star} with $R\ge2B_0$ gives
\[
  \Prob[L_j\mid\tau]\le\frac{2(8B_0+4B_1)}{3KR}=O\Bigl(\frac1{KR}\Bigr).
\]
Averaging over $\tau$ gives $\Prob[L_j]=O(1/(KR))$, and summing over the $q$
queries shows that the first stop is lucky with probability $O(q/(KR))$.
\end{proof}

\subsection{The coupling: bounded reuse of weights}\label{sec:coupling}

\Cref{lem:screening} shows that stops other than closure overflow are
unlikely. In this subsection we show that closure overflow is unlikely too, and
we compare the two experiments up to their first stop. The two experiments
differ only in the weights, and the weights enter only through the potential
stars. Before the first stop, the answers depend on the potential stars only
through the stars \emph{visible} from the exposed set $Q$: those with at least
three members in $Q$, including both keys. We show that the visible stars
have almost the same distribution in both experiments, in four steps.
\begin{itemize}
  \item \Cref{sec:simulate} computes the transcript, up to the first stop,
        from weight-independent data and the stars visible from $Q$.
  \item \Cref{sec:sparsity} shows that, with high probability, every small
        set of special terminals contains few filler pairs. This must hold
        for all sets at once, since $Q$ depends on the run.
  \item \Cref{sec:moment} proves a general lemma. Suppose that each
        possible support is first proposed with a probability that does not
        depend on the weights, and that each proposed support is then retained
        with probability equal to the product of its normalized weights. Then
        the two weight distributions give exactly the same distribution of
        what is revealed until some terminal lies in $s+1$ revealed proposals,
        and this is unlikely.
  \item \Cref{sec:visible} applies this lemma to the visible stars and
        bounds the probability of closure overflow.
\end{itemize}

\subsubsection{Simulating the experiment with star requests}\label{sec:simulate}

We first reduce the comparison of the two experiments to a comparison
of the potential stars visible from the exposed set.
We call the bijection $\pi$ together with all fillers $F_c$ the
\emph{common data}, and denote it by $\Gamma$.
They have the same distribution in both worlds and are independent of the weights.
For $Q\subseteq[m]$, a potential star $H_c$ is \emph{visible} from $Q$ if
\[
  |F_c|=2,\qquad F_c\subseteq Q,\qquad |H_c\cap Q|\ge3.
\]
A \emph{star request} for $Q$ returns exactly the existing potential
stars visible from $Q$. Its reply includes the center labels and entire
supports, even for discarded stars. An empty reply also counts as an answer.
The \emph{star record} is the sequence of requests and replies.
Recall from \cref{sec:bookkeeping} that $j_*$ is the first query at
which a stop occurs (a cluster stop, a star-counter stop, a lucky query,
notice overflow, or closure overflow), and $j_*=q+1$ if no stop occurs.

\begin{lemma}[Simulation]\label{lem:simulation}
There is a deterministic procedure, the \emph{simulator}, that uses
$\Gamma$ and star requests and whose answers agree with the two-key
experiment through every query before $j_*$. Its requests are for
increasing sets $Q$ of size at most $2R$, each determined by $\Gamma$ and
earlier replies.
\end{lemma}

\begin{proof}
The simulator maintains the transcript, activation states, the sets
$B_c$, the counters $\deg,b^F$, and $\mathcal N,Q$. It adds notices using
the simulated answers and uses star requests to implement closure.
If a notice or closure update would cross either size cap, it halts, and
its remaining transcript is completed by any fixed answer rule, without
further requests. Otherwise it repeats closure to completion.

For an inactive center it answers from $F_c$. For an active center it
also uses every revealed potential star disjoint from $B_c$.
Inductively these answers agree before $j_*$. Indeed, both keys
of an active center lie in $Q$. If a queried non-key $u\in Q$ belongs
to $H_c$, then three star members lie in $Q$, and a request has
revealed it. If $u\notin Q$, the query is not lucky, so $u\notin H_c$ and
the answer is $2$. All other queries are determined by $\Gamma$ alone. The same
induction identifies the maintained $Q$ with the exposed set. It also
covers query $j_*$ itself unless that query is lucky: the answer to it and
the resulting updates of $\mathcal N$ and $Q$ agree as well.
\end{proof}

The simulator is only a device for comparing distributions. In
particular, it need not know when an unobservable star-counter stop or lucky
query occurs. We charge these events using \cref{lem:screening}.

\subsubsection{Uniform sparsity of the filler pairs}\label{sec:sparsity}

We view each pair filler $F_c=\{u,v\}$ as an edge between $u$ and $v$,
labeled by $c$. Since several centers may have the same pair, these edges
form a multigraph on $[m]$. For $Q\subseteq[m]$, let
$g(Q)=\#\{c:|F_c|=2,\ F_c\subseteq Q\}$ be the number of these edges
inside $Q$.

\begin{lemma}[Filler sparsity]\label{lem:sparsity}
If $R\ge40\log m$, then with probability at least $1-m^{-10}$ every
terminal belongs to at most $40R$ pair fillers and, simultaneously for
every $Q\subseteq[m]$,
\begin{equation}\label{eq:filler-sparsity}
  g(Q)\le40\left(\frac{R|Q|^2}{m}+|Q|\log m\right).
\end{equation}
\end{lemma}

\begin{proof}
We use the following tail bound, whose proof is the standard derivation
of the Chernoff bound. If $X$ is a sum of independent Bernoulli
variables and $\Ex[X]\le\mu$, then $\Ex[e^X]\le e^{(e-1)\mu}$
(\cref{sec:prob}), and Markov's inequality applied to $e^X$ gives
\begin{equation}\label{eq:chernoff-simple}
  \Prob[X\ge t]\le e^{(e-1)\mu-t}\le e^{-t/2}
  \qquad\text{for every }t\ge4\mu .
\end{equation}
There are $mR$ centers, their fillers are independent, and each filler
equals a given pair with probability $\beta=1/(2m(m-1))$.

A fixed terminal lies in $m-1$ pairs. So the number of pair fillers
containing it is a sum of $mR$ independent Bernoulli variables, each with
mean $(m-1)\beta=1/(2m)$, so its expectation is $R/2$. By
\eqref{eq:chernoff-simple} with $t=40R$, it is at least $40R$ with
probability at most $e^{-20R}\le m^{-20}$, as $R\ge40\log m$. A union bound
over the $m$ terminals proves the degree statement, except with probability
$m^{-19}$.

Now fix $Q$ of size $h\ge1$. Then $g(Q)$ is a sum of $mR$ independent
Bernoulli variables, each with mean $\binom h2\beta$, so
$\Ex[g(Q)]=Rh(h-1)/(4(m-1))\le Rh^2/(2m)$. Apply
\eqref{eq:chernoff-simple} with $t=40(Rh^2/m+h\log m)\ge4\Ex[g(Q)]$: the
bound \eqref{eq:filler-sparsity} fails with probability at most
$e^{-20h\log m}=m^{-20h}$. There are at most $m^h$ sets of size $h$, so
\eqref{eq:filler-sparsity} fails for some $Q$ with probability at most
$\sum_{h\ge1}m^{-19h}\le2m^{-19}$. Altogether, the failure probability is
at most $3m^{-19}\le m^{-10}$.
\end{proof}

The uniformity is essential: $Q$ depends on the transcript and is not
an independently sampled set. We call the common data $\Gamma$
\emph{good} if every terminal belongs to at most $40R$ pair fillers and
\eqref{eq:filler-sparsity} holds for every $Q\subseteq[m]$. By
\cref{lem:sparsity}, $\Gamma$ is good with probability at least $1-m^{-10}$.

\subsubsection{A discrete moment-matching lemma}\label{sec:moment}

The goal of this subsection is a general lemma behind the moment
matching. We generate random supports in two stages: each possible support is
first \emph{proposed} with a probability that does not depend on the weights,
and a proposed support is then \emph{retained} with probability equal to the
product of its normalized weights. A protocol may observe the proposals
adaptively, the rejected ones included. We first show (\cref{cl:moment}) that a fixed pattern of
retained and rejected supports has the same probability in both worlds, on
average over the weights, as long as no terminal lies in more than $s$ of
these supports. \Cref{lem:visibility} extends this to adaptive protocols and
shows that some terminal lies in $s+1$ revealed proposals only with small
probability. \Cref{sec:visible} applies the lemma to the potential stars
visible from the exposed set.

For a support $H\subseteq[m]$, write $p_H(W)=\prod_{u\in H}(W_u/M)\in[0,1]$.

\begin{claim}[Moment identity]\label{cl:moment}
Let $H_1,\dots,H_\ell$ be supports such that every terminal lies in at most
$s$ of them. Then for all $b_1,\dots,b_\ell\in\{0,1\}$, the expectation
\begin{equation}\label{eq:moment-identity}
  \Ex\left[
     \prod_{i:b_i=1}p_{H_i}(W)
     \prod_{i:b_i=0}(1-p_{H_i}(W))
  \right]
\end{equation}
has the same value under the two weight distributions.
\end{claim}

\begin{proof}
Let $I=\{i:b_i=1\}$. Expanding each factor $1-p_{H_i}(W)$ gives
\[
  \prod_{i\in I}p_{H_i}(W)\prod_{i\notin I}\bigl(1-p_{H_i}(W)\bigr)
  =\sum_{J\subseteq[\ell]\setminus I}(-1)^{|J|}\prod_{i\in I\cup J}p_{H_i}(W)
  =\sum_{J\subseteq[\ell]\setminus I}(-1)^{|J|}\prod_{u\in[m]}
     \Bigl(\frac{W_u}M\Bigr)^{k_u(J)},
\]
where $k_u(J)=\#\{i\in I\cup J:u\in H_i\}\le s$. As the weights are
independent, the term for $J$ has expectation
$\prod_u\Ex[(W_u/M)^{k_u(J)}]$, and by \eqref{eq:moments} each factor is the
same under the two distributions.
\end{proof}

\Cref{cl:moment} concerns a fixed collection of supports. We now
describe the random two-stage process that generates the supports, and how
an adaptive protocol observes it. The protocol is only an analysis device.
It answers no distance queries and only sees which supports are proposed
and which are retained. In \cref{sec:visible}, it will be the simulator of
\cref{lem:simulation}, whose star requests ask for the supports visible from
the exposed set. There the potential stars are drawn by this two-stage
process, so a center may have several retained supports, and
\cref{lem:loads} shows that this is close to the two-key experiment, in which
every center has at most one potential star.

\paragraph{Proposals.}
Fix a finite set of \emph{types}. A type $x$ is an abstract label that comes
with a support $H_x\subseteq[m]$ of size at most $d$ and a \emph{proposal
probability} $\lambda_x\in[0,1]$. Different types may have the same support
and the same proposal probability. Independently of the weights, and
independently across types, each type $x$ is \emph{proposed} with
probability $\lambda_x$. A proposed type is called a \emph{proposal}.
Conditional on the weights, each proposal $x$ is then \emph{retained}
independently with probability $p_{H_x}(W)$, and \emph{rejected} otherwise.
In \cref{sec:visible}, the types will be the pairs $(c,H)$ of a center $c$
and a possible support $H$ of its potential star. The factors for $b_i=0$
in \eqref{eq:moment-identity} account for rejected proposals: rejection also
depends on the weights.

\paragraph{Adaptive requests.}
A \emph{region} is a set of types. A \emph{protocol} observes the proposals
through \emph{requests}. It requests regions
$\mathcal D_1\subseteq\mathcal D_2\subseteq\cdots$ one by one, and the answer
to the request for $\mathcal D_k$ reveals which types in
$\mathcal D_k\setminus\mathcal D_{k-1}$ are proposed, together with the
retention bits of these proposals (where $\mathcal D_0=\varnothing$). Each
region is chosen using only the earlier answers and fixed side information,
which is independent of the weights and of the proposal and retention bits.
The \emph{record} is the sequence of requests and answers, and the
\emph{revealed proposals} are the proposals in the requested regions.

\begin{lemma}[Adaptive Bernoulli proposals]\label{lem:visibility}
Suppose that every region $\mathcal D$ requested by the protocol
satisfies
\begin{equation}\label{eq:load-budgets}
  \sum_{x\in\mathcal D}\lambda_x\le L_*,
  \qquad\text{and}\qquad
  \sum_{\substack{x\in\mathcal D\\u\in H_x}}\lambda_x\le G_*
       \quad\text{for every }u.
\end{equation}
Then the total variation distance between the distributions of the record
in the two worlds is at most $dL_*G_*^s$. In either world, the expected
number of revealed proposals is at most $L_*$.
\end{lemma}

\begin{proof}
Fix the side information, that is, the data that the protocol knows in
advance (in \cref{sec:visible}, these are the common data $\Gamma$). We may
assume that the protocol reads the types of
each newly requested region one at a time, in a fixed order: for a type $x$
it first reads the proposal bit $a_x$ ($a_x=1$ if $x$ is proposed) and then,
if $a_x=1$, the retention bit $b_x$ ($b_x=1$ if $x$ is retained). Which types
are read, and in which order, depends only on the bits read so far. For a
terminal $u$, let $N_u$ be the number of proposed types read so far whose
support contains $u$.

\emph{Step 1: equal distributions until some weight is used $s+1$ times.}
Stop reading just before the retention bit of the first proposed type $x$
whose proposal makes $N_u=s+1$ for some $u\in H_x$. Consider a possible
stopped record $r$. Let $X_r$ be the set of types whose proposal bits are
read in $r$, and $X_r^+\subseteq X_r$ the set of types whose retention bits
are read in $r$. Both sets are determined by $r$, and
\[
  \Prob[r]
  =\prod_{x\in X_r}\lambda_x^{a_x}(1-\lambda_x)^{1-a_x}
   \cdot\Ex\Bigl[\prod_{x\in X_r^+}
     p_{H_x}(W)^{b_x}\bigl(1-p_{H_x}(W)\bigr)^{1-b_x}\Bigr].
\]
The first factor does not depend on the weights. By the choice of the
stopping time, every terminal lies in at most $s$ of the supports $H_x$ with
$x\in X_r^+$, so by \cref{cl:moment} the second factor is the same
under both weight distributions. Hence the stopped record has the same
distribution in the two worlds. Couple the two runs so that their stopped
records coincide, and continue them independently. If the stopping rule
never applies, the stopped record is the full record. So the full records
can differ only if, in the full run, $N_u\ge s+1$ at the end for some $u$.
By \cref{lem:prob}(a), the total variation distance is at most the
probability of this event.

\emph{Step 2: some weight is rarely used $s+1$ times.}
Consider the full run, without stopping. For a read type $x$, let
$N_u^-(x)$ be the value of $N_u$ just before $x$ is read. When $N_u$
increases from $k$ to $k+1$, the value of $\binom{N_u}{j+1}$ increases by
$\binom kj$, by Pascal's rule. Hence, at the end of the run, for every
integer $j\ge0$,
\[
  \binom{N_u}{j+1}
  =\sum_{\substack{x\text{ read}\\u\in H_x}}a_x\binom{N_u^-(x)}j.
\]
Just before $x$ is read, $a_x$ is a $\Ber(\lambda_x)$ variable independent
of everything read so far, and in particular of whether $x$ is read and of
$N_u^-(x)$. Taking expectations, we get
\[
  \Ex\Bigl[\binom{N_u}{j+1}\Bigr]
   =\Ex\Bigl[\sum_{\substack{x\text{ read}\\u\in H_x}}
           \lambda_x\binom{N_u^-(x)}j\Bigr]
   \le G_*\,\Ex\Bigl[\binom{N_u}j\Bigr].
\]
The inequality uses $N_u^-(x)\le N_u$ and the fact that the read types form
the last requested region, so by \eqref{eq:load-budgets} their values
$\lambda_x$ with $u\in H_x$ sum to at most $G_*$. Applying this for
$j=s,s-1,\dots,1$ gives $\Ex[\binom{N_u}{s+1}]\le G_*^s\Ex[N_u]$. Let $Y$ be
the number of proposed types read, that is, of revealed proposals. Then
$\sum_uN_u\le dY$, and the same argument with all read types in place of
those containing $u$ gives
$\Ex[Y]=\Ex[\sum_{x\text{ read}}\lambda_x]\le L_*$, which is the last
statement of the lemma. Therefore
\[
  \Prob[\exists u:N_u\ge s+1]
  \le\sum_u\Ex\Bigl[\binom{N_u}{s+1}\Bigr]
  \le G_*^s\sum_u\Ex[N_u]
  \le dL_*G_*^s .\qedhere
\]
\end{proof}

\subsubsection{Applying the lemma to visible stars}\label{sec:visible}

Condition on good common data $\Gamma$. For each center
with pair filler and each $d$-set $H\supseteq F_c$, introduce the type
$x=(c,H)$, with support $H$ and proposal probability
\begin{equation}\label{eq:candidate-rate}
  \lambda_*=\frac{H_0M^d}{dR\beta\binom md}.
\end{equation}
For large parameters $\lambda_*<1$. Retaining a proposal with probability
$p_H(W)$ gives probability
\[
  r_H:={}\lambda_*p_H(W)=\frac{a_W(H)}{\beta},
\]
exactly the mass of $H$ in \eqref{eq:potential-star}. The types whose
supports are visible from $Q$ form the region
\[
  \mathcal D(Q)=\{(c,H): |F_c|=2,\ F_c\subseteq Q,\ |H\cap Q|\ge3\}.
\]
These regions grow with $Q$.

\paragraph{Budgets.}
Assume $R\ge40\log m$ and $2R\le m$, and put
\begin{equation}\label{eq:visible-budgets}
  B_*=160\left(\frac{R^3}{m}+R\log m\right),\qquad
  L_*=\frac{2B_1B_*}m,\qquad
  G_*=\frac{240B_1R}m ,
\end{equation}
with $B_1$ as in \eqref{eq:mark-bounds}. We show that, for every $Q$ with
$|Q|\le2R$, at most $B_*$ centers have their filler pair in $Q$, and the
region $\mathcal D(Q)$ satisfies \eqref{eq:load-budgets} with these $L_*$
and $G_*$.

The first claim is \eqref{eq:filler-sparsity} with $|Q|\le2R$. For the
others, fix a pair $e$ and a terminal $u\notin e$. There are
$\binom{m-3}{d-3}$ sets $H\supseteq e\cup\{u\}$ of size $d$, and, as in the
derivation of \eqref{eq:mark-bounds},
\begin{equation}\label{eq:third-member}
  \lambda_*\binom{m-3}{d-3}
  =\frac{H_0M^d}{dR\beta}\cdot\frac{\binom{m-3}{d-3}}{\binom md}
  =\frac{(d-2)B_0}{R(m-2)}
  \le\frac{B_1}{Rm}.
\end{equation}
Consider a center $c$ with $F_c\subseteq Q$. Every type $(c,H)$ in
$\mathcal D(Q)$ has a member of $Q\setminus F_c$ in $H$. So, by
\eqref{eq:third-member}, the types of $c$ in $\mathcal D(Q)$ have total
proposal probability at most $|Q|B_1/(Rm)\le2B_1/m$. Summing over the at
most $B_*$ such centers gives $L_*$. Now fix a terminal $u$. By
\cref{lem:sparsity}, at most $40R$ centers have $u\in F_c$, and they
contribute at most $40R\cdot2B_1/m$. Every other center with
$F_c\subseteq Q$ contributes at most $B_1/(Rm)$, by \eqref{eq:third-member}
with $e=F_c$. Together, these centers contribute at most $B_*B_1/(Rm)\le160B_1R/m$,
since $B_*/R\le160(R/2+R/40)\le160R$. The total is at most $G_*$.

\paragraph{Two models for the stars.}
We call the way the two-key experiment draws potential stars,
\eqref{eq:potential-star}, the \emph{star model}: given its keys and the
weights, a center has potential star $H$ with probability $r_H$ and none
with the remaining probability, so it has at most one potential star. In the
\emph{proposal model}, every type $(c,H)$ is proposed and retained
independently, so $H$ is retained at $c$ with probability $r_H$, and a
center may have several retained supports. The simulator then treats each
retained support as a potential star of $c$.

\begin{lemma}[Visible-star coupling]\label{lem:loads}
Given good common data, the distributions of
the star record in the two worlds differ by at most
\begin{equation}\label{eq:visible-tv}
  \frac{2B_0^2B_*}{R^2}+dL_*G_*^s.
\end{equation}
In either world, the probability that closure overflow is the first
stop is at most
\begin{equation}\label{eq:closure-prob}
  \frac{B_0^2B_*}{R^2}+\frac{dL_*}{R}.
\end{equation}
\end{lemma}

\begin{proof}
Fix one world. For a center $c$ with a pair filler, let
$v_c=\sum_{H\supseteq F_c}r_H$. Then $v_c\le B_0/R$ by
\eqref{eq:mark-bounds}.

\emph{One center.} Given the weights, the set of supports at $c$ has
distributions at total variation distance at most $v_c^2$ in the two models.
Indeed, the star model gives probability $r_H$ to each singleton $\{H\}$ and
$1-v_c$ to the empty set. The proposal model gives no less probability to the
empty set, since $\prod_H(1-r_H)\ge1-v_c$, and no more to any singleton. Its
other outcomes have at least two supports. So the distance is the total
singleton deficit
\[
  \sum_H r_H\Bigl(1-\prod_{H'\ne H}(1-r_{H'})\Bigr)
  \le\sum_Hr_H\sum_{H'\ne H}r_{H'}\le v_c^2 .
\]

\emph{From the star model to the proposal model.} Call a center
\emph{opened} at the first request whose set contains its filler pair,
and order the opened centers by this time, breaking ties by a fixed order.
Before $c$ is opened, no request has involved a support at $c$, so the
supports at $c$ can be sampled when $c$ is opened. Every opened center has
its filler pair in the last requested set, which has size at most $2R$, so
at most $B_*$ centers are opened. For $0\le k\le B_*$, let $\mathsf M_k$ be
the run in which the first $k$ opened centers use the proposal model and all
other centers use the star model. Thus $\mathsf M_0$ uses the star model,
and $\mathsf M_{B_*}$ has the same star record as the proposal model. The
runs $\mathsf M_{k-1}$ and $\mathsf M_k$ agree until the $k$th center is
opened. At that moment, given the weights and the record so far, the
supports at that center still have their original distribution in either
model, since different centers are independent given the weights. Couple
them as in the previous paragraph, and use the same randomness for
everything else. Then the two star records differ with probability at most
$B_0^2/R^2$. By \cref{lem:prob}(a) and the triangle inequality, the star
records of the two models are within total variation distance
$B_0^2B_*/R^2$.

\emph{Comparing the worlds.} In the proposal model, a star request for $Q$
amounts to requesting the region $\mathcal D(Q)$, where $|Q|\le2R$ and $Q$
is chosen using only the common data and earlier replies, so the common
data are the side information of \cref{lem:visibility}. Give the simulator
the stronger record of all proposals in the requested regions and their
retention bits. It uses only the retained supports. By the budgets and
\cref{lem:visibility}, the distributions of this stronger record in the two
worlds are within $dL_*G_*^s$, and hence so are those of the star record,
which is a function of it (\cref{lem:prob}(b)). Adding the error
$B_0^2B_*/R^2$ in each world proves \eqref{eq:visible-tv}.

\emph{Closure overflow.} Suppose that closure overflow is the first stop.
Since closure overflow comes last in the tie order, at most $R$ terminals
are noticed at that query, so closure has added more than $R$ terminals to
$Q$. Each closure step adds at most $d$ terminals, all from one star visible
from the current $Q$, which has size at most $2R$. By the proof of
\cref{lem:simulation}, the simulator performs the same closure steps, so its
star record reveals more than $R/d$ stars. In the proposal model, revealed
supports are revealed proposals, whose expected number is at most $L_*$ by
\cref{lem:visibility}. By Markov's inequality, more than $R/d$ of them are
revealed with probability at most $dL_*/R$. Adding the error
$B_0^2B_*/R^2$ between the two models proves \eqref{eq:closure-prob}.
\end{proof}

\subsection{Proof of \texorpdfstring{\cref{thm:lower}}{Theorem 1.2}}\label{sec:lb-proof}

We assemble the argument before choosing the constants in
\eqref{eq:lb-sizes}. The resulting bound displays the two main costs:
finding hidden special terminals, and reusing a weight beyond the
matched moment order.

\begin{lemma}[The two transcripts are close]\label{lem:tv}
Suppose $C\log m\le R$, $2R\le m$, $R\log m\le m$, and $q\le KR$.
For every deterministic algorithm making at most $q$ queries,
\begin{equation}\label{eq:master}
\begin{split}
  &\dtv\bigl(\trans(\mathcal A,\Dyes),\trans(\mathcal A,\Dno)\bigr)\\
  &\qquad\le
  C\left(
       \sqrt{\frac q{mR^2}}+\frac q{KR}+\frac q{AK}
       +\frac{R^{s+3}}{m^{s+2}}+\frac Rm+\frac{\log m}{R}
    \right)+m^{-10}.
\end{split}
\end{equation}
\end{lemma}

\begin{proof}
For $\bullet\in\{\mathsf y,\mathsf n\}$, write
$T_\bullet=\trans(\mathcal A,\mathcal D_\bullet)$ for the transcript
distribution in the metric and $E_\bullet=\trans(\mathcal A,\mathcal G_\bullet)$
for that in the two-key experiment. By the triangle inequality,
\begin{equation}\label{eq:tv-chain}
  \dtv(T_{\mathsf y},T_{\mathsf n})
  \le\dtv(T_{\mathsf y},E_{\mathsf y})+\dtv(E_{\mathsf y},E_{\mathsf n})
     +\dtv(E_{\mathsf n},T_{\mathsf n}).
\end{equation}
By \cref{lem:gated}, the first and last terms are at most
$2C\sqrt{q/(mR^2)}$ each, since $q/(mR^2)=q/(dKR)\le1$.

For the middle term, the common data $\Gamma$ are good except with
probability at most $m^{-10}$ (\cref{lem:sparsity}). Since
$\Gamma$ has the same distribution in both experiments, \cref{lem:prob}(c)
bounds $\dtv(E_{\mathsf y},E_{\mathsf n})$ by the average over $\Gamma$ of the
distance between the conditional distributions given $\Gamma$. Bad $\Gamma$
contribute at most $m^{-10}$ to this average. Now fix good $\Gamma$. In each
experiment, the output of the simulator is a function of the star record,
and by \cref{lem:simulation} it equals the transcript unless a stop occurs.
So, by \cref{lem:prob}(a,b) and \cref{lem:loads}, the conditional distance
given $\Gamma$ is at most
\[
  \Prob_{\mathsf y}[\text{a stop occurs}\mid\Gamma]
  +\Prob_{\mathsf n}[\text{a stop occurs}\mid\Gamma]
  +\frac{2B_0^2B_*}{R^2}+dL_*G_*^s .
\]
If a stop occurs, then either the first stop is not closure overflow, or it
is closure overflow. After averaging over $\Gamma$, the first case has
probability at most $C(q/(AK)+q/(KR))$ in each experiment, by
\cref{lem:screening}. The second case has probability at most
$B_0^2B_*/R^2+dL_*/R$, by \cref{lem:loads}. Altogether,
\begin{equation}\label{eq:middle}
  \dtv(E_{\mathsf y},E_{\mathsf n})
  \le m^{-10}+2C\Bigl(\frac q{AK}+\frac q{KR}\Bigr)
     +\frac{4B_0^2B_*}{R^2}+dL_*G_*^s+\frac{2dL_*}R .
\end{equation}
By \eqref{eq:visible-budgets}, the last three terms satisfy
\begin{align*}
  \frac{B_*}{R^2}
    &=O\left(\frac Rm+\frac{\log m}{R}\right),\\
  dL_*G_*^s
    &=O\left(\frac{R^{s+3}}{m^{s+2}}
               +\left(\frac Rm\right)^{s+1}\log m\right),\\
  \frac{dL_*}{R}
    &=O\left(\frac{R^2}{m^2}+\frac{\log m}{m}\right).
\end{align*}
As $s\ge1$, $R\log m\le m$, and $\log m\le R$, the secondary term
in the second line and both terms in the third are $O(R/m)$.
Inserting these bounds and \eqref{eq:middle} into \eqref{eq:tv-chain}
gives \eqref{eq:master}.
\end{proof}

\paragraph{Choosing the parameters.}
Recall
$A=t^{s+3}$, $R=\lfloor\gamma_0t^{s+2}\rfloor$, $K=AR$, and $m=dA$.
Take $q=\lfloor c_0KR\rfloor$ for a small $c_0\in(0,1)$. Then
\[
  \frac q{mR^2}\le\frac{c_0}{d},\qquad
  \frac q{KR}\le c_0,\qquad
  \frac q{AK}\le\frac{c_0R}{A}=o(1),\qquad
  \frac{R^{s+3}}{m^{s+2}}
       \le\frac{\gamma_0^{s+3}}{d^{s+2}}.
\]
The remaining terms of \eqref{eq:master} tend to zero, and all its side
conditions hold for large $t$. Choose $\gamma_0$ and then $c_0$ small
enough that the transcript distance is at most $1/10$ for all large $t$.
The sizes satisfy
\begin{equation}\label{eq:balance}
  n=\Theta(t^{2s+5}),\qquad
  KR=\Theta(t^{3s+7})
     =\Theta\left(n^{(3s+7)/(2s+5)}\right).
\end{equation}

\begin{proof}[Proof of \cref{thm:lower}]
Draw the input from $\mu=\tfrac12\Dyes+\tfrac12\Dno$.
By \cref{prop:lecam,lem:tv}, any deterministic algorithm with at most
$q$ queries guesses its world correctly with probability at most
$(1+1/10)/2=11/20$.

Turn an estimator into a guess by answering \yes{} precisely when
$\widehat C\le(2-\eta)U_m$. Whenever the estimator is correct and the
cost bound of \cref{lem:gap} holds, this guess is correct: in \yes{},
$\widehat C\le(2-\eta)\OPT\le(2-\eta)U_m$, whereas in \no{},
$\widehat C\ge\OPT\ge V_m>(2-\eta)U_m$. Thus any such estimator succeeds
under $\mu$ with probability at most
$11/20+e^{-\Omega(m)}<17/30$ for large $t$. Averaging over random bits
gives the same bound for randomized algorithms with at most $q$ queries.

\emph{All input sizes.}
Let $n_t$ be the size in \eqref{eq:lb-n}. Since $n_{t+1}/n_t\to1$, for
any sufficiently large $n$ choose the largest $t$ with $n_t\le n$ and
add $n-n_t$ publicly named nonterminals at distance $2$ from all other
points. The result is a metric and the optimum is unchanged. Indeed,
a new nonterminal of degree $j\ge2$ can be removed by joining its former
neighbors with a path of $j-1$ edges of weight at most $2$, while a
new leaf can simply be deleted. Queries involving these points have
fixed answers and give no information. The lower bound therefore
applies to every sufficiently large $n$, with $n=\Theta(n_t)$,
$\Theta(n)$ terminals, and
$\lambda=K^{-3}=\Theta(n^{-3})$.

\emph{Expected query complexity.}
Suppose an algorithm succeeds with probability at least $2/3$ on every
instance and uses at most $q/10$ queries in expectation on every
instance. Truncating it after $q$ queries loses at most $1/10$ success
probability, by Markov's inequality. Its success under $\mu$ would then
be at least $2/3-1/10=17/30$, contradicting the preceding bound.
Some instance therefore requires more than
$q/10=\Omega(n^{b_s})$ expected queries, with $b_s$ as in
\eqref{eq:balance}. Finally,
$b_s=3/2-1/(4s+10)$ tends to $3/2$, proving \cref{cor:exponent}.
\end{proof}

\section{Conclusion and AI Disclosure}
\label{sec:conclusion}

In this work we showed that the query complexity of better-than-2 estimation of metric Steiner tree cost is $\Theta(n^{3/2})$, up to subpolynomial factors.
The most interesting follow-up question is be to understand the frontier of query complexity versus estimability
for smaller approximation ratios.
Specifically, \cite{CKT23} showed that estimation within a factor better than $5/3$ is impossible in sublinear time,
whereas
the algorithms given in~\cite{CKT23,MRTV25} and this paper
give $(2-\eta)$-approximation for a very small constant $\eta$.
Is it still possible to get sublinear query complexity
while achieving a significantly better approximation ratio, say $1.9$?

\paragraph{AI Disclosure.}
The algorithm and the lower bound were discovered using a custom agentic harness developed by the authors,
which used GPT 6 Astra, Claude Opus 5, Gemini 3.8 Flash, Kimi K3, and MAI-Thinking-1. The main ideas behind both results can be largely attributed to GPT 6 Astra.
AI tools were also used for proofreading and polishing.
The authors assume all responsibility for the paper's content and its correctness.
GPT 6.1 Sol (in GitHub Copilot) was used to develop the Lean formalization.

\bibliographystyle{alpha}
\bibliography{ref}

\end{document}